\documentclass[11pt,preprint]{aastex}
\newcommand{\hst}{{\it HST}}
\newcommand{\hut}{{\it HUT}}
\newcommand{\iue}{{\it IUE}}
\newcommand{\parcsec}{\mbox{$\stackrel{\prime\prime}{\textstyle .}$}}
\newcommand{\kms}{\ifmmode {\rm km\,s}^{-1} \else km\,s$^{-1}$\fi}
\newcommand{\lsim}{\stackrel{\scriptscriptstyle <}{\scriptstyle {}_\sim}}
\newcommand{\gsim}{\stackrel{\scriptscriptstyle >}{\scriptstyle {}_\sim}}
\newcommand{\et}{\mbox{et~al.}\ }
\newcommand{\eg}{\mbox{e.g.,}\ }
\newcommand{\ie}{\mbox{i.e.,}\ }
\newcommand{\hb}{H$\beta$}
\newcommand{\Hbeta}{\ifmmode {\rm H}\beta \else H$\beta$\fi}
\newcommand{\oiii}{[O\,{\sc iii}]} 
\newcommand{\niv}{N\,{\sc iv}}
\newcommand{\civ}{C\,{\sc iv}} 
\newcommand{\heii}{He\,{\sc ii}}
\newcommand{\mgii}{Mg\,{\sc ii}}
\newcommand{\feii}{Fe\,{\sc ii}} 
\newcommand{\lam}{$\lambda$}
\newcommand{\Msigma}{\ifmmode M_{\rm BH} - \sigma \else $M_{\rm BH} - \sigma$\fi}
\newcommand{\lol}{$L_{\rm bol}/L_{\rm Edd}$}

\begin{document}

\shorttitle{Central Masses of Distant Active Galaxies and Quasars}
\shortauthors{Vestergaard \& Peterson}

\title{Determining Central Black Hole Masses in Distant Active
Galaxies and Quasars.\ II.\ Improved Optical and UV Scaling
Relationships.$^1$}

\author{Marianne Vestergaard\altaffilmark{2} and 
Bradley M.\ Peterson\altaffilmark{3}}

\altaffiltext{1}{ Based in part on observations made with the
NASA/ESA Hubble Space Telescope, obtained from the Data Archive at the
Space Telescope Science Institute, which is operated by the
Association of Universities for Research in Astronomy, Inc., under
NASA contract NAS 5-26555. }
\altaffiltext{2}{Steward Observatory, University of Arizona,
	933 N. Cherry Avenue, Tucson, AZ 85718.
	Email: mvestergaard@as.arizona.edu}
\altaffiltext{3}{Department of Astronomy, The Ohio State
        University, 140 West 18th Avenue, Columbus, OH 43210-1773.
        Email: peterson@astronomy.ohio-state.edu
}

\begin{abstract}
We present four improved empirical relationships useful for estimating the
central black hole mass in nearby AGNs and distant luminous quasars alike
using either optical or UV single-epoch spectroscopy. 
These mass-scaling relationships
between line widths and luminosity 
are based on recently improved empirical
relationships between the broad-line region
size and luminosities in various energy bands and are calibrated to the 
improved mass measurements of nearby AGNs 
based on emission-line reverberation mapping.
The mass-scaling relationship based on the \hb\ line luminosity allows 
mass estimates for low-redshift sources with strong contamination of the 
optical continuum luminosity by stellar or non-thermal emission, while that 
based on the \civ\,$\lambda$1549 line dispersion allows mass estimates in 
cases where only the line dispersion (as opposed to the FWHM) can be reliably 
determined. We estimate that the
absolute uncertainties in masses given by these mass-scaling
relationships are typically around a factor of 4.
We include in an Appendix mass estimates for all
the Bright Quasar Survey (PG) quasars for which direct reverberation-based
mass measurements are not available.
\end{abstract}


\setcounter{footnote}{3}

\section{Introduction}

A problem of current interest is determination of the mass function
of the central black holes in active galactic nuclei (AGNs) and quasars 
over the history of the Universe in order to determine 
how these black holes evolve with time. 
Unfortunately, measurement
of black hole masses by direct methods such as modeling
of stellar or gas dynamics requires high spatial resolution and
is thus limited to relatively nearby galaxies. Moreover, the
brightness of the AGN itself makes it extremely difficult to
observe suitable stellar absorption lines for dynamical studies within
the black hole radius of influence, and the complex gas dynamics in
AGNs frustrate attempts to disentangle the emission-line kinematics.
Megamaser dynamics have been used successfully to measure the
black hole mass in NGC 4258 (Miyoshi et al.\ 1995),
but this required particular fortunate circumstances
that do not seem to be generally realized. Thus, for AGNs and quasars,
the most promising method for measuring black hole masses
is reverberation mapping of the broad emission lines. The advantages
of this technique are that (a) it does not depend on angular resolution
and (b) it yields straightforward empirical relationships that 
provide effective secondary indicators that can be used to estimate
the masses of large numbers of AGNs and quasars based on 
single observations. The disadvantages of the technique are that 
(a) the accuracy of reverberation-based black hole masses are
fundamentally limited by our lack of knowledge of the detailed
structure and kinematics of the BLR and 
(b) it is observationally demanding.

Reverberation-based black hole masses are computed from the virial
equation
\begin{equation}
\label{eq:virial}
M_{\rm BH} = \frac{f\,R \Delta V^2}{G},
\end{equation}
where $R$ is the size of the region as estimated by
the mean emission-line lag $\tau$ (time delay relative to continuum
variations), i.e.,  $R = c\tau$, $\Delta V$ is the emission-line width
(preferably the width of the {\em variable} part of the 
emission line), and $f$ is a scale factor of order unity
that depends on the structure and geometry of the BLR.
Two lines of evidence suggest that these masses have
some validity:
\begin{enumerate}
\item In AGNs for which time delays have been measured
for multiple lines, there appears to be a virial relationship
between time delay and line width, i.e., 
$\tau \propto \Delta V^{-2}$ (Peterson \& Wandel 1999, 2000;
Onken \& Peterson 2002; Kollatschny 2003).
\item In reverberation-mapped AGNs for which host galaxy bulge
velocity dispersions $\sigma_*$ are available, the reverberation-based
masses $M_{\rm BH}$ are consistent with the $\Msigma$ relationship
seen in quiescent galaxies (Gebhardt et al.\ 2000;
Ferrarese et al.\ 2001; Onken et al.\ 2004; Nelson et al.\ 2004).
\end{enumerate}

Reverberation mapping also shows that there is a simple
relationship between the size of the BLR and the continuum
luminosity $L$ of the AGN of the form $R \propto L^\gamma$
(Kaspi et al.\ 2000; 2005). This is an important result,
not only because it constrains the physics of the BLR, but also because
it provides a secondary method of estimating the black hole
masses by using $L^\gamma$ as a surrogate for $R$
in eq.\ (\ref{eq:virial}). Since a single spectrum of
an object in principle yields both $L$ and a line
width $\Delta V$, we have a powerful tool for estimating
the masses of large populations of quasars.
Wandel, Peterson, \& Malkan (1999) carried out some preliminary
tests of this method using the \hb\ emission line.
Vestergaard (2002; hereafter Paper I) used
the \civ\,$\lambda1549$ emission line to probe 
much higher redshifts, up to $z \approx 6$. There have
been several other extensions of this methodology.
McLure \& Jarvis (2002) used the 
Mg\,{\sc ii}\,$\lambda2798$ emission line in a similar study.
Wu et al.\ (2004) suggested that recombination-line luminosities should be used
since they are a better measure of the ionizing continuum
that drives the line variations than the longer-wavelength
continuum, which may be contaminated by 
hard-to-quantify jet emission or
host-galaxy starlight, depending on the wavelength at
which the continuum is measured.

Since the original papers appeared, there have been a number
of significant developments that have led us to decide to revisit
the mass-scaling relationships based on the \hb\ 
and \civ\ emission lines. Specifically,
\begin{enumerate}
\item The reverberation-mapping database that provides
the fundamental calibration for the mass-scaling relationships has been
completely reanalyzed (Peterson et al.\ 2004). Of
particular relevance here is that some inadequate or poor
data were identified and removed from the database.
\item The reverberation-based masses have now been
empirically scaled to the quiescent galaxy black hole mass scale
through use of the \Msigma\ relationship
(Onken et al.\ 2004). The zeropoint of the
AGN \Msigma\ relationship was adjusted to that
of quiescent galaxies by determining a mean
value for the scale factor $f$.
\item The radius--luminosity ($R-L$) relationship
between the broad-line region size and continuum luminosity
has been updated based on the reanalyzed reverberation data 
(Kaspi et al.\ 2005), and new {\it HST} imaging of reverberation-mapped
AGNs enable us to correct 
for host galaxy contamination of the optical continuum luminosity 
measured from spectra (Bentz \et 2006).
\item Additional spectra of reverberation-mapped AGNs
have become available in the public domain, making
it possible to improve the calibration and better quantify 
the uncertainties of mass estimates based on 
single-epoch spectra.
\end{enumerate}
In contrast to Paper I, we have also adopted 
the current benchmark cosmology with
$H_0$ = 70 ${\rm km~ s^{-1} Mpc^{-1}}$,
$\Omega_{\Lambda} = 0.7$, and $\Omega_m = 0.3$; in Paper I,
we used  $H_0$ = 75 ${\rm km~ s^{-1} Mpc^{-1}}$, $q_0 = 0.5$, and
$\Lambda = 0$ to effect more direct
comparisons of quasar luminosities with those in previous
work. 

In the following,  we describe the data and the spectral measurements
(\S\ref{data}), perform the calibration of the single-epoch unscaled
mass estimates (\S\ref{Mcal}) and determine the inherent statistical
uncertainties of these relationships (\S\ref{errors}). 
In \S\ref{discussion}, we
briefly discuss (1) the improvements in the updated mass-scaling 
relationships and (2) the appropriateness of using the \civ\ emission
line for mass estimates.
Our main results are summarized in \S\ref{summary}.

Also, there has been a controversy over the use of the \civ\
emission line in particular (\eg Baskin \& Laor 2005)
and we wish to address that issue as well. 
In Appendix A, we scrutinize the data used by Baskin \& Laor
that led them to challenge the validity of \civ-based mass estimates
and we conclude that a more suitable selection of data largely 
removes the problems that they identified.
Finally, in Appendix B, we
provide a complete list of estimated masses for
all those quasars from the Bright Quasar Survey
(the Palomar--Green or ``PG'' quasars; Schmidt \& Green 1983)
for which reverberation-based masses are not available.

\section{Sample and Data \label{data}}

We base our study on the 32 AGNs for which reliable
reverberation-based mass estimates\footnote{Specifically, PG\,1211$+$143, 
NGC\,4593, and IC\,4329A were excluded from the sample of Peterson \et (2004)
on grounds of having unreliable or low-precision
reverberation-based mass estimates.} 
were calculated by Peterson et al.\ (2004); 
we hereafter refer to this sample as the ``reverberation sample.'' 
We obtain from independent sources emission-line widths and fluxes 
and continuum fluxes for these same objects and calibrate
mass-scaling relationships.
The 28 objects with optical spectra and the 27 objects with UV spectra 
that are used in this study are listed in Tables 1 and 2, respectively;
column (1) of both tables gives the commonly used name of the object,
and frequently used alternative names appear in column (2) of Table 1.
Column (3) gives the redshift of each object.

\subsection{Optical Data}

Optical spectral measurements for a large fraction of the
reverberation-mapped AGNs are available from large compilations
by Boroson \& Green (1992) and by Marziani et al.\ (2003). 
Both of these data sets are particularly suitable for
determination of the \hb\ emission-line width because
in both cases (a) the optical \feii\ emission was accounted for
by fitting the spectrum with a suitable template based on the spectrum 
of I Zw 1, and (b) an attempt was made to 
remove the \hb\ narrow-line component.
In addition to the line-width measurements, we use the
\hb\ equivalent widths from  Boroson \& Green (1992) 
and the continuum and \hb\ line flux densities from 
Marziani et al.\ (2003).

There is some overlap between the objects observed by Boroson \& 
Green and Marziani et al.:
specifically, of the 32 sources in the reverberation sample, 16 
were observed by Boroson \& Green and 28 were observed by Marziani 
et al. Fourteen of the sources were included in both studies,
although the two studies did not always provide the same type of
data.
Collectively, for 25 objects in the reverberation
sample, there are a total of 34 individual pairs of 
reliable FWHM(\hb) and 
$L_{\lambda}$(5100\AA) measurements, plus
pairs of FWHM(\hb) and $L$(\hb) measurements for
28 objects in the reverberation sample, as listed in Table 1.
These are the two optical data samples analyzed in this work. The
individual measurements comprising these samples are described below.

\subsubsection{Optical Luminosity Measurements}

\paragraph{Continuum Luminosities.}
Most of the monochromatic continuum luminosities $L_{\lambda}$(5100\AA) 
for the PG quasars 
{in the reverberation sample (14 out of 16) are }
computed from the specific fluxes and spectral indices
measured by Neugebauer \et (1987); in Paper I, a continuum slope of 
$\alpha_{\nu} = -0.5$, as is commonly adopted, was used instead.
As in Paper I, continuum measurements for Mrk\,110 and Mrk\,335 are 
based on the $B$-band photometry presented by Kellermann \et (1989) and
Schmidt \& Green (1983)
with corrections described by Schmidt, Schneider, \& Gunn (1995). 
Continuum luminosities {for 28 of the objects (mostly Seyfert galaxies)} are
based on the continuum flux densities given by Marziani \et (2003).

\begin{figure}[h]
\plottwo{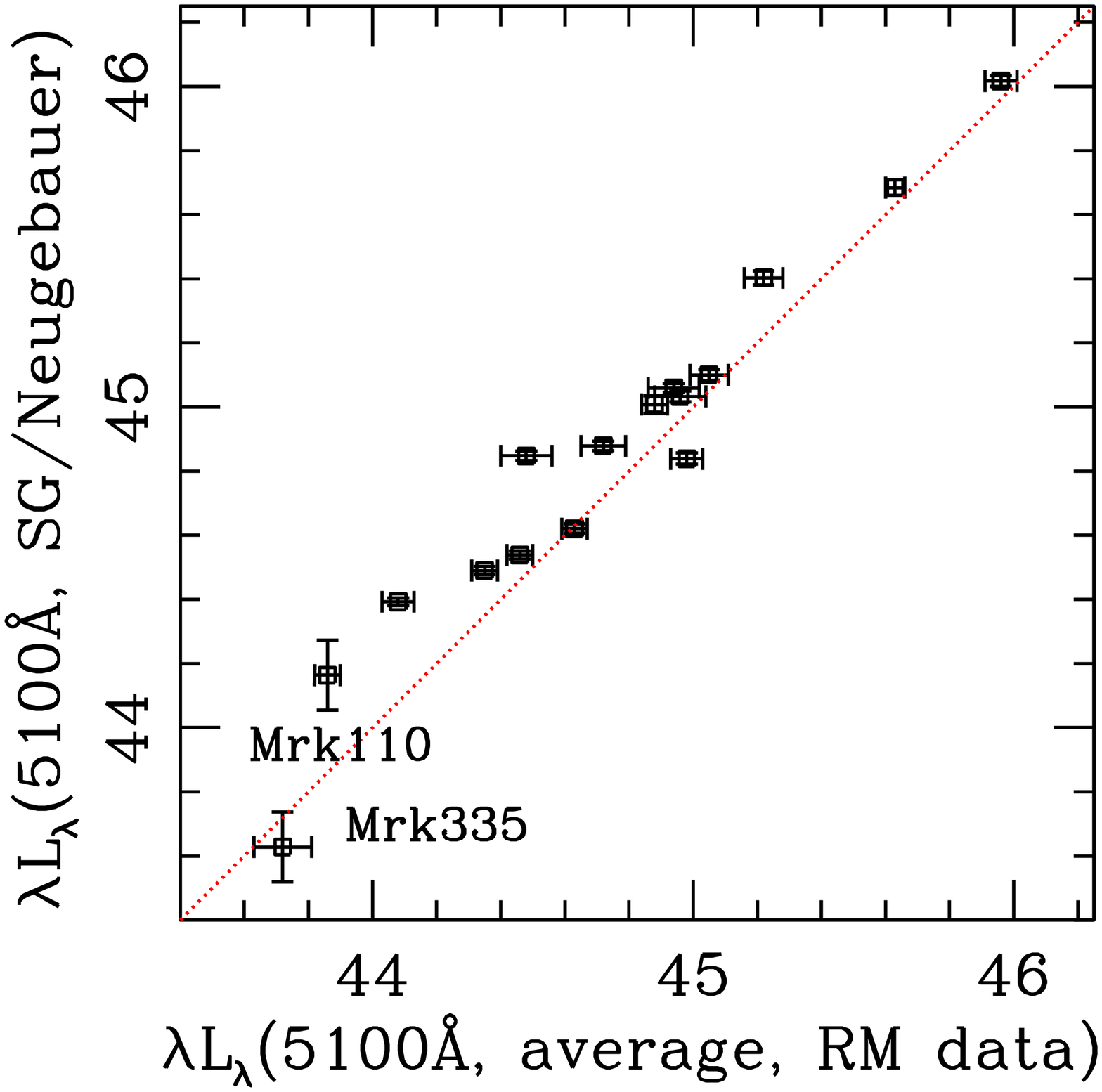}{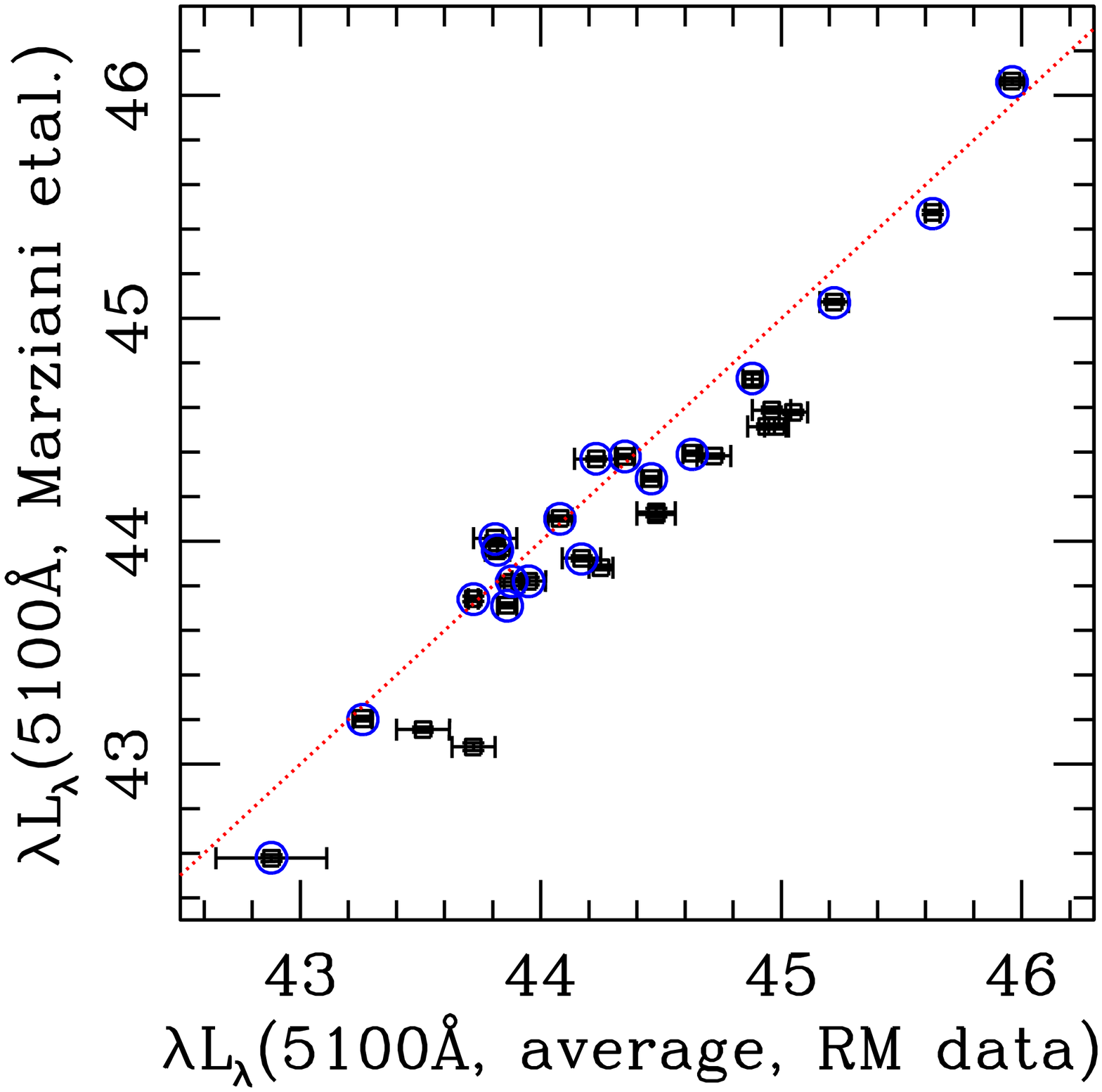}
\caption[]{
Comparison of the average $L_{\lambda}$(5100\,\AA) values determined from the 
monitoring data of Peterson \et (2004) with the single-epoch 5100\,\AA{} 
continuum luminosities $L_{\lambda}$(5100\,\AA) based on ({\it Left}) 
Neugebauer \et (1987) plus Schmidt \& Green's (1983) 
measurements of Mrk\,110 and Mrk\,335, and
({\it Right}) Marziani \et (2003).
The data points marked by (blue) circles in the right panel deviate less 
than 0.3\,dex from the average monitoring luminosity. These data points 
are selected for calibration of eq.\ (1).
The (red) dotted lines denote unity relationships.
\label{LumLum.fig}}
\end{figure}

In Figure~\ref{LumLum.fig}, we compare the single-epoch luminosities 
with the mean 
source luminosities of Peterson \et (2004). 
The Neugebauer \et  
luminosities are seen to scatter mostly within about 0.2--0.3\,dex of the mean 
monitoring luminosities. While the Neugebauer \et values tend to be slightly 
higher, the effect is minor ($\lsim$0.1\,dex) for many objects.  A similar 
offset was found and discussed in \S~4.2 of Paper~I,  and was also noted
by Maoz (2002).  A likely cause was 
considered to be slightly different absolute flux calibration scales, although
Maoz suggests that imperfect sky subtraction in the Neugebauer \et data 
may also be a source of error. The somewhat larger aperture 
(15\arcsec) used by Neugebauer \et compared to that used during the monitoring 
campaigns (typically 4\arcsec -- 5\arcsec) likely provides a sizable 
contribution to the offset.
For lower luminosity objects where the relative contribution of the host galaxy
is stronger, the luminosity difference is expected the largest, and indeed,
the largest deviations are seen
for sources with $\log L_{\lambda}{\rm (5100\,\AA)} \lsim 44.5$.
The systematic offset is small and indeed is well within
the envelope expected simply from source variability.
We adopt these luminosities without further correction.

The spectra of Marziani et al.\ (2003) are not necessarily expected to be
of photometric quality in part because the 
narrow slit (1\parcsec5 $-$ 2\arcsec) used in the observations was not always
aligned at the parallactic angle and in part because of variable sky conditions. 
Nevertheless, we find the luminosities based on these data to be 
generally consistent with the monitoring data to within the
factor 2 (0.3\,dex) or so allowed by source variability.
The spectra of the sources for which the Marziani et al.\ 
luminosities deviate by 0.3\,dex or less from the mean 
(monitoring) luminosities are included in this study.

A correction for Galactic extinction is applied to all the flux density 
measurements in the observed frame of reference for each object. 
The extinction values we use are those listed by 
Kaspi et al.\ (2005) and are based on the Galactic extinction curve of Cardelli, 
Clayton, \& Mathis (1989). 

\begin{figure}[h]
\plottwo{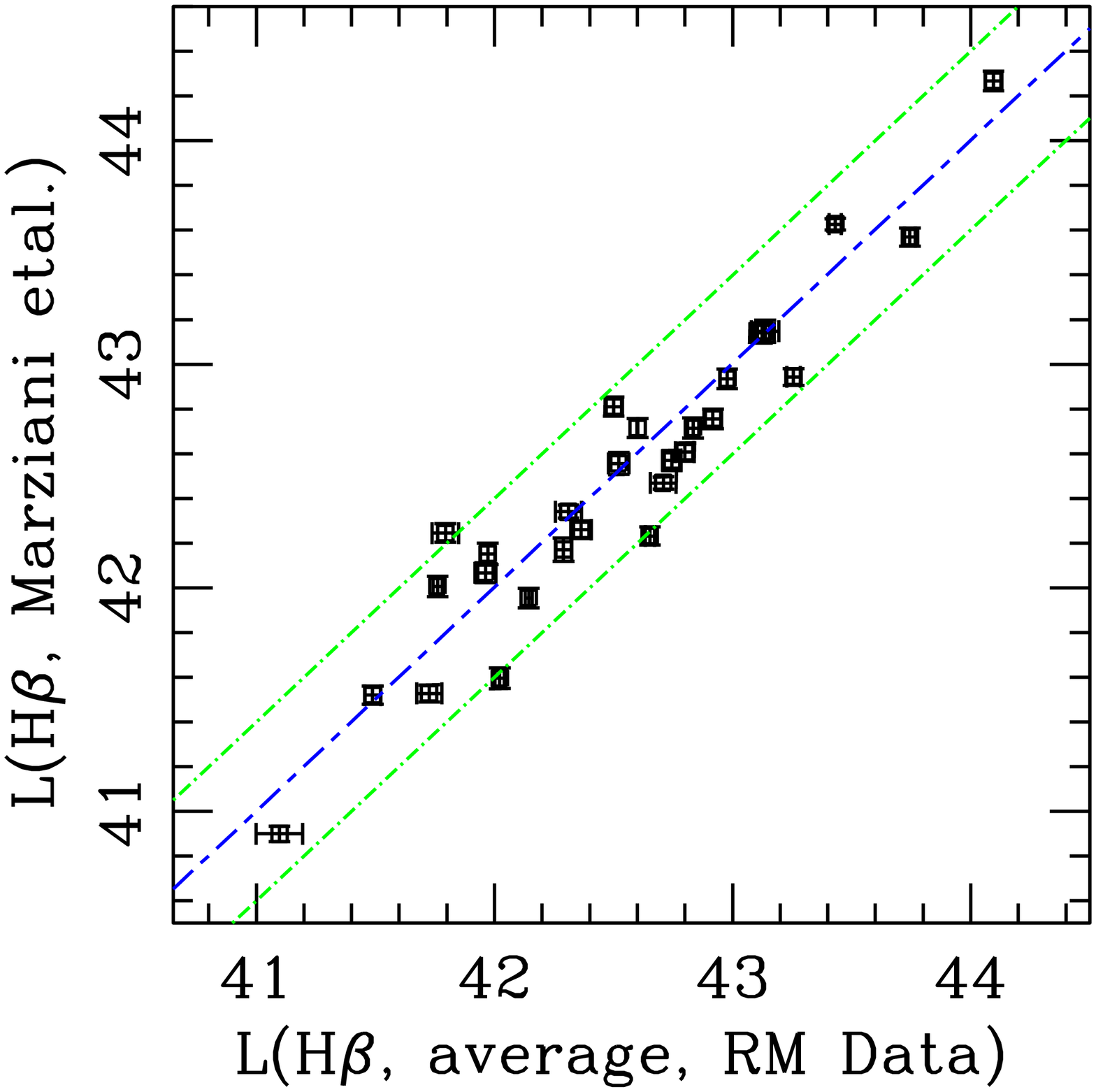}{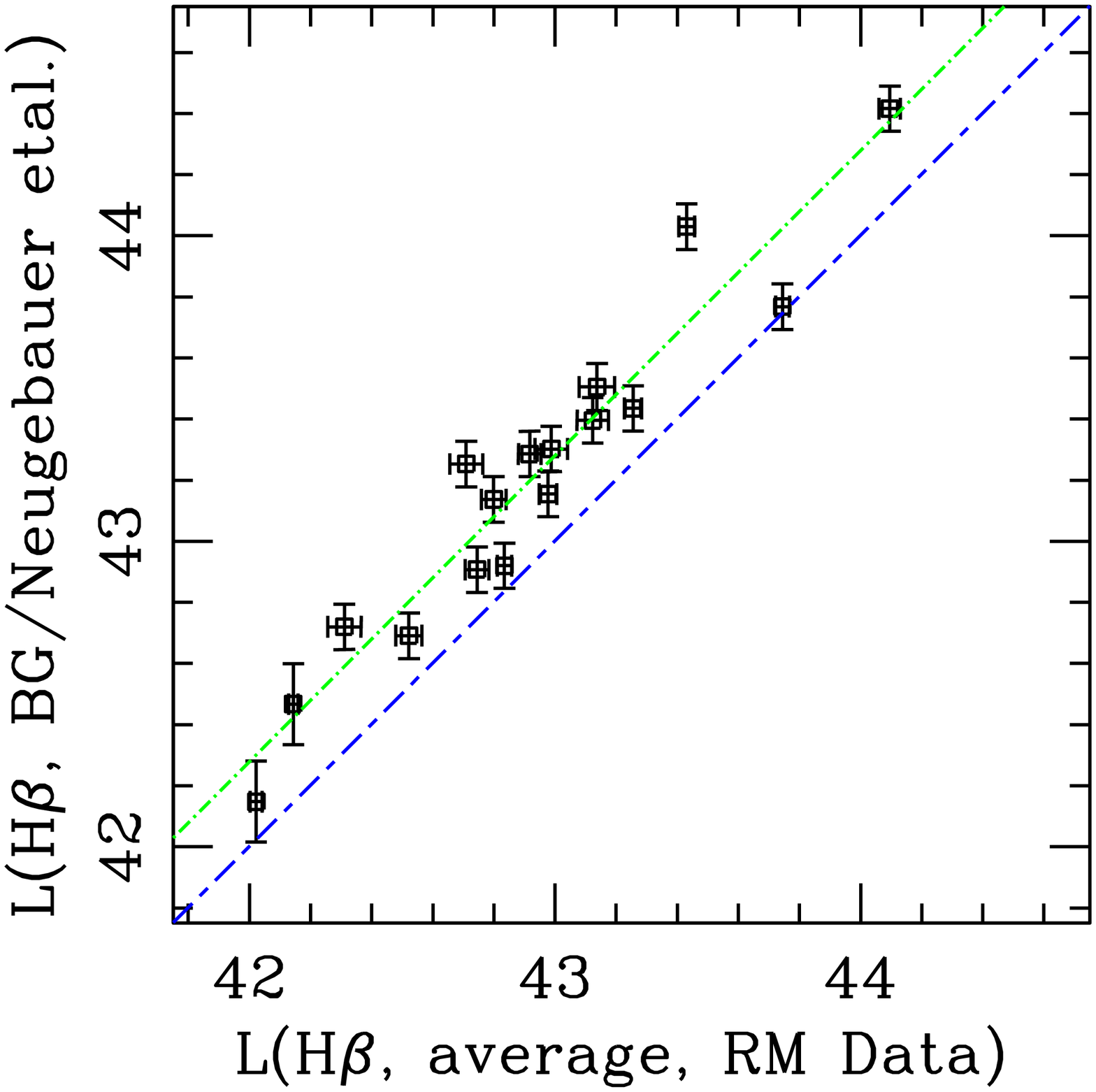}
\caption[]{
Comparison of the mean \Hbeta\ line luminosities $L(\rm H\beta)$ based on
monitoring data with the single-epoch $L(\rm H\beta)$ values determined 
based on ({\it left}) the Marziani \et (2003) data, and ({\it right}) the 
combination of Boroson \& Green (1992),
Neugebauer \et (1987), and Schmidt \& Green (1992) data.
({\it Left}) The Marziani \et line luminosities 
scatter well around a one-to-one relationship (blue short-long dashed line) 
and within approximately 0.4\,dex as indicated by the (green) dot-dashed lines.
The mean offset of the Marziani \et measurements is $-$0.04$\pm$0.21\,dex. Combined
with the even scatter around the monitoring values this indicates that the 
offsets are very likely to be due to variability (contrary to the Boroson \& 
Green values; right panel).
({\it Right})  The line luminosities obtained based on the Boroson \& Green 
and Neugebauer
\et measurements scatter evenly around a systematic offset of about 0.28$\pm$0.15\,dex
(green dot-dashed line) from a one-to-one relationship (blue short-long dashed 
line). Origin of this systematic significant offset is unknown. Therefore, these data 
are not included in the analysis.
\label{LineLum.fig}}
\end{figure}

\paragraph{Line Luminosities.}
We determined the emission-line luminosities of the 
broad-line component of \hb\ by multiplying the equivalent width
by the extinction-corrected monochromatic continuum
luminosity at the position of the \hb\ line. 
This luminosity was determined by converting the observed 4861\,\AA{}
flux densities provided by Marziani \et and by extrapolating the 
Neugebauer \et $L_{\lambda}$(4416\,\AA) and Schmidt \& Green 
$L_{\lambda}$(4400\,\AA) values to 4861\,\AA{} using the 
continuum slopes measured by Neugebauer et al. For the two objects with
Schmidt \& Green flux densities only, the average continuum slope of $-$0.2
measured by Neugebauer \et for the PG sample was adopted.
In the left panel in Fig.\ \ref{LineLum.fig}, 
we compare the values of $L({\rm H}\beta)$ 
determined from the Marziani \et (2003) measurements with the mean values
from the monitoring programs. We find good agreement between
these two sets of measurements, with scatter
consistent with intrinsic variability.
However, the line luminosities based on the Boroson \& Green (1992)
and Neugebauer \et (1987) data systematically overestimate 
the line luminosities relative to the values obtained from the
monitoring data by an average of 0.28$\pm$0.15\,dex, i.e., almost a factor of two
(Fig.~\ref{LineLum.fig}, right). We have been unable to identify
the origin of this offset, although we 
have been able to eliminate some of the more obvious potential sources
of error, such as the above mentioned extrapolation of the continuum 
from 4416\,\AA\ or 4400\,\AA\ to \hb{} (maximum effect less than 0.05\,dex).
Given this lack of agreement, we omit the data shown in the 
right diagram in Fig.\ \ref{LineLum.fig} from further consideration.

We give the extinction-corrected 5100\,\AA\ continuum luminosities 
$\lambda L_{\lambda}$(5100\AA) and line luminosities $L(\Hbeta)$ 
in columns (6) and (7), respectively, of Table~\ref{Opt_pars.tab}. 
Uncertainties in flux densities are taken directly from
Neugebauer \et (1987) when available, and errors are propagated 
assuming that the uncertainties in the spectral slopes 
are $\sigma (\alpha)= 0.2$ (see Paper~I).
In the case of the Marziani \et (2003) data, we
estimated the uncertainty in the flux from their 
quoted signal-to-noise ratios. In the case of the PG
quasars not observed by Neugebauer et al., the $B$-band 
magnitude uncertainty of 0.27\,mag (Schmidt \& Green 1983) is adopted. 

\subsubsection{\hb\ Line Width Measurements}

\begin{figure}[]
\plotone{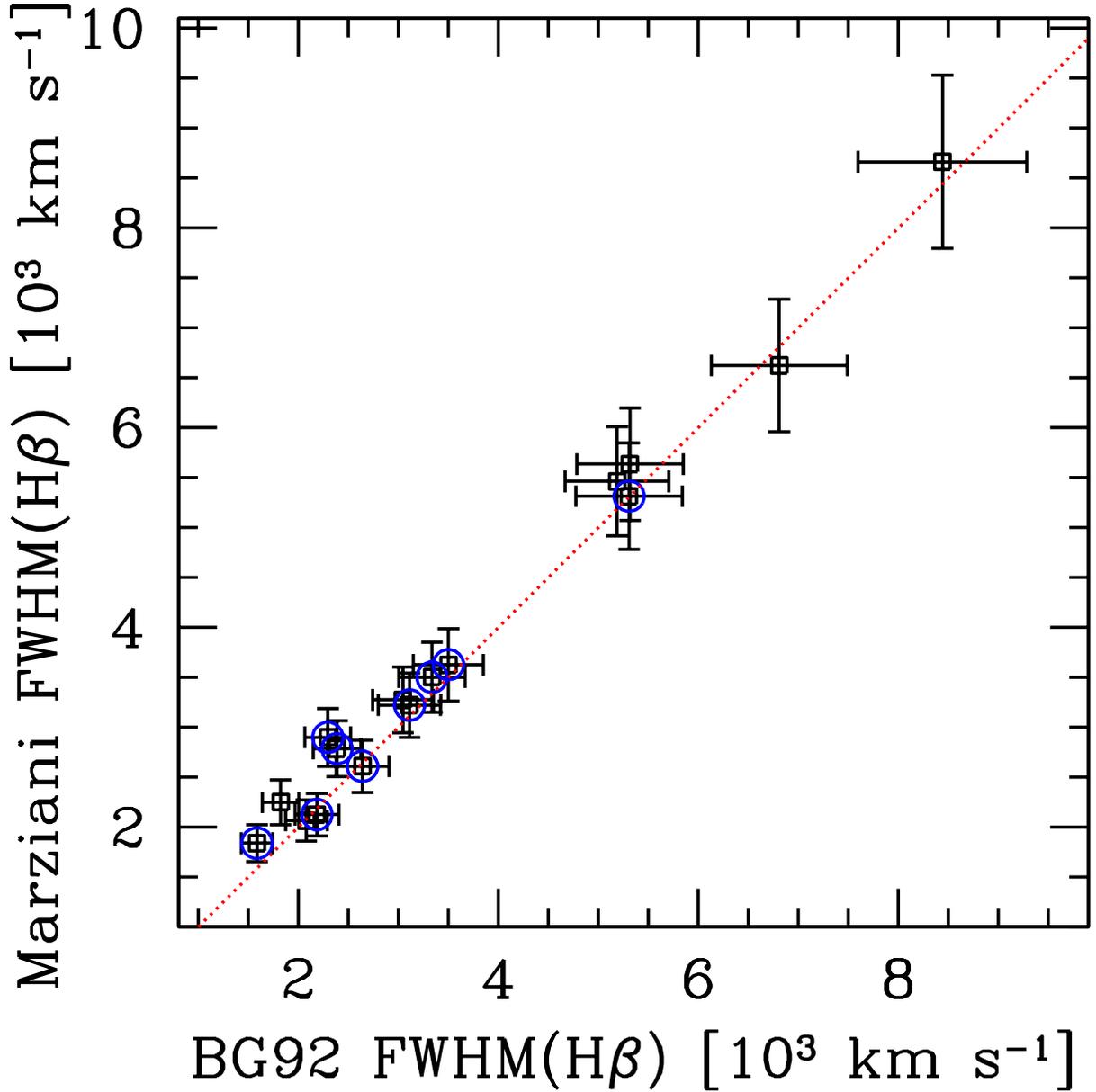}
\caption[]{
Comparison of the FWHM(\hb) measurements by Marziani \et (2003) and
Boroson \& Green (1992) for the objects common to both studies.
The dotted (red) line indicate a one-to-one relationship. The measurements
are essentially consistent to within the (expected) errors. The data
points marked by (blue) circles mark those objects for which the
Marziani \et 5100\,\AA{} continuum luminosity deviate by less than 
0.3\,dex from the average luminosity based on the monitoring data 
of Peterson \et (2004). This shows that a more deviant continuum 
luminosity does not affect the line widths for the Marziani 
\et sample.
\label{FWHM.fig}}
\end{figure}

For the PG objects,
we use the FWHM(\Hbeta) measurements of the broad-line component
from Boroson \& Green (1992), except for the corrected value  
for PG 1307$+$085 (FWHM = 5320\,\kms) from Laor (2000). Broad-component
widths were also taken  from Marziani \et (2003).
In Fig.\ \ref{FWHM.fig}, we show that for the objects 
common to both studies the FWHM measurements of the Boroson \& Green
and the Marziani \et studies are consistent within the errors. 
The circled data points are those for which the
Marziani \et luminosities differ from those measured during the
reverberation-mapping monitoring program by less than 0.3\,dex.
We are thus reassured that larger changes in luminosity do not
strongly affect the line width.

We note an important change from Paper~I is
that we now measure the width of the broad component only\footnote
{In Paper~I, we noted that the narrow-component of \Hbeta\ did not vanish
in the rms spectrum formed from the monitoring data in the
case of PG 1704$+$608 (Kaspi \et 2000); it is now clear that this 
result was spurious (see Boroson 2003; Peterson \et 2004).}.

We correct the line width measurements for spectral resolution
following the procedure of Peterson \et (2004). For the
Boroson \& Green (1992) measurements, we adopt a value of the
spectral resolution of FWHM = 7\,\AA.
The resolution of each of the
Marziani \et (2003) spectra is given in Table 1 of their paper.

The FWHM line width of \Hbeta, corrected for spectral resolution, is listed in 
column (4) of Table~\ref{Opt_pars.tab}.
Measurement uncertainties for the FWHM values are not quoted by 
Boroson \& Green (1992) or Marziani et al.\ (2003) and therefore a 10\% error is 
adopted, similar to our approach in Paper~I; this is likely a lower limit,
especially for low values of FWHM.

\subsection{UV Data \label{uvdata}}

We retrieved all the UV spectra of the 32 reverberation sample AGNs
that were available as of 2004 May from the {\em IUE, HUT}, 
and {\em HST} archives; one source (PG\,0844$+$349) has
no \civ\ data in the archives and another four objects 
(PG\,0804$+$761, NGC\,3227, PG\,1411$+$442, PG\,1700$+$518) were later omitted
(see \S~\ref{omitteddata}), leaving a final sample of 27 objects with UV data.
We processed and measured each spectrum to ensure that 
the data were treated in a consistent manner. 
{\em IUE} spectra of AGNs are of widely varying quality owing to 
the small size of the
telescope and limitations of the detector; 
quasar spectra, in particular,  can often be of low quality.
We therefore use {\em IUE} spectra of only the brighter AGNs and 
quasars in this analysis (Table~\ref{UV_pars.tab}).
Of the {\em HUT} spectra available in the MAST\footnote{Multi-Mission Archive
at Space Telescope; http://archive.stsci.edu/.} archives,
only those we deem to be of sufficient quality are selected.
The spectra are corrected as needed for photometric calibration following
the prescription at the \hut{} website\footnote{http://archive.stsci.edu/hut}.

The {\em HST} spectra were observed with a variety of grating settings
and we reprocessed only those covering the \civ\ region of each spectrum.
For PG\,0953$+$414 and 3C\,273 some of the FOS spectra were obtained in 
``rapid'' mode (45 and 384 spectra, respectively),
that is, all the individual subexposures of a given observation are 
preserved and a final combined spectrum is yet to be produced.
For these spectra obtained in ``rapid'' mode, the 
individual spectra are thus
extracted and combined, weighted by their variance spectra so to maintain 
their optimal signal-to-noise ratio; this procedure is analogous to that 
described by Horne (1986).
For each object in our sample,
spectra obtained on the same day that show no significant 
differences in continuum level or line profile are similarly 
combined by variance weighting. In a few cases, especially for {\em IUE} 
data, spectra spanning weeks to months with no noticeable difference in 
continuum level or line profile are similarly combined. 
Specific potentially problematic data sets that are 
flagged or omitted are briefly discussed in 
\S~\ref{omitteddata}.

The final sample of 27 objects for which the UV spectra retrieved 
from the public archives are used in this study are listed in 
column (1) of Table~\ref{UV_pars.tab}.
Column (2) lists the date of observation, column (3) gives
the redshift, and column (4) lists the telescope and
instrument used. 
Owing to the availability of multiple spectra for many
of the objects, we have a total of 85 individual estimates 
of black hole masses.

\subsubsection{UV Continuum Luminosities}

We fitted the rest-frame UV spectra  with a power-law continuum
in nominally line-free windows typically in the wavelength
ranges 1265 -- 1290\,\AA, 1340 -- 1375\,\AA,
1425 -- 1470\,\AA, 1680 -- 1705\,\AA, and 1950 -- 2050\,\AA, 
but slightly adjusted interactively for each individual spectrum 
in order to avoid broad absorption features or extended wings of 
emission lines; the fitting 
algorithm we employed automatically excludes strong, 
narrow absorption lines. We computed the
root-mean-square (rms) continuum flux density 
(hereafter simply called ``continuum rms'') relative to the best-fit 
continuum within the continuum windows.
In addition to the best-fit continuum, 
we also generated four extreme continua using the best fit continuum
and the continuum rms: a high-level continuum, 
a low-level continuum, and two continua with extreme blue and red slopes.
These four extreme continuum settings were used to estimate
uncertainties in the line measurements ascribable to the 
choice of continuum; this is typically the largest source of error in 
AGN spectral analysis, but is seldom well-quantified. The
uncertainty in the line-width measurement is of particular importance
in this context on account of the sensitivity of the inferred black hole
masses to the line width measurements.

\begin{figure}[h]
\plotone{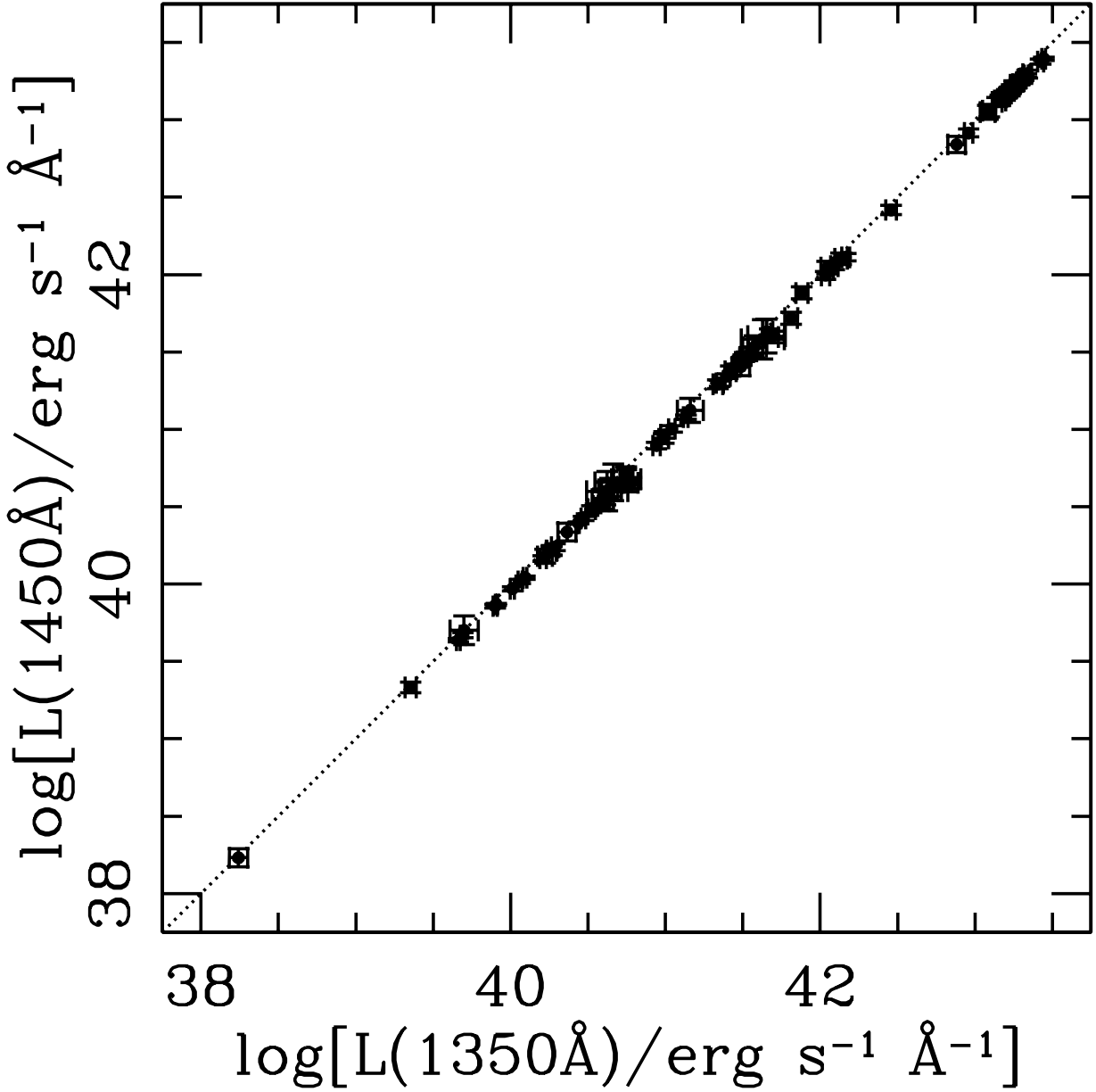}
\caption[]{
Comparison of the 1350\AA{} and 1450\AA{} continuum luminosities, $L_{\lambda}$(1350)
and $L_{\lambda}$(1450) measured for the UV sample described in \S~\ref{uvdata}. 
The two luminosities are so tightly correlated (scatter is
within the measurement uncertainties) that separate scaling relationships for
$L_{\lambda}$(1450) are not needed. The two luminosity measures are interchangeable
in the calibrated mass estimation (eqs.\ 5 and 6).
\label{L13L14.fig}}
\end{figure}

We compute 
monochromatic continuum luminosities $L_{\lambda}$(1350\,\AA)
and $L_{\lambda}$(1450\AA) for each spectrum from the observed continuum flux density
of the best-fit continuum. The uncertainty in the continuum
luminosity is determined from the continuum rms. The luminosities are 
extinction-corrected in the same manner as the optical luminosities
discussed earlier. Figure~\ref{L13L14.fig} shows that the 
$L_{\lambda}$(1350\AA) and $L_{\lambda}$(1450\AA)
values are essentially the same to within the uncertainties, so we
list only the single-epoch $\lambda L_{\lambda}$(1350\AA) values
in column (8) of Table~\ref{UV_pars.tab}.

\subsubsection{C\,{\footnotesize\it IV}  Line Width Measurements}

We measured two line-width parameters,
FWHM and line dispersion $\sigma_l$, using the methodology 
described by Peterson et al.\ (2004), in particular to deal with double-peaked
emission lines\footnote{Peterson et al.\ (2004) note a number of
advantages to using $\sigma_l$ rather than FWHM as the line-width
measure. We did not do this here with \Hbeta\ because we are relying
completely on published line measurements, which only give FWHM.
An \Hbeta\ mass-scaling relationship based on $\sigma_l$
will be explored in a separate paper.}.
For practical purposes, the line limits for
$\sigma_l$, the second moment of the emission-line profile, are
set to $\pm$10,000 \kms{} of the rest-frame line center, since
in every case  either the profile 
and the best-fit continuum merged between $\sim$9,000
\kms{} and 10,000 \kms{} or the \niv] \lam 1486 line contributes blueward of 
$-$10,000 \kms{} from the \civ{} profile center.  
For spectra with a strong contribution from \heii\ \lam 1640 the observed 
\civ{} red wing lies above the best-fit continuum.  A red wing limit of 10,000 \kms{} 
is adopted nonetheless based on the assumption that this provides a reasonable
compromise estimate of the red wing flux, some of which will extend into and blend 
with the blue wing of \heii{} which also may contribute a fraction of the flux in
the extreme red wing of \civ{}; the 10,000 \kms{} mark falls approximately halfway 
between the \civ{} and \heii{} lines. For those spectra, the line flux and line
dispersion may be slightly overestimated, especially if \feii{} emission is present
(Marziani \et 1996).

Some of the objects have absorption superposed on the \civ{} emission lines. 
We experimented with different ways of correcting for mild absorption in the profile
and concluded that a simple interpolation across the absorption does
a reasonably good job of approximating the (non-absorbed) line profile,
as long as the absorption is not too close to the line center.
We discarded from the sample objects
for which the absorption is so strong that it is clear that a simple 
interpolation is misleading (see \S~\ref{omitteddata}). 

Both of the line width parameters were measured relative to each of the five
continuum settings described above. We adopt the line
width measurements based on the best-fit continuum.
We then compute the difference between this measurement and the other four
and, to be conservative, adopt as the uncertainty the largest difference.

For consistency, 
we correct the line-width measurements for spectral resolution effects
in the same manner as the optical measurements (following Peterson \et 2004), even
though the corrections are typically insignificant for the
high-resolution \hst{} and \hut{} data.
We adopt a resolution of FWHM = 6\,\AA\ for the {\em IUE} spectra\footnote{The 
resolution of the 
SWP {\it IUE} spectra is listed on the MAST web page: http://archive.stsci.edu/iue/.}.
The MAST web page\footnote{http://archive.stsci.edu/hut} gives 
FWHM = 3\,\AA\ for {\it HUT1} spectra and 2\,\AA\ at 1600\,\AA\ for the {\it HUT2} 
spectra.  Based on the {\it HST} FOS 
Handbook\footnote{http://www.stsci.edu/hst/HST\_overview/documents},
the approximate resolution 
of both pre-costar and post-costar spectra were estimated for each of the
G130H, G190H, and G270H gratings.
These estimates are generally
consistent with an average FOS resolution of 1.9\,\AA\ (across the three grating 
spectra) adopted by Peterson \et  The {\it HST} GHRS spectral resolution
was likewise estimated to be 0.65\,\AA\ for the G140L grating.  The spectral 
resolution of {\it HST} STIS spectra vary with the specific instrument 
configuration. The resolution adopted for each UV spectrum is listed in 
column (5) of Table~\ref{UV_pars.tab}. The resolution-corrected FWHM(\civ)
and $\sigma_l$(\civ) values are listed in columns (6) and (7), respectively.

\subsubsection{Flagged and Omitted Data Sets \label{omitteddata}}

\paragraph{Flagged Data.}
Data sets that we consider to be of questionable or marginal quality are 
flagged so to check whether they distribute differently than the 
higher-quality data.
The flagged data sets are marked in column (11) of Table~\ref{UV_pars.tab}.
The three \iue{} spectra of Mrk\,79 are of borderline quality. The NGC\,3516 
\hst{} spectrum of 1996 November 28 exhibits a large deviation in the unscaled
mass estimate ($v^2 R \propto {\rm FWHM}^2 L^{\gamma}$, where $\gamma$ is the slope of 
the $R - L$ relationship) from the remaining data sets for unknown reasons.  
The PG\,0026$+$129 \hst{} spectrum has an odd appearance and this data point is
an outlier relative to the cluster of PG quasars. As there is only one spectrum
available of this source, and we have no independent means of assessing
the quality of these data, we leave this data point in the database,
but flag it.
The \iue{} spectrum of PG\,1617$+$175
is of marginal quality, but the large uncertainties we find seem to be
commensurate with the data quality.
Given the low number of 
luminous quasars in our sample, this data point is not omitted, but flagged.

\paragraph{Omitted Data.}
For PG\,0804$+$761, only \iue\ data are available, but the quality is 
very poor and this target was thus excluded. 
In spite of the good quality of the \hst{} spectrum of NGC\,3227 (Crenshaw
\et 2001), these data are omitted from the analysis owing to the strong 
internal reddening in this source, which cannot be reliably corrected and which
affects the spectral measurements (see also Kaspi \et 2005).
The \iue{} spectra of NGC\,4051 are omitted as the spectral resolution is too
low for objects with such strong absorption features;
the \iue{} line profiles are not
representative of the intrinsically emitted profiles.
We use the \hst{} echelle spectrum of NGC 4051 
(Collinge \et 2001) instead. For NGC\,4151, the \hut{} spectrum 
of 1990 December 8 is 
contaminated by a high background, rendering the data unusable. 
The 2000 May 28 \hst{} spectrum was obtained during a very 
low-luminosity state of
NGC\,4151. At this flux state the low resolution of the G140L grating does not 
allow the \civ{} profile to be well defined owing to the strong absorption in 
the blue profile wing. The \hut{} spectra, and the 1992 April, 1992 May, and 
1995 December 18 \iue{} spectra of 3C\,273 are of insufficient quality.
While PG\,1411$+$442 was observed with both \iue{} and \hst,
strong absorption centered on the \civ{} profile render the
data unusable for our purposes.
Similarly, PG\,1700$+$518 is excluded as it is a broad absorption-line quasar.

\section{Calibration of Single-Epoch Mass Estimates \label{Mcal}}

As in Paper~I, we calibrate the single-epoch mass estimate
by using the reverberation-based mass measurements. 
We compute for each spectrum (i.e., for each single epoch) a measure of the 
unscaled mass $\mu$ defined as the product 
\begin{equation}
\mu = 
  \left( \frac{\Delta V}{1000\,{\rm km~s^{-1}}} \right)^2 ~
  \left( \frac{\lambda L_{\lambda}}{10^{44} \rm ~erg~s^{-1}}\right)^{\gamma} 
\label{muLlam}
\end{equation}
where $\Delta V$ is the line-width measurement (either FWHM or $\sigma_l$)
for either \civ\,$\lambda1549$ or \Hbeta, $\gamma$ is the slope of
the relevant $R-L$ relationship, and
and $L_{\lambda}$ is the continuum luminosity (at either 1350\,\AA\ or
5100\,\AA). We will also use the broad component of \Hbeta\ itself as
a luminosity measure, i.e., 
\begin{equation}
\mu = 
   \left(\rm \frac{FWHM(H\beta)}{1000~km~s^{-1}} \right)^2 ~
   \left( \frac{L(\rm H\beta)}{10^{42} \rm ~erg~s^{-1}}\right)^{\gamma}.
\label{muLHb}
\end{equation}
The unscaled mass estimate $\mu$ should be proportional to the black hole mass
(eq.\ \ref{eq:virial}) obtained by reverberation mapping $M_{\rm BH}$(RM).
We thus first check this assumption, because if it holds, 
the problem of calibrating 
the unscaled masses reduces to a simple relationship with only 
one degree of freedom,
\begin{equation}
\label{eq:mass_scale}
\log M_{\rm BH}{\rm (RM)} = \log \mu + a.
\end{equation}
Following Paper~I, the constant zero-point offset
$a$ between the unscaled mass estimate and the black hole mass is thus 
the weighted average of the difference between these two measures 
obtained for each spectrum of each object.

We use two independent algorithms, FITEXY (Press \et 1992) and BCES 
(Akritas \& Bershady 1996), in our regression analysis. 
Since the error bars in the reverberation-mapping mass estimates are
often asymmetric, we adopt the same method as Kaspi \et (2005) of
using the error value for a given point that is in the direction of 
the best-fit line.  This requires a few iterations 
of the regression analysis and the error selection 
(i.e., either the positive or the negative error). 
We incorporate intrinsic scatter in the FITEXY relationships 
in the same way as Kaspi \et (2005).

\subsection{Radius -- Luminosity Relationships}

Kaspi \et (2005) recently updated the $R-L$ relationships
for the Balmer lines based on the revised and improved reverberation
database of Peterson \et (2004). 
Kaspi \et also used the same fitting algorithms that we use
here, which give slightly different results from one another on account of
slightly different underlying assumptions. Since both methods
have merit and one is not obviously more appropriate than
the other, we form a weighted average of the 
{slopes $\gamma$ } 
obtained from the FITEXY and BCES regressions. In each case, we 
conservatively adopt a final uncertainty in the exponent of 
$\Delta \gamma = 0.06$, which is typical of the maximum
differences in the index $\gamma$ yielded by the two
algorithms and slightly larger than the typical uncertainty
resulting from the individual algorithms and the errors on the
weighted mean slopes (i.e., $\sigma(\gamma) \approx$ 0.03--0.05).

For the relationship between $\Hbeta$ radius and the optical 
continuum, we use the Kaspi \et FITEXY and BCES slopes 
based on each individual reverberation measurement of the 
\Hbeta\ line for all AGNs in the sample,
and modified by Bentz et al. (2006) to include 
corrections for host galaxy stellar light
contribution to the $L_{\lambda}$(5100\AA) values for most
of the low-luminosity AGNs in the sample,
i.e., $ R \propto L_{\lambda}(5100{\rm \AA})^{0.50\pm0.06}$.
Use of the alternative bases (e.g., the results based
on averaging all results for a single object) changes the
results only slightly.

For the relationship between the \Hbeta\ radius and
the \Hbeta\ broad component luminosity, we use 
$R \propto L(H\beta)^{0.63\pm0.06}$,
based on the Kaspi \et (2005) results as explained above.

For the relationship between  the \Hbeta\ radius and 
the UV continuum luminosity, we use 
$ R \propto L_{\lambda}(1350{\rm \AA})^{0.53\pm0.06}$,
and note that use of $L_{\lambda}$(1450\AA) 
instead of $L_{\lambda}$(1350\AA) gives an identical
result. We also note that $\gamma = 0.53$ is a distinctly shallower slope than
we used in Paper~I ($\gamma = 0.7$), 
but is much closer
to that expected for a photoionized BLR (see below).

Selecting a slope to use for the \civ\ $R-L$ relationship
is more difficult because there are so few actual
measurements of the \civ\ response time and most of
these measurements are over a very narrow range in
luminosity. Recently, however, Peterson \et (2005) measured
the \civ\ response in NGC~4395, the least luminous
known Seyfert 1 galaxy. This result shows that 
the size of the \civ\ emitting region is about as
expected if the $R$(\civ) -- $L_{\lambda}$(1350\,\AA) 
relationship has a slope similar to that of 
$R$(\Hbeta) -- $L_{\lambda}$(1350\,\AA).
Peterson \et find that the slope of
\civ\ $R - L$ relationship to  be $\gamma = 0.61 \pm 0.05$,
although this result is based heavily on 
the NGC\,4395 reverberation measurement\footnote{This
value of the slope of the \civ\ $R - L$ relationship
appears in an erratum to this paper.}.
The \civ\ slope  is generally consistent with the
the slope of the 
$R$(\Hbeta) -- $L_{\lambda}$(1350\,\AA) relationship,
$\gamma = 0.56 \pm 0.05$ (Kaspi et\ 2005), as well
as the relationship between 
$R$(\Hbeta) and the starlight-corrected
optical luminosity $L_{\lambda}$(5100\,\AA),
$\gamma = 0.54 \pm 0.04$ (Bentz \et 2006).

\subsection{Optical Mass Relationships \label{optcal}}

\begin{figure}[ht]
\plottwo{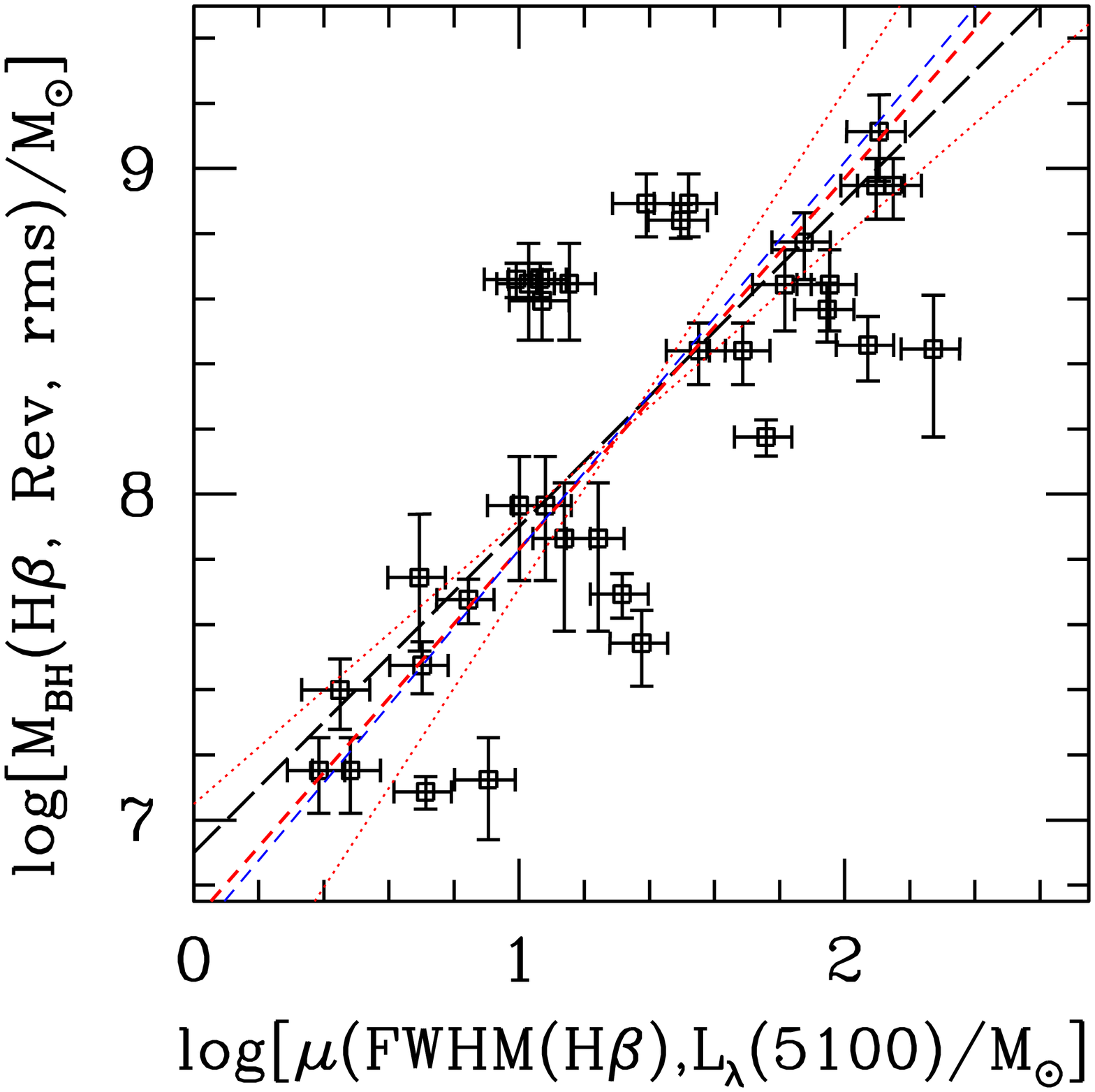}{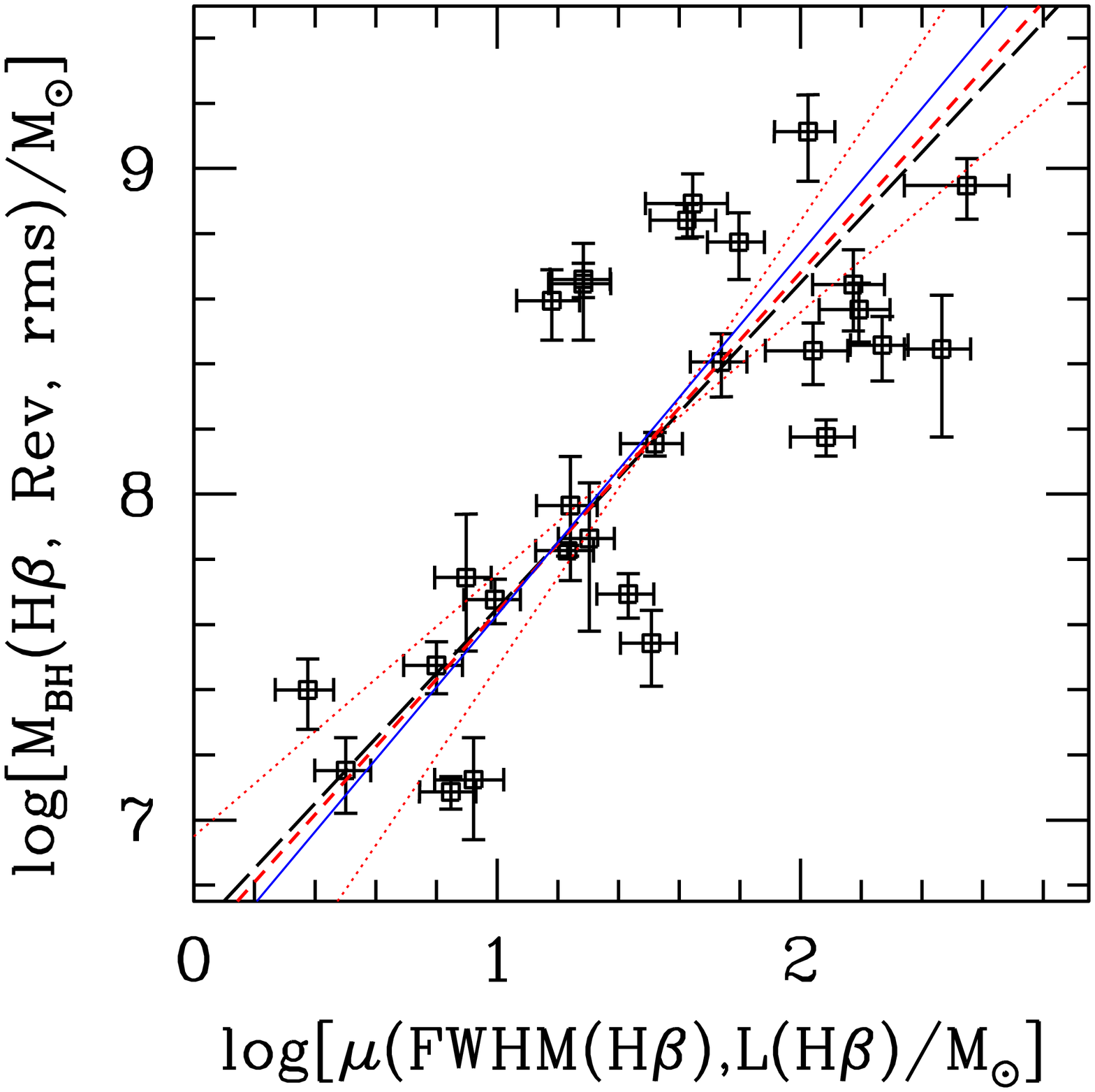}
\caption[]{
Distribution of normalized single-epoch unscaled mass estimates based on optical 
data with the reverberation mapping black hole masses. 
({\it Left}) Unscaled mass estimates based on FWHM(\hb) and the continuum luminosity 
$L_{\lambda}$(5100\AA). 
({\it Right}) Unscaled mass estimates based on FWHM(\hb) 
and the \hb{} line luminosity
$L(\rm H\beta)$. 
In both cases are the slopes consistent with unity (within the errors).
{\it Symbols:} The (black) long-dashed line shows a slope of 1.0, the 
(red) short-dashed line denotes the BCES bisector, the (red) dotted lines 
show the BCES(Y$|$X) and BCES(X$|$Y) fits, and the (blue) short-long dashed 
line represent the FITEXY fit. 
\label{M-Mopt.fig}}
\end{figure}

There are two luminosity surrogates for $R$ in the optical regime,
the 5100\AA{} continuum luminosity $L_{\lambda}$(5100\,\AA) and
the \Hbeta\ line luminosity $L(\Hbeta)$. The associated unscaled 
masses $\mu$[FWHM(\Hbeta), $L_{\lambda}$] and $\mu$[FWHM(\Hbeta), 
$L$(\Hbeta)] are computed according to eqs.\ (\ref{muLlam}) and
(\ref{muLHb}), respectively.
In Fig.\ \ref{M-Mopt.fig}, we compare directly 
our single-epoch unscaled mass estimates with the calibrated black hole 
masses determined by reverberation mapping, and we perform
a regression analysis on these data, the results of which 
are given in Table~\ref{regres.tab}. 
In the case of the BCES regressions, we list only the bootstrapped bisector 
slopes, intercept, and related errors as these results are relatively 
insensitive 
to outliers. A few thousand realizations have been made
in each regression analysis. For the FITEXY regressions, we apply
for convenience an equal amount of intrinsic scatter to both the
masses from reverberation mapping $M_{\rm BH}$(RM) and the single-epoch 
unscaled mass estimates $\mu$.
The level of scatter needed to obtain a reduced $\chi^2$ value of unity
is a little higher ($\sim$50\%) than obtained by Kaspi \et (2005) for the
$R-L$ relationships ($\sim$40\% -- 45\%), but is consistent.
The increased level of scatter obtained for the mass relationships is
entirely expected because single-epoch spectra are not 
necessarily obtained in an average state, which the reverberation mapping 
data somewhat represent. 
Both unscaled mass estimates are strongly correlated with the black hole
masses with a regression slope consistent with a value of 1.0 to within
the quoted uncertainties.  The scatter in the relationships is similar.

The excellent correlation between $M_{\rm BH}$(RM) and $\mu$ justifies the
use of eq.\ (\ref{eq:mass_scale}), thus requiring only that 
we establish the offset between these two quantities.
The results are listed in Table~\ref{zeropt.tab}.
The calibrated single-epoch black hole mass estimates based on the
unscaled masses $\mu$[FWHM(\Hbeta),$L_{\lambda}$] or 
$\mu$[FWHM(\Hbeta),$L$(\Hbeta)],
are listed in columns (9) and (10), respectively, of Table~\ref{Opt_pars.tab}.
The reverberation-based black hole mass measurements
used to effect this calibration are listed in column (11).

\subsection{UV Mass Relationships \label{uvcal}}

In the UV regime, we have two measures of the line width, FWHM 
and $\sigma_l$. We therefore have two separate sets of 
$\mu$ values to calibrate. The zero-point of these unscaled mass 
estimates is not expected to be the same
(\eg Onken \et 2004; Peterson \et 2004). In addition, the sample of UV
measurements is so large that it is worth testing whether we get a 
significantly different result if the weighted mean of the individual 
measurements obtained at different epochs are analyzed instead. This
approach is similar to that adopted by Kaspi \et (2005).
When computing the mean weighted by the measurement uncertainties, we
exclude the few (7) entries that are based on mean data of monitoring campaigns
{(marked in Table~2)} owing to the unnecessary complications involved in computing 
the true
weighted mean if they are included. The unscaled mass estimates based on the
mean monitoring data are included in the analysis as seven separate data points;
hence the full sample of weighted averages counts 34 entries.

\begin{figure}[ht]
\plottwo{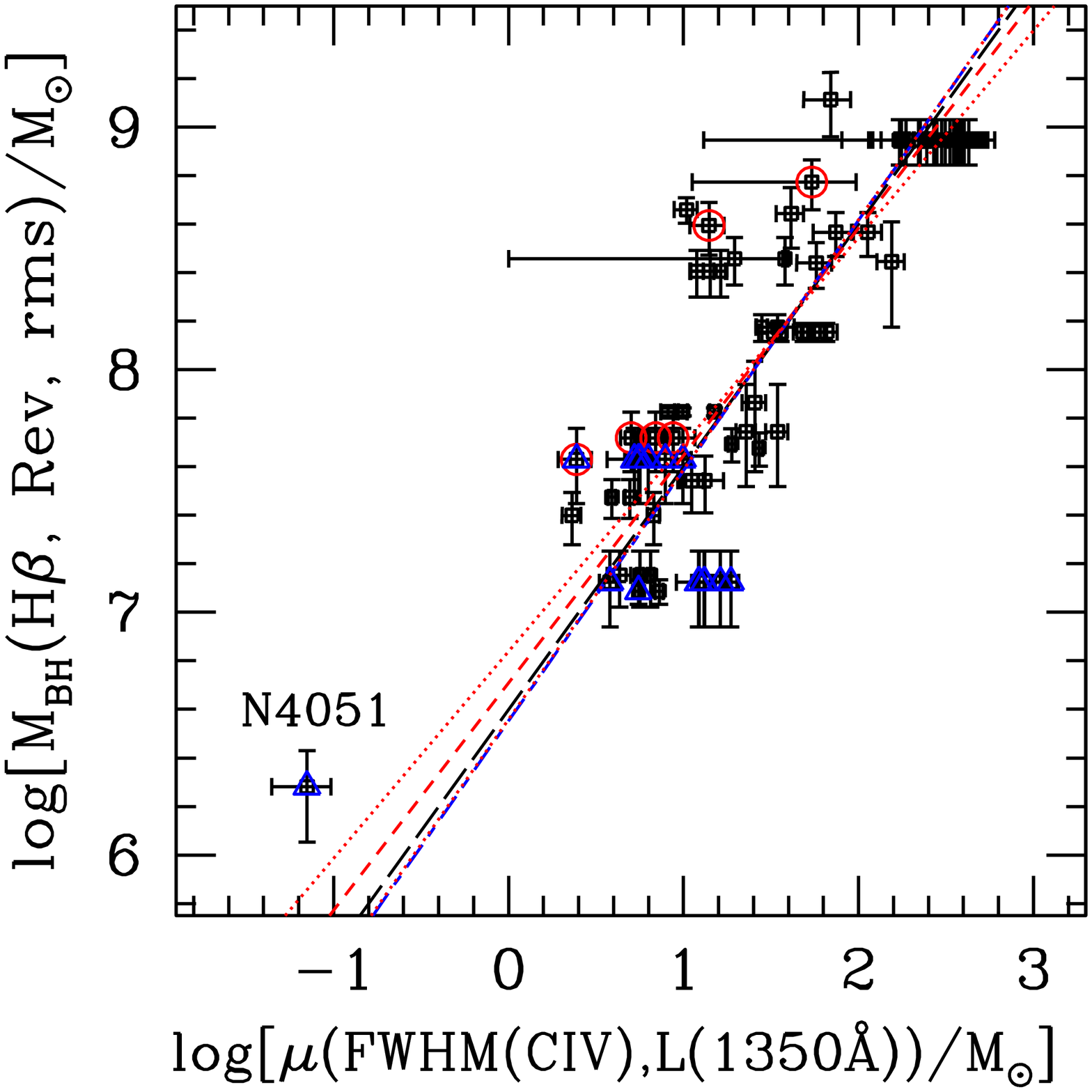}{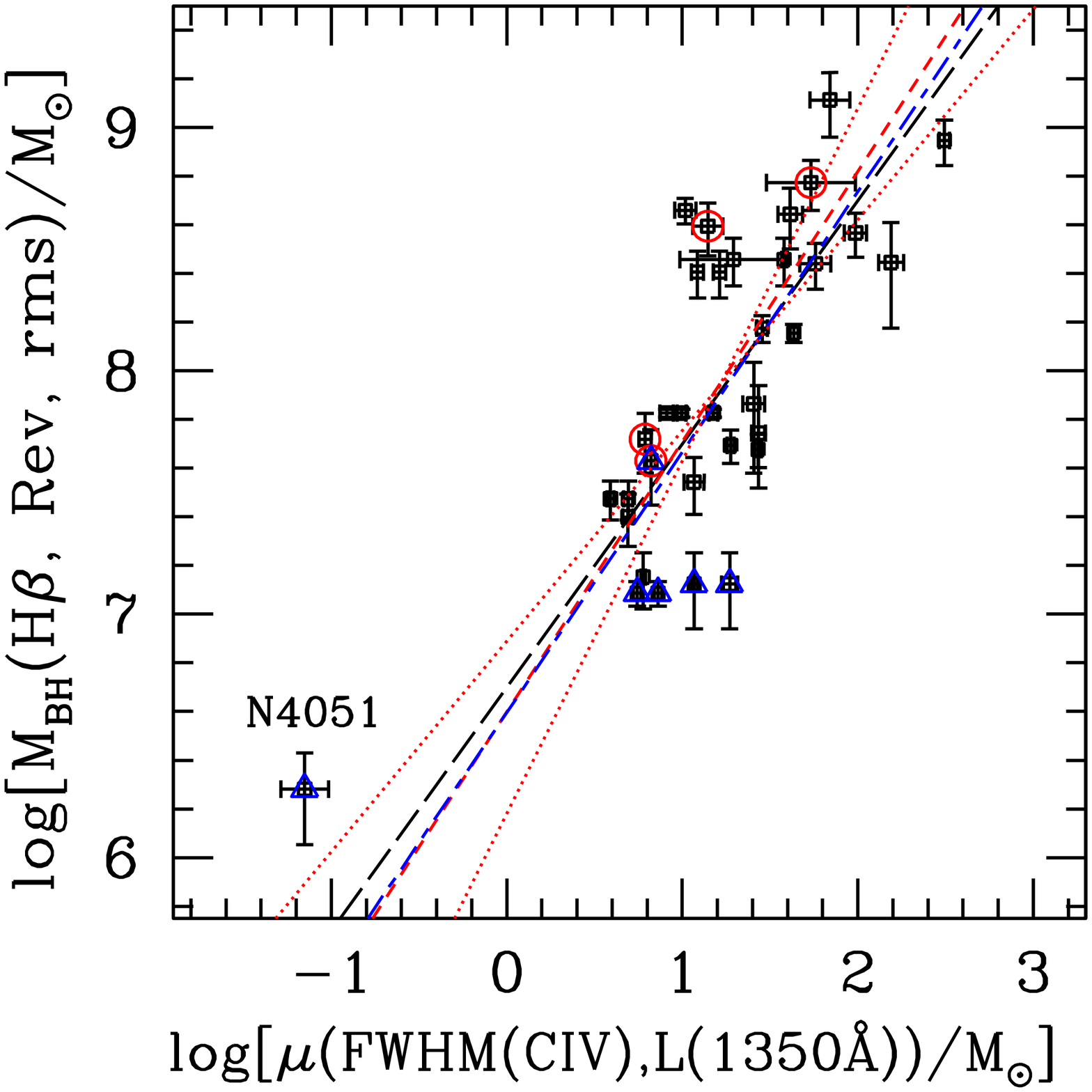}
\caption[]{ 
Distribution of normalized single-epoch unscaled mass estimates $\mu$ 
based on UV data, 
specifically the FWHM(\civ) and $L_{\lambda}$(1350\AA), with the 
reverberation mapping black hole masses: 
({\it Left}) All individual measurements for each object in the sample;
({\it Right}) the weighted mean $\mu$ of individual (non-monitoring) measurements of 
a given object and the weighted mean of monitoring data when available.
In both cases are the slopes consistent with unity (within the errors).
{\it Symbols:} (red) encircled data points denote spectra flagged for
being of borderline quality, and (blue) triangles mark spectra which
were corrected for mild absorption in the \civ{} profile. 
See Fig.~\ref{M-Mopt.fig} for line codes.
\label{Muvfw-Mrms.fig}}
\end{figure}

In Fig.\ \ref{Muvfw-Mrms.fig}, we compare the unscaled mass estimates
$\mu$[FWHM(\civ), $L_{\lambda}$(1350\AA)] (hereafter $\mu$[FWHM(\civ)]; 
eq.\ \ref{muLlam})
determined from the full sample of individual entries 
(left panel) and the sample 
of weighted averages (right panel) with the reverberation-based masses.
The low-luminosity Seyfert NGC 4051 is labeled. Flagged objects
(\S\ref{omitteddata}) are marked 
by red circles and objects with mild absorption corrected for in the 
\civ\ line profile are marked with blue triangles. In both cases, the 
unscaled mass correlates well with the black hole mass. In addition, 
neither the flagged nor marked objects are conspicuous outliers, and
therefore none of these measurements will be omitted from the analysis. 
These regression results are also listed in Table~\ref{regres.tab}. 
We note in particular that the relationship between $\mu$[FWHM(\civ)]
and the reverberation-based masses is consistent with a linear
relationship. The uncertainty in the fitted slope is higher for
the sample based on weighted means, and the estimated intrinsic
scatter increases from 42\% to $\sim$52\%.
Given the somewhat isolated position of NGC 4051, we repeated 
the regression analysis 
with this source omitted to test the sensitivity of the 
regression fits to this data point. 
The sample of weighted means is more sensitive to whether or not NGC 4051 is 
included, but in neither case does the slope deviate from a value of
unity by more than  $2\sigma$. 

\begin{figure}[ht]
\plottwo{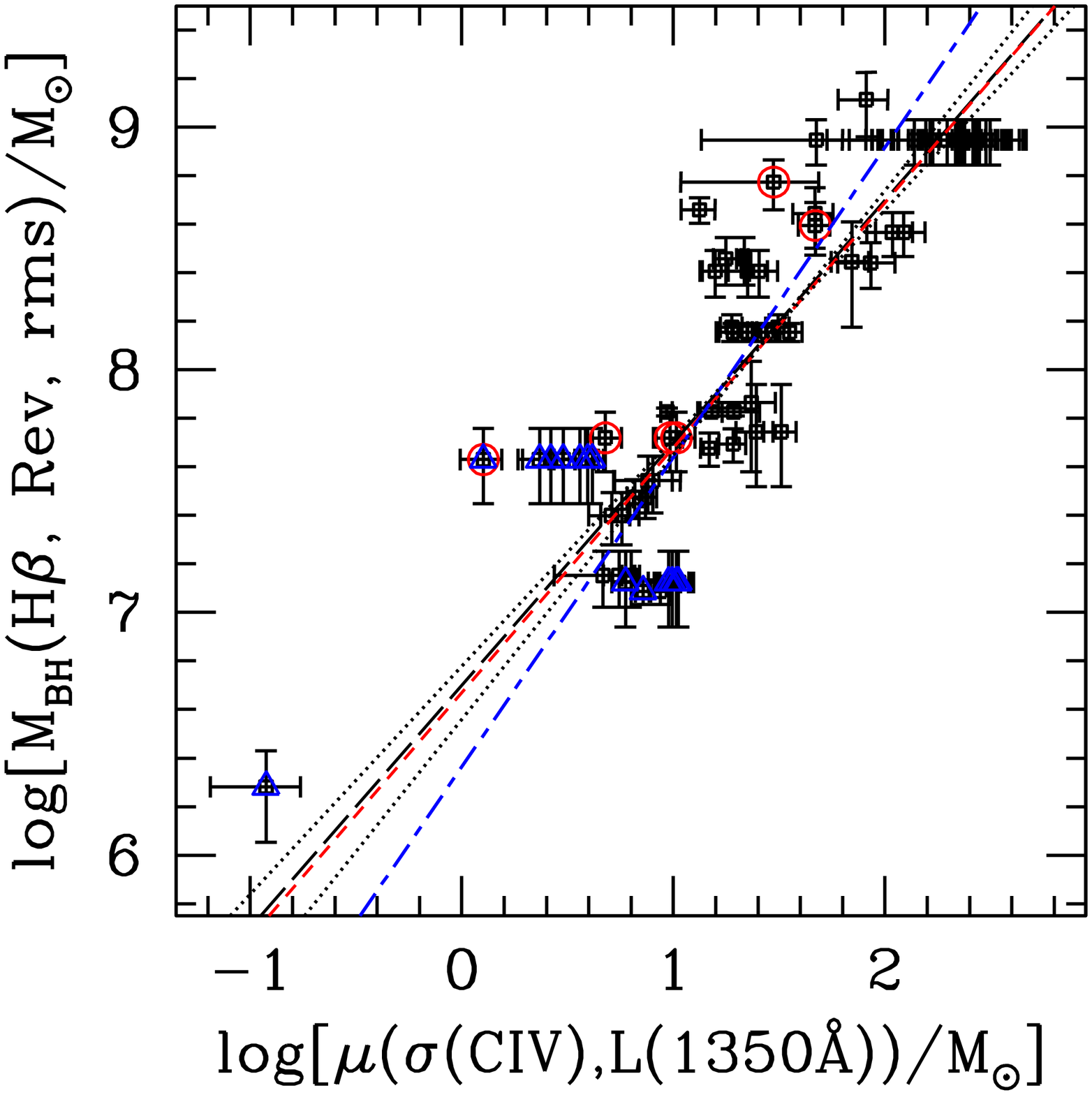}{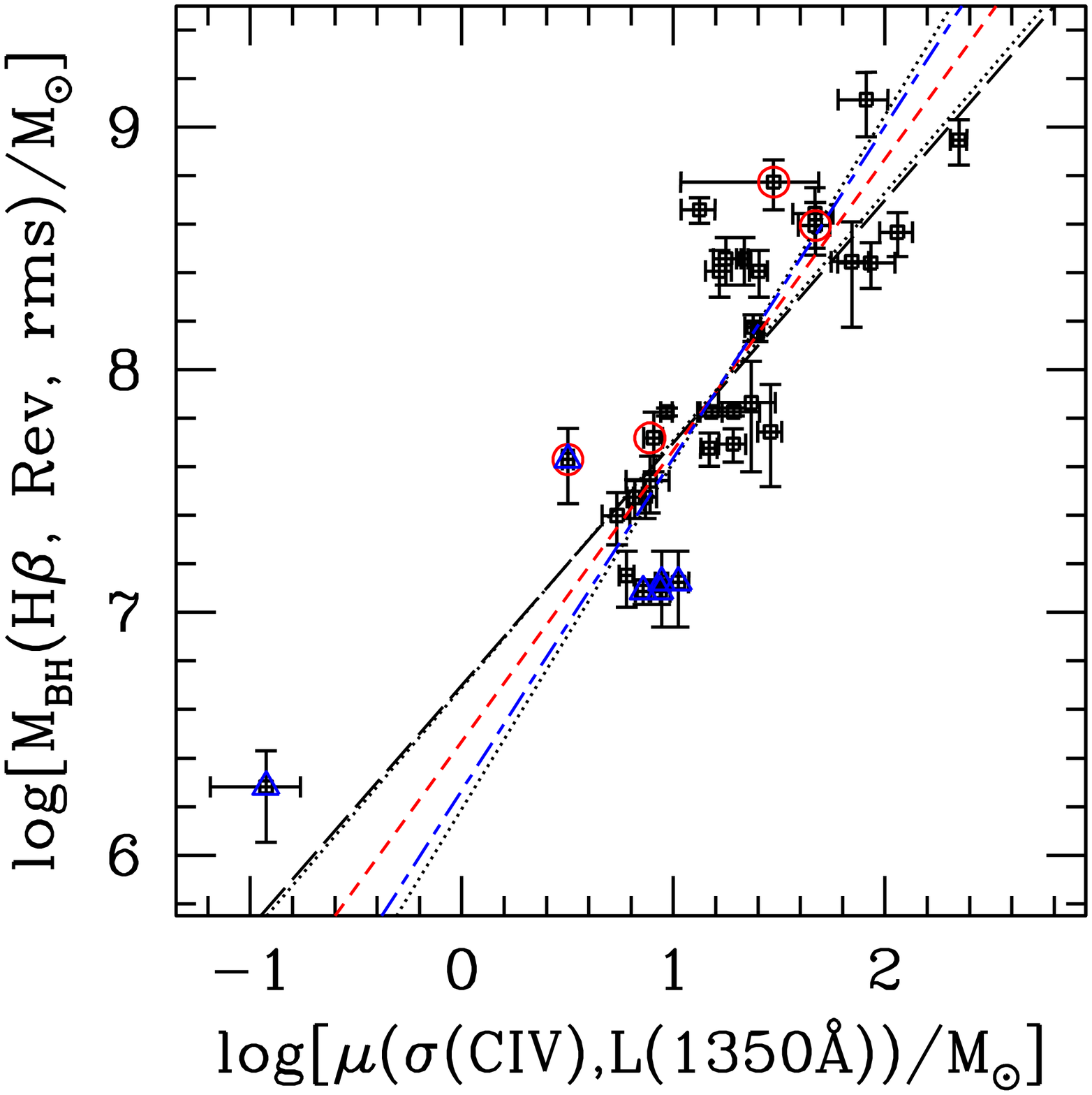}
\caption[]{
Distribution of single-epoch unscaled mass estimates $\mu$ based on UV data, specifically
the line dispersion $\sigma_l$(\civ) and $L_{\lambda}$(1350\AA), with the 
reverberation mapping black hole masses: 
({\it Left}) All individual measurements for each object in the sample;
({\it Right}) the weighted mean $\mu$ of individual (non-monitoring) measurements of a 
given object and the weighted mean of monitoring data when available.
In both cases are the slopes consistent with unity (within the errors).
For symbols, see Fig.~\ref{Muvfw-Mrms.fig}.
\label{MuvSigma-Mrms.fig}}
\end{figure}

We repeat this analysis for the UV unscaled mass based on the \civ{} line
dispersion $\mu$[$\sigma_l$(\civ), $L_{\lambda}$(1350\AA)] (hereafter
$\mu$[$\sigma_l$(\civ)], eq.\ \ref{muLlam}) instead of FWHM.
We compare the unscaled masses $\mu$[$\sigma_l$(\civ)] and reverberation-based
masses in Fig.\ \ref{MuvSigma-Mrms.fig} for both the sample of individual
measurements (left panel) and the sample of weighted means (right
panel). Again, the
measurements exhibit a strong correlation with similar scatter in the
two samples and the flagged measurements show no particular bias.
The BCES regression slopes are also here consistent with a slope of unity,
typically to within 1$\sigma$.

In this case, the FITEXY regression analysis yields slopes that tend to be
steeper than unity, but are nevertheless consistent with a value of unity to
within 3$\sigma$. The estimated intrinsic scatter is here some 10\% lower
than for the $\mu$[FWHM(\civ)] relationship.
Again, the weighted mean sample tends to yield steeper slopes than those 
of the sample of individual measurements, especially when NGC\,4051 is
excluded, but the errors are correspondingly larger.

Since each of the BCES and FITEXY regression fits are consistent with one 
another and with a linear relationship between the unscaled single-epoch
mass estimates and the reverberation-based masses, 
we again conclude that it is appropriate to estimate the black hole mass
through eq.\ (\ref{eq:mass_scale}) and that all we need to do
is establish the offset $a$.
The zero-points and errors we compute are listed in Table~\ref{zeropt.tab}. 
The final calibrated single-epoch black hole mass estimates based on 
FWHM(\civ) and on $\sigma_l$(\civ) are listed in columns (9) and (10) of 
Table~\ref{UV_pars.tab}, respectively.

\subsection{Summary of the Calibrated Mass Scaling Relationships \label{calsummary}}

We conclude this section with a summary of the mass-scaling relationships
for obtaining black hole masses estimates from single-epoch spectra.
\begin{enumerate}
\item {\bf FWHM(\mbox{\boldmath \Hbeta}) and 
\mbox{\boldmath $L_{\lambda}$(5100\,\AA)}.} For the optical 
continuum luminosity and FWHM of the \Hbeta\ broad component, 
\begin{equation}
\log \,M_{\rm BH} (\rm H\beta) = 
   \log \,\left[ \left(\frac{\rm FWHM(H\beta)}{1000~km~s^{-1}} \right)^2 ~ 
   \left( \frac{\lambda \it L_{\lambda} {\rm (5100\,\AA)}}{10^{44} \rm erg~s^{
-1}}\right)^{0.50} 
	\right]	+ (6.91 \pm 0.02).
\label{logMopt_L51.eq}
\end{equation}
The sample standard deviation of the weighted average zeropoint offset, 
which shows the intrinsic scatter in the sample, is $\pm 0.43$ dex.
This value is more representative of the uncertainty in the zero-point than
is the formal error. 

\item {\bf FWHM(\mbox{\boldmath \Hbeta}) and 
\mbox{\boldmath $L(\Hbeta)$}.} For the \Hbeta\ broad-component luminosity
and FWHM,
\begin{equation}
\log \,M_{\rm BH} (\rm H\beta) = 
   \log \,\left[ \left( \frac{\rm FWHM(H\beta)}{1000~km~s^{-1}} \right)^2 ~ 
   \left( \frac{\it L(\rm H\beta)}{10^{42} \rm erg~s^{-
1}}\right)^{0.63} 
	\right]	+ (6.67 \pm 0.03).
\label{logMopt_LHb.eq}
\end{equation}
The sample standard deviation of the weighted average zeropoint offset 
is $\pm 0.43$ dex.

\item {\bf FWHM(\mbox{\boldmath \civ}) and 
\mbox{\boldmath $L_{\lambda}$(1350\,\AA)}.} For the ultraviolet
continuum luminosity and the FWHM of the \civ\ line,
\begin{equation}
\log \,M_{\rm BH} \mbox{(\rm \civ)} = 
   \log \,\left[ \left(\rm \frac{FWHM\mbox{(\rm \civ)}}{1000~km~s^{-1}} \right)^2 ~ 
   \left( \frac{\lambda L_{\lambda} (1350\,{\rm \AA})}{10^{44} \rm erg~s^{-
1}}\right)^{0.53} 
	\right]	+ (6.66 \pm 0.01).
\label{logMuv_fw.eq}
\end{equation}
The sample standard deviation of the weighted average zeropoint offset 
is $\pm0.36$ dex.

\item {\bf \mbox{\boldmath $\sigma_l$(\mbox{\boldmath \civ})} and 
\mbox{\boldmath $L_{\lambda}$(1350\,\AA)}.} For the ultraviolet
continuum luminosity and the dispersion of the \civ\
emission line,
\begin{equation}
\log \,M_{\rm BH} \mbox{(\rm \civ)} = 
   \log \,\left[ \left(\frac{\sigma_l\mbox{(\rm \civ)}}
{\rm 1000\,~km~s^{-1}} \right)^2 ~ 
   \left( \frac{\lambda L_{\lambda} (1350\,{\rm\AA})}{10^{44} \rm erg~s^{-
1}}\right)^{0.53} 
	\right]	+ (6.73 \pm 0.01).
\label{logMuv_sigma.eq}
\end{equation}
The sample standard deviation of the weighted average zeropoint offset 
is $\pm0.33$ dex.
\end{enumerate}

As noted earlier, the $L_{\lambda}$(1450\,\AA) luminosity
can be straightforwardly be substituted for $L_{\lambda}$(1350\,\AA)
without error or penalty in precision.

\section{Accuracy of the Estimated Masses \label{errors}} 

\begin{figure}[ht]
\plottwo{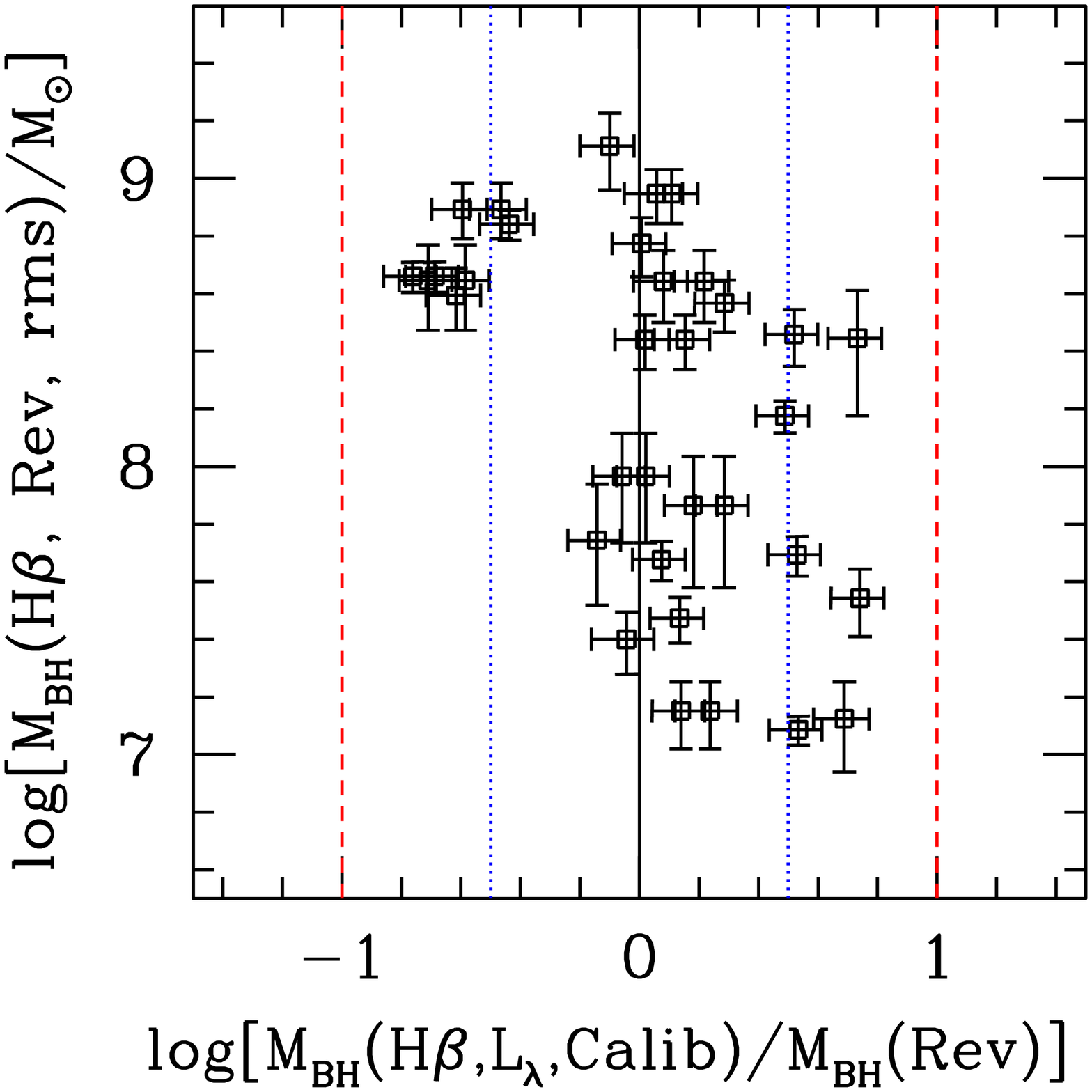}{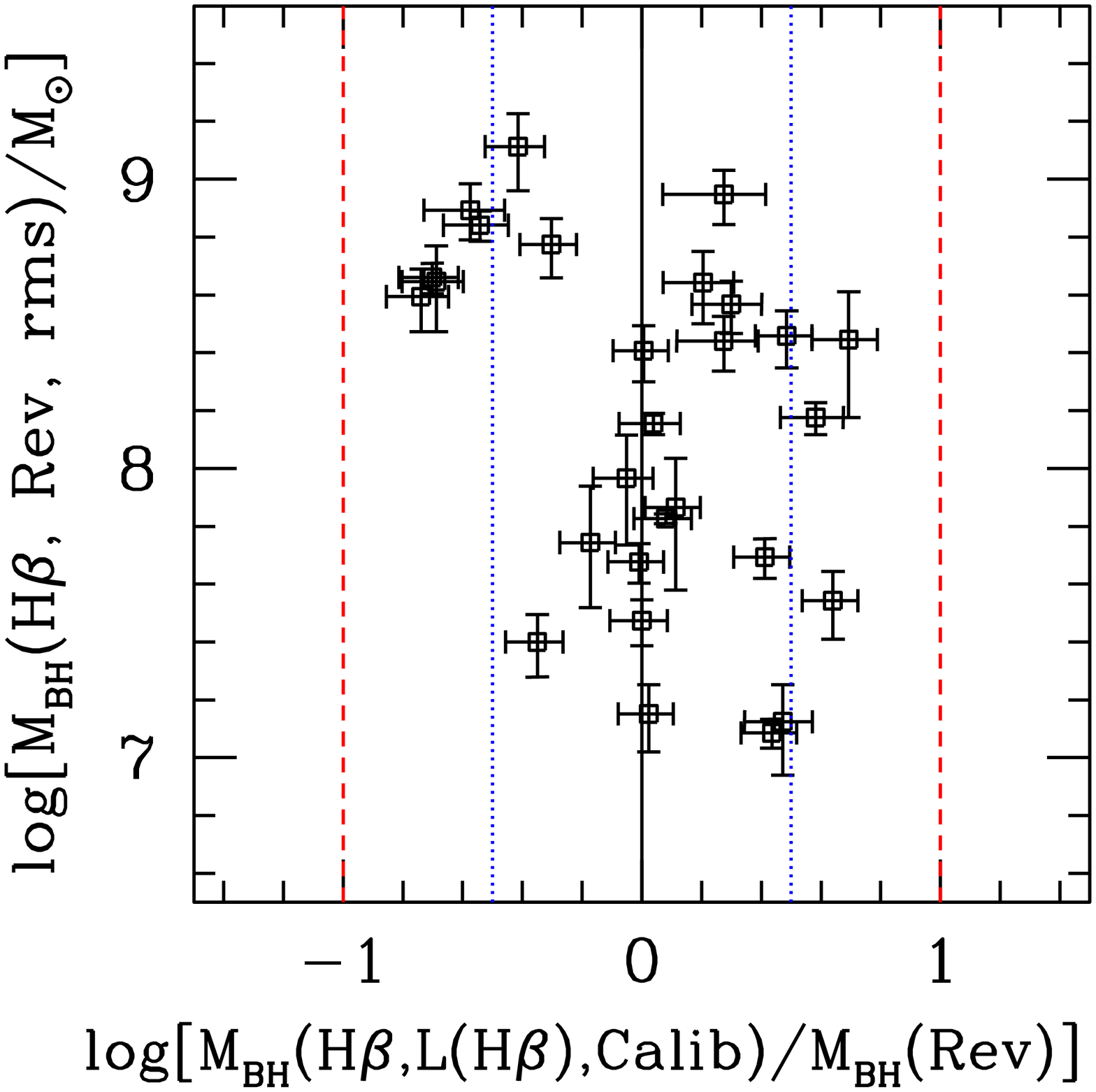}
\caption[]{
Deviation of the optical single-epoch black hole mass estimates from the reverberation
mapping established mass $M_{\rm BH}$(Rev) plotted versus $M_{\rm BH}$(Rev).
({\it Left}) Single-epoch mass estimates based on FWHM(\hb) and $L_{\lambda}$(5100\AA).
({\it Right}) Single-epoch mass estimates based on FWHM(\hb) and $L(\rm H\beta)$.
The uncertainties in the abscissa are the (propagated) uncertainties in
the single-epoch mass estimates (\ie {\it not} the error in the mass deviation).
A strictly unity relationship is indicated by the (black) solid line. Offsets of
$\pm$0.5\,dex and $\pm$1\,dex are indicated by the (blue) dotted and (red) 
dashed lines, respectively.
\label{MoptCal-problty.fig}}
\end{figure}

\begin{figure}[ht]
\plottwo{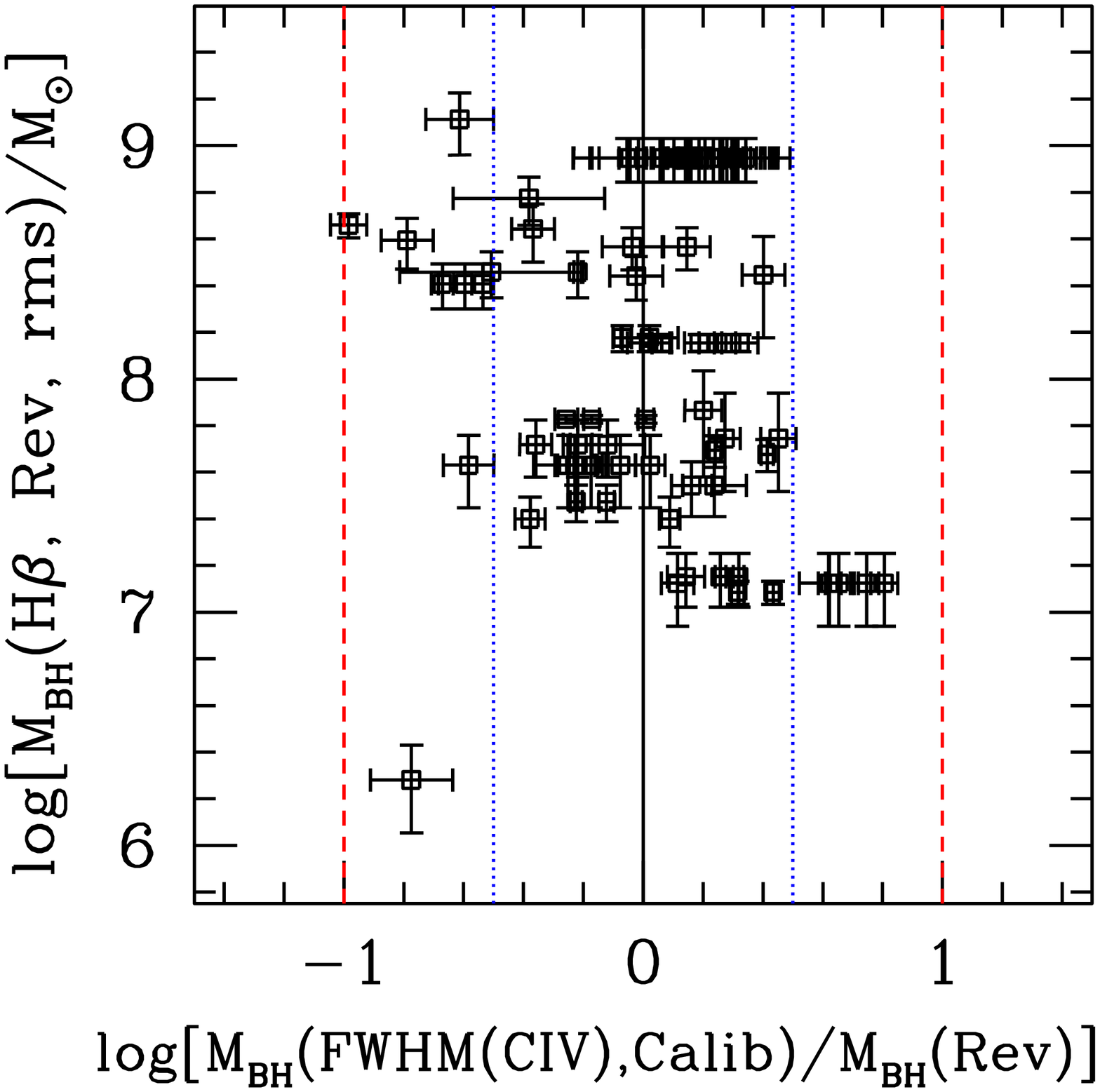}{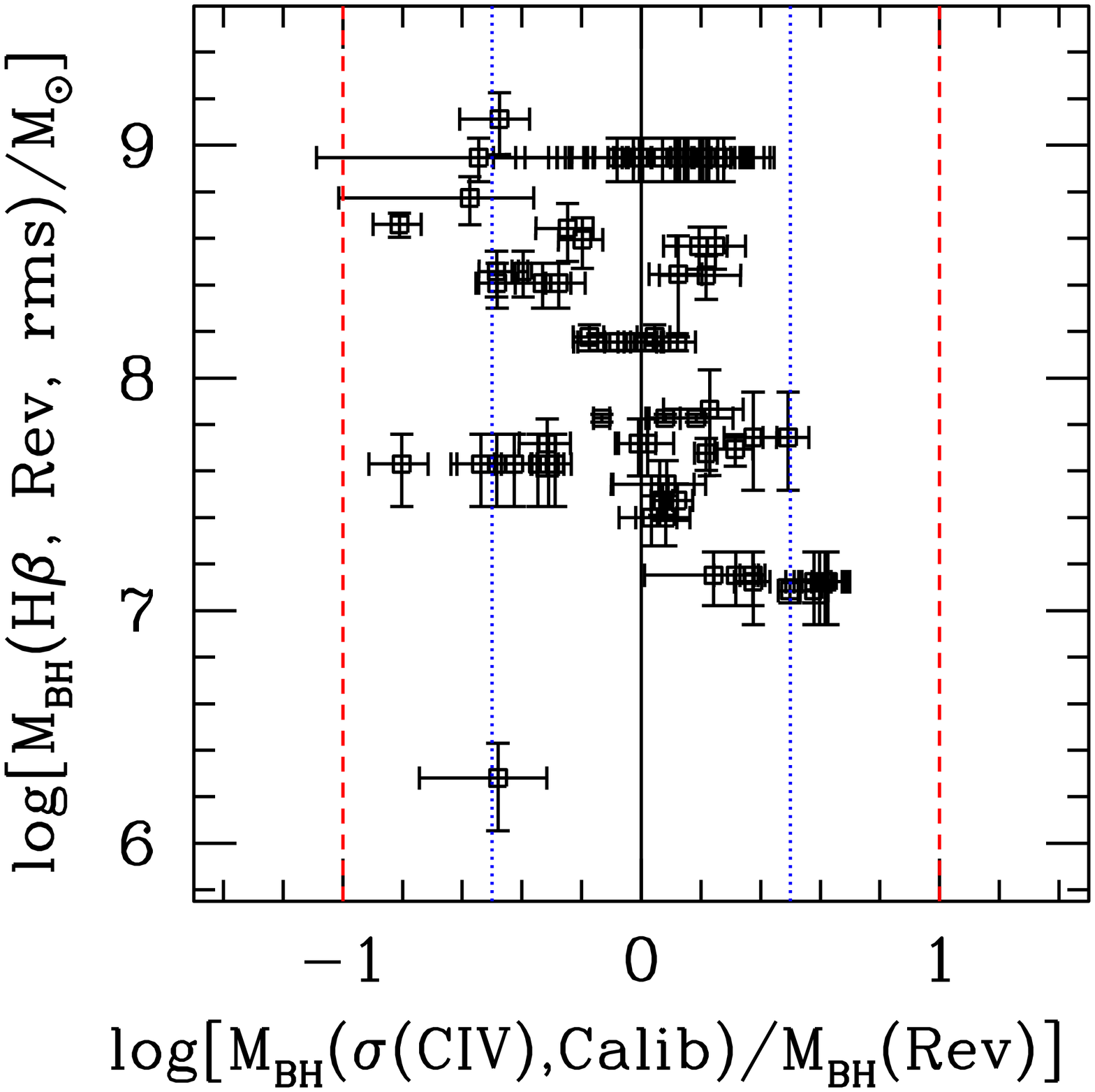}
\caption[]{
Deviation of the UV single-epoch black hole mass estimates from the reverberation
mapping established mass $M_{\rm BH}$(Rev) plotted versus $M_{\rm BH}$(Rev).
({\it Left}) Single-epoch mass estimates based on FWHM(\civ) and $L_{\lambda}$(1350\AA).
({\it Right}) Single-epoch mass estimates based on the line dispersion $\sigma_l$(\civ) 
and $L_{\lambda}$(1350\AA). 
The uncertainties in the abscissa are the (propagated) uncertainties in
the single-epoch mass estimates (\ie {\it not} the error in the mass deviation).
A strictly unity relationship is indicated by the (black) solid line. Offsets of
$\pm$0.5\,dex and $\pm$1\,dex are indicated by the (blue) dotted and (red) 
dashed lines, respectively.
\label{MuvCal-problty.fig}}
\end{figure}

Following the approach of Paper~I, we use simple counting statistics to 
evaluate the {\it statistical} uncertainty in the calibrated 
mass-scaling relationships. In Figs.\ \ref{MoptCal-problty.fig} and 
\ref{MuvCal-problty.fig}, we show 
the deviations of the calibrated single-epoch optical and UV mass estimates
(eqs.\ \ref{logMopt_L51.eq} -- \ref{logMuv_sigma.eq})
from the reverberation-based black hole mass 
as a function of the mass. 
The typical dispersion of 
the distributions of optical mass estimates is about 0.5\,dex.
The UV single-epoch mass estimates exhibit a slightly larger range in
deviations,
but all measurements remain within 1\,dex of the reverberation masses
(Fig.~\ref{MuvCal-problty.fig}).

In Table~\ref{probabilities.tab}, we show the probability that
mass estimates based on
eqs.\ (\ref{logMopt_L51.eq}) -- (\ref{logMuv_sigma.eq})
will reproduce the reverberation-based mass to a specified level
of accuracy.
Table~\ref{probabilities.tab} also
lists the 1$\sigma$ and 2$\sigma$ uncertainties in the mass estimates
for each mass-scaling relationship. We see that
there is a fairly high probability that the mass estimates
are good to within a factor of about 3. Interestingly, the UV relationships
have a slightly higher probability ($\sim$85\%) of being accurate to a factor 
3 than do the optical relationships ($\sim$70\%), while the latter show no
objects deviating by a factor 6 or more, contrary to the UV relationships.
This is also reflected in the scatter around zero deviation: the UV
relationships display a 1$\sigma$ scatter of only $\sim$0.3\,dex, while
the optical relationships have a 1$\sigma$ scatter of 
$\sim$0.5\,dex. 

One source of additional scatter in the optical relationship is contamination 
of the single-epoch $L_{\lambda}$(5100\AA) by host galaxy light. This is 
typically strongest for the lower luminosity AGNs but is not corrected for in 
the single-epoch luminosities on the grounds that for typical applications of
this relationship a correction for host galaxy light, which requires additional 
data and non-trivial analysis, will not commonly be performed.
The error made in not correcting for the host galaxy luminosity will be
explored elsewhere.
 
An important point to make is that this discussion refers to
how accurately the mass-scaling laws reproduce the reverberation-based
masses. The reverberation-based masses are themselves uncertain
typically by a factor of $\sim2.9$, based on the scatter of the
reverberation-based masses around the 
$\Msigma$ relationship (Onken \et 2004).
To obtain the absolute uncertainties, we need to fold in the absolute 
accuracy of the reverberation mapping-mass measurements, and
these values are given in column (7) of Table~\ref{probabilities.tab}.
We estimate that the  absolute accuracy of the single-epoch mass estimates 
range between a factor 3.6 and 4.6.

In closing, it is worth re-emphasizing that these uncertainties
are only of statistical nature; any given estimate from a
mass-scaling relationship can be off by up to an order of magnitude
and should not be trusted in applications where high accuracy is
critical. 
These mass estimates are, however, suitable
for application to large statistical samples.

\section{Discussion \label{discussion}}

In Paper~I, we found that the predominant source of scatter in
the mass-scaling relationships is traceable to the relatively
large scatter in the $R - L$ relationship (Kaspi \et 2000). 
The recent improvements in both the $R - L$ relationships
(Kaspi \et 2005) and in the reverberation-based mass measurements
(Peterson \et 2004) have correspondingly decreased the scatter in the
mass estimates, at least for the estimates based on UV data. The 1$\sigma$
scatter is now only a factor of about 2, compared to the 
factor of 3.2 found in Paper~I. Implicit in this estimate is that
the reverberation-mapped AGNs are reasonably
representative of the AGN and quasar population as a whole.
 
It appears that no improvement has been achieved for the optical 
relationship based on FWHM(\Hbeta) and $L_{\lambda}$(5100\,\AA). However,
this relationship was, in fact, {\it not calibrated} in Paper~I; the
single-epoch estimates based on the relationships quoted by Kaspi \et
(2000) were merely confirmed to be consistent with the 
reverberation-based masses. In the current calibration, these optical mass estimates 
now distribute more evenly around zero offset 
(Fig.~\ref{MoptCal-problty.fig}, left panel)
compared to those of Paper~I (cf.\ Fig.\ 6a in Paper~I).

There are a number of advantages to the use of the \civ\ line width as 
a mass indicator (eq.\ \ref{logMuv_fw.eq}), as argued in Paper~I and by 
Warner, Hamann, \& Dietrich (2003) and Vestergaard (2004).
It has been argued, however, that \civ\ may be an inappropriate choice 
for this application on the grounds that the dynamics of the 
\civ-emitting gas may not be determined primarily by gravitation; 
in particular, it has been suggested (1) that the line profiles
might be affected by obscuration within the line-emitting region,
and (2) that there are reasons to believe that a significant
fraction of the \civ\ line arises in an outflowing wind.
We address these two issues in turn.

Richards \et (2002) used a large sample of SDSS quasars to investigate 
the observed blueshift of the peak of the \civ\ profile relative to the 
low-ionization lines, which is a well-known phenomenon (\eg Gaskell 
1982; Wilkes 1984; Tytler \& Fan 1992). They suggest that the apparent 
blueshift may actually be a lack of emission in the red profile wing, 
suggestive of occultation or obscuration, i.e., a non-gravitational 
alteration of the profile. Richards \et divide their database into four 
groups based on the magnitude of the \civ\ blueshifts (their Fig.~5 and 
Table~2) and form composite spectra for each of these subsets. They find 
that the most blueshifted profile typically has FWHM(\civ) = $1600 \pm 
300\,\kms$, which is only 15\% broader than the \civ\ profile from the 
least blueshifted composite. There is thus a bias in the sense that the 
subset with the most strongly affected profiles will have masses 
overestimated by $\sim$30\%. However, this effect is considerably 
smaller than the typical uncertainties even in the reverberation-based 
masses, which as noted earlier 
are uncertain by a factor of  $\sim2.9$ (Onken \et 2004).

It is sometimes stated (\eg Dunlop 2004; Bachev \et 2004; Shemmer \et 2004; 
Baskin \& Laor 2005) that the \civ\ profile is unsuitable for estimating AGN 
black hole masses from single-epoch spectra because the component of 
\civ\ that arises in an outflowing wind, as is the case with other 
higher-ionization lines, is much more significant than in, say, lower 
ionization lines such as \mgii\ or \Hbeta. The \civ\ profile of some 
narrow-line Seyfert~1 galaxies (NLS1s) may contain a significant 
contribution from a wind, and indeed Vestergaard (2004) cautioned against 
using \civ\ to estimate the masses of black holes in this type of 
object. It should be pointed out, however, that not all NLS1s exhibit 
this behavior. But there does seem to be a tendency for 
higher-luminosity NLS1s to exhibit more highly asymmetric profiles (Leighly 
2000), and there is thus a concern that the masses of luminous quasars 
may be overestimated by failing to account for a strong wind 
contribution. Vestergaard (2004) examined in detail
the \civ\ line asymmetries of high-luminosity quasars in the range 
$1.5 \leq z \leq 5$ and found that none of the profiles resembled in 
any way the triangular shape seen in the spectra of some luminous NLS1 
galaxies, like I\,Zw\,1 (e.g., Vestergaard \& Wilkes 2001). 
Also, such profiles are not seen by Richards et al.\ (2002).
With a conservative selection of objects with 
possible blue wing asymmetries, Vestergaard (2004) determined the typical mass 
and luminosity of the quasars with and without these asymmetries and 
found insignificant differences between the two subsets.

Baskin \& Laor (2005) find that the \civ\ equivalent width EW(\civ) 
correlates with the Eddington luminosity ratio computed from the 
\Hbeta-based masses, \lol(\Hbeta). Since they do not find a similarly 
strong correlation between EW(\civ) and \lol(\civ) for their sample of 
BQS sources, they argue that the \civ\ profile may yield biased black 
hole mass estimates.  As described in Appendix~\ref{Laor.app}, we 
have undertaken a reanalysis of the Baskin \& Laor sample in which
we removed (a) the NLS1s, (b) low-quality {\em IUE} spectra of PG 
quasars, (c) objects with strong absorption near the \civ\
emission-line peak, and (d) 
measurements based on what we consider to be a dubious 
assumption about a narrow-line component of the \civ\ emission line, 
and we find that the problems cited by Baskin \& Laor are much less 
severe. The remaining \civ\ FWHM values scatter within $\pm0.2$\,dex of 
FWHM(\Hbeta), thereby showing reasonable consistency. This result is 
also consistent with that of Warner \et (2003), who find, on average, 
reasonable agreement between the mass estimates based on \Hbeta\ and 
those based on \civ.

Indeed, the most compelling reason to have some confidence 
in the \civ-based mass estimates is because in the handful of objects in which 
reverberation results are available for multiple emission lines, the 
virial products are consistent for {\em all} the measured emission 
lines, including \civ\ (Peterson \& Wandel 1999, 2000; Onken \& 
Peterson 2002). 

\section{Summary \label{summary}}

We present improved relationships between spectrophotometric
parameters, specifically luminosities and emission-line widths,
that allow us to estimate the masses of the central
black hole in AGNs. The calibration of these
relationships is based on reverberation results for 32 AGNs.
These relationships allow black hole mass
estimates of large samples of AGNs to be obtained easily and within
a short period of time and are thus particularly useful for distant
AGNs and quasars where more direct mass measurement techniques
are impractical or unfeasible.

The new mass-scaling relationships, presented here in
\S~\ref{calsummary}, 
supersede those presented by
Kaspi \et (2000), Vestergaard (2002; Paper~I), and Wu \et (2004) on
account of significant improvements, beginning with the
reanalysis of the reverberation database (Peterson \et 2004), which
yield improved BLR sizes and masses. Also, the reverberation-based mass scale has
now been empirically calibrated for the first time 
through the $\Msigma$ relationship (Onken \et 2004), 
the BLR $R-L$ relationships have been recomputed
(Kaspi \et 2005; Bentz \et 2006), and there are more and better
optical and UV spectra available to establish these scaling laws
and assess their uncertainties. Moreover, these mass-scaling relationships
have also been updated to the current $\Lambda$CDM benchmark
cosmology.

The updated mass-scaling relationships show considerable 
improvement in the internal scatter compared to the 
same relationships in Paper~I. The 
1$\sigma$ uncertainty is of order of a factor of 2 to 2.5 relative 
to the reverberation-based masses and 
we estimate that the absolute accuracy 
of the masses from the scaling relationships
is a factor of $\sim4$
(Table~\ref{probabilities.tab}). We emphasize, however,
that the uncertainties quoted here are only 
applicable to statistical samples. The uncertainty associated with a single 
mass estimate may be much higher; 
possibly these relationships may only 
yield order of magnitude mass estimates for individual measurements.

\acknowledgments

The anonymous referee is thanked for careful reading of the
manuscript and for helpful comments.
We are grateful for support of this research through grant
HST-AR-10691 from NASA through the Space Telescope Science Institute,
which is operated by the Association of Universities for Research in
Astronomy, Incorporated, under NASA contract NAS5-26555, and through
grant AST-0205964 from the National Science Foundation.
The optical and UV {\it HST} STIS spectra of NGC~3227 and NGC~4051 and the
STIS echelle spectrum of NGC~4051 were all kindly provided by Mike Crenshaw. 
Table~1 of Baskin \& Laor (2005) was kindly provided in electronic form by 
Alexei Baskin. 
This research has made use of the NASA/IPAC Extragalactic Database (NED)
which is operated by the Jet Propulsion Laboratory, California Institute
of Technology, under contract with the National Aeronautics and Space
Administration.


\appendix
\section{Reanalysis of Baskin \& Laor Data \label{Laor.app}}

As noted in \S\ref{discussion}, Baskin \& Laor (2005)
caution against the use of the \civ\ emission line for
estimating black hole masses through mass-scaling relationships
of the type derived in this paper. As their conclusion was
contrary to that reached in Paper~I and by Warner \et (2003)
and Vestergaard (2004), we felt that is was important for
us to examine the same data as Baskin \& Laor and attempt
to understand the origin of these different results.

Fundamentally, Baskin \& Laor (2005) were concerned about
apparent differences between the profiles of \Hbeta\ and
\civ. To compare these profiles for a large number of objects,
they  used data on 81 of the 87 BQS (PG) quasars from the
study of Boroson \& Green (1992). The \civ\ profile measurements
were based on  \hst\ FOS and \iue\ archival spectra, and these
were compared with the \Hbeta\ measurements from Boroson \&
Green. For the purpose of this discussion, their main conclusions
were (1) that FWHM(\civ) is not always larger than FWHM(\Hbeta) as expected 
if \civ\ is emitted closer to the center than \Hbeta\ which 
photoionization models and monitoring data indicate, and 
(2) that ``\civ\ appears to provide a less accurate and possibly 
biased estimate of the black hole mass in AGNs, compared to \Hbeta.'' 

Upon investigating this issue ourselves, we have concluded that 
the poor correlation between FWHM(\civ) and FWHM(\Hbeta) that
concerned Baskin \& Laor (2005) is due in large part to the database
they used, both in terms of sample bias and screening of
sources and secondarily to their assumption about a \civ\
narrow component.

The first problem is inclusion of NLS1s and similar
AGNs in the Baskin \& Laor (2005) sample\footnote{Despite its 
many virtues, the BQS is limited
in some applications because it is neither 
a complete sample nor representative of the typical quasar population. 
Compared with the SDSS sample, Jester \et (2005) find that  
the BQS quasars tend to be bluer and brighter than typical quasars 
and that the BQS sample is incomplete at the 50\% level. 
In addition, 
the PG sample of Boroson \& Green contains a larger fraction of 
specific subgroups of objects such as NLS1s and narrow-lined quasars,
as well as more powerful radio sources than typical quasar samples.}.
There are 17 NLS1s and narrow-lined quasars 
(i.e., with FWHM(\Hbeta) $\leq 2000$\,\kms) in the Baskin \& Laor
analysis. As already noted in this paper and by Vestergaard (2004), 
we agree that the \civ\ profiles in NLS1s are not suitable for 
the purpose of estimating virial masses on account of the
probability that there is a strong component from an outflowing
wind. We thus removed these objects from the sample.

A second problem is data screening, principally of the 
{\em IUE} spectra. The quality of {\em IUE} data on objects
as faint as BQS quasars is low on account of the small aperture
(0.45 m) of the {\em IUE} telescope  and, 
by current standards, relatively insensitive detectors.
In our opinion, only few of the \iue\ spectra of 
quasars are of adequate quality for mass estimates as we describe
here. Also, some objects have relatively strong absorption 
superposed on the \civ\ profile, sometimes very close to line
center. This makes it very challenging, or even impossible,
to establish the intrinsic profile, especially 
at the spectral resolution of \iue\ and even for certain of the
\hst\ spectra (\S~\ref{uvdata}).

A final problem we identify is subtraction of a narrow-line
component from the \civ\ profile. Narrow-component removal
is important in the case of \Hbeta\ because it is a significant
contributor to the total flux. It is also {\em possible} because
the adjacent \oiii\,$\lambda5007$ line can be used as a template
profile. While Baskin \& Laor are not the first to
attempt to remove a narrow \civ\ component
(cf.\ Bachev \et 2004), there is little evidence that it is
in fact necessary since the UV narrow-line components seem to be
very weak (Wills \et 1993). Moreover, it is not clear how to
effect this since there are no isolated narrow lines in the
UV to use as templates, so both the narrow line width and strength
have to be guessed.

In addition to removing the 17 NLS1s from the sample, on the
basis of inspection of the \civ\ line profiles shown in
Fig.\ 1 of Baskin \& Laor (2005), we have removed several
additional spectra from the database because of 
insufficient quality, problematic narrow-component subtraction,
or strong absorption. 
Omission of these objects and the 17 NLS1s and narrow-line
quasars leaves us with 46 objects, which is 57\% of the original
sample. Our modified sample is listed in Table~\ref{LaorModSample.tab}.

\begin{figure}[ht]
\plottwo{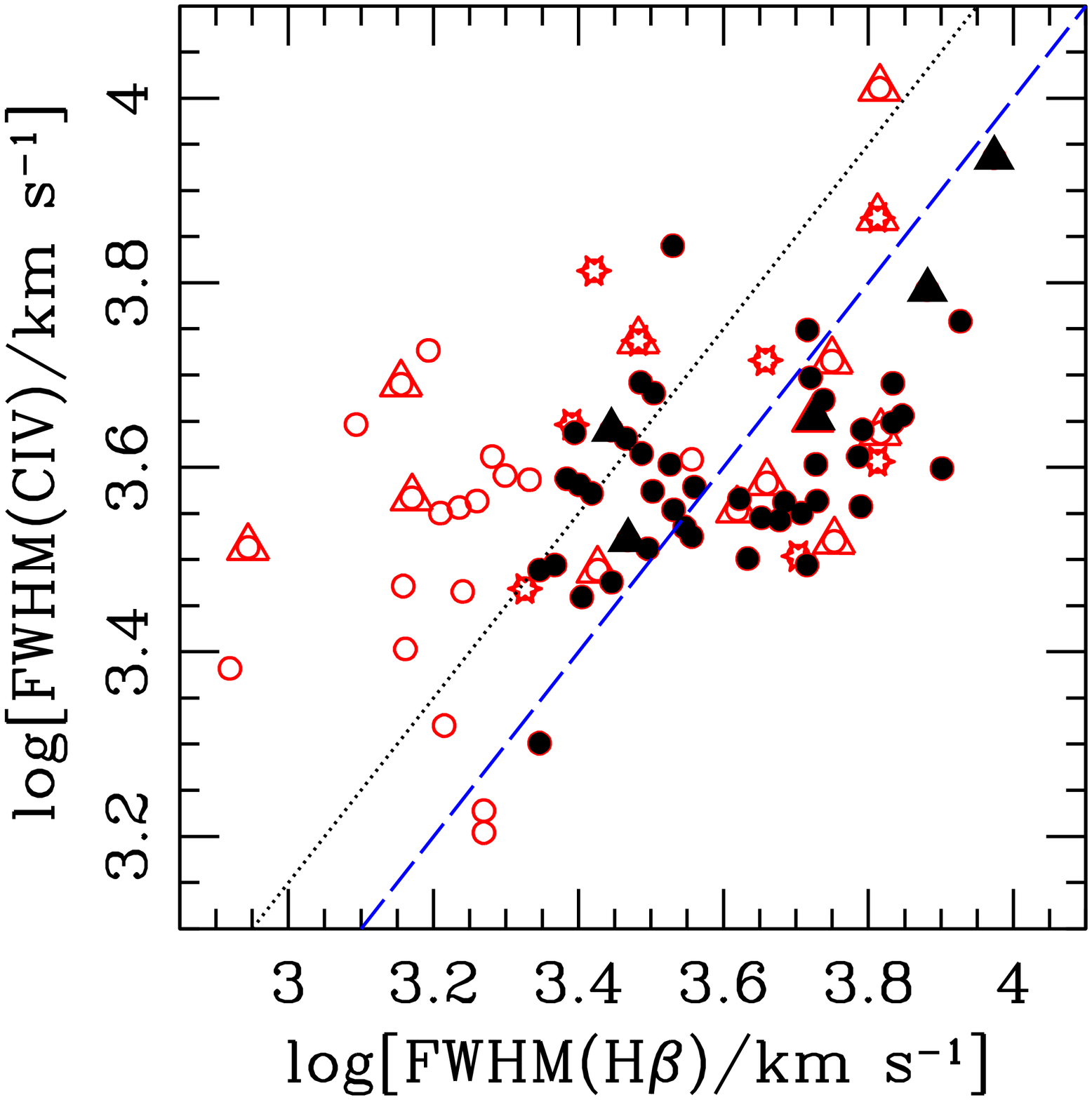}{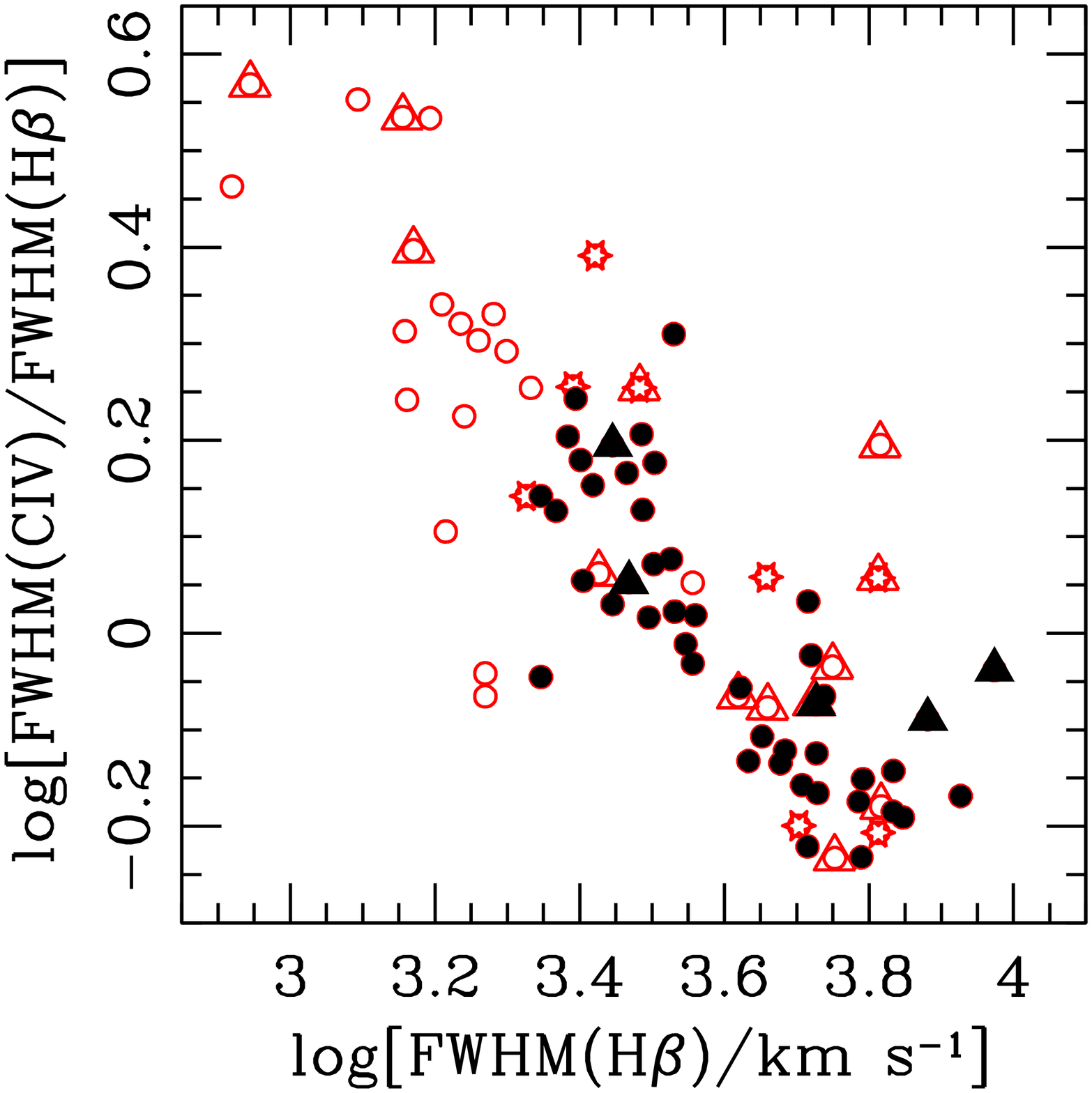}
\caption[]{
Relationship between the \civ{} and \hb{} line widths for the BQS sample. 
({\it Left}) The FWHM(\civ) and FWHM(\hb) are compared directly.
The unity relationship (blue dashed line) is shown for comparison.
The (black) dotted line shows the expected relationship if FWHM(\civ)
is always $\sqrt2$ larger than FWHM(\hb).
({\it Right}) The ratio of line widths are compared to FWHM(\hb).
Symbols: The original sample analyzed by Baskin \& Laor (2005) is shown 
with (red) open symbols. The objects for which a NLR contribution stronger
than 2\AA{} is subtracted from the \civ{} profile are marked by open (red)
stars. Entries that are based on bad profile fits owing to low quality
\iue{} data or bad absorption are marked with open (red) triangles superposed
on the (red) open circle.
The modified and reanalyzed sample (see text) is shown with filled symbols. 
Entries among those with borderline quality \iue{} data or \civ{} profile
absorption are marked with solid triangles.
In the left diagram, the rms around the (blue) dashed line for the original 
sample is 0.22\,dex and 0.15\,dex for the modified sample.
\label{Laorf2.fig}}
\end{figure}

In Fig.\ \ref{Laorf2.fig},
we show how our modified sample selection affects the parameter
distributions presented by Baskin \& Laor (2005). The full sample
analyzed by Baskin \& Laor is shown as open symbols, while the
filled symbols denote the measurements which we consider to
be more reliable, based entirely on
Figure~1 and column (16) of Table~1 in their paper.
The full (original) sample has a standard deviation (rms)
of 0.22\,dex around the unity line (hereafter, ``case 1'')
in the left panel of Fig.~\ref{Laorf2.fig}.
Assuming that \civ\ is always emitted from a distance half that 
of \hb\ as seen for NGC\,5548 (Korista \et 1995), the rms 
is 0.23\,dex (hereafter, ``case 2'').  However, the significance
of this is hard to judge since Baskin \& Laor do not quote 
errors on their parameters.
We examined to what extent the exclusion of the NLS1s
and each additional subset (A: low quality \iue\ data; B: strong
absorption; and C: subtraction of a strong ($\gsim 2$\AA) narrow 
\civ\ component) changes the rms in each case. The most significant 
changes are seen when the NLS1s alone are excluded. In this case,
the rms reduces 
to 0.16\,dex for case 1 and to 0.22\,dex for case 2. When, in 
addition, the objects in subset C are also excluded,
the rms reduces to 0.15\,dex (case 1) and 0.21\,dex (case 2), 
very similar to the rms of the modified sample (i.e., all data 
deemed to be unreliable are excluded; Table~\ref{LaorModSample.tab}), 
namely 0.15\,dex (case 1) and 0.22\,dex (case 2).
We conclude that for the modified sample, the FWHM(\civ) and FWHM(\Hbeta) 
values are correlated,  albeit with some 
scatter (Fig~\ref{Laorf2.fig}, left panel ). 

\begin{figure}[ht]
\plottwo{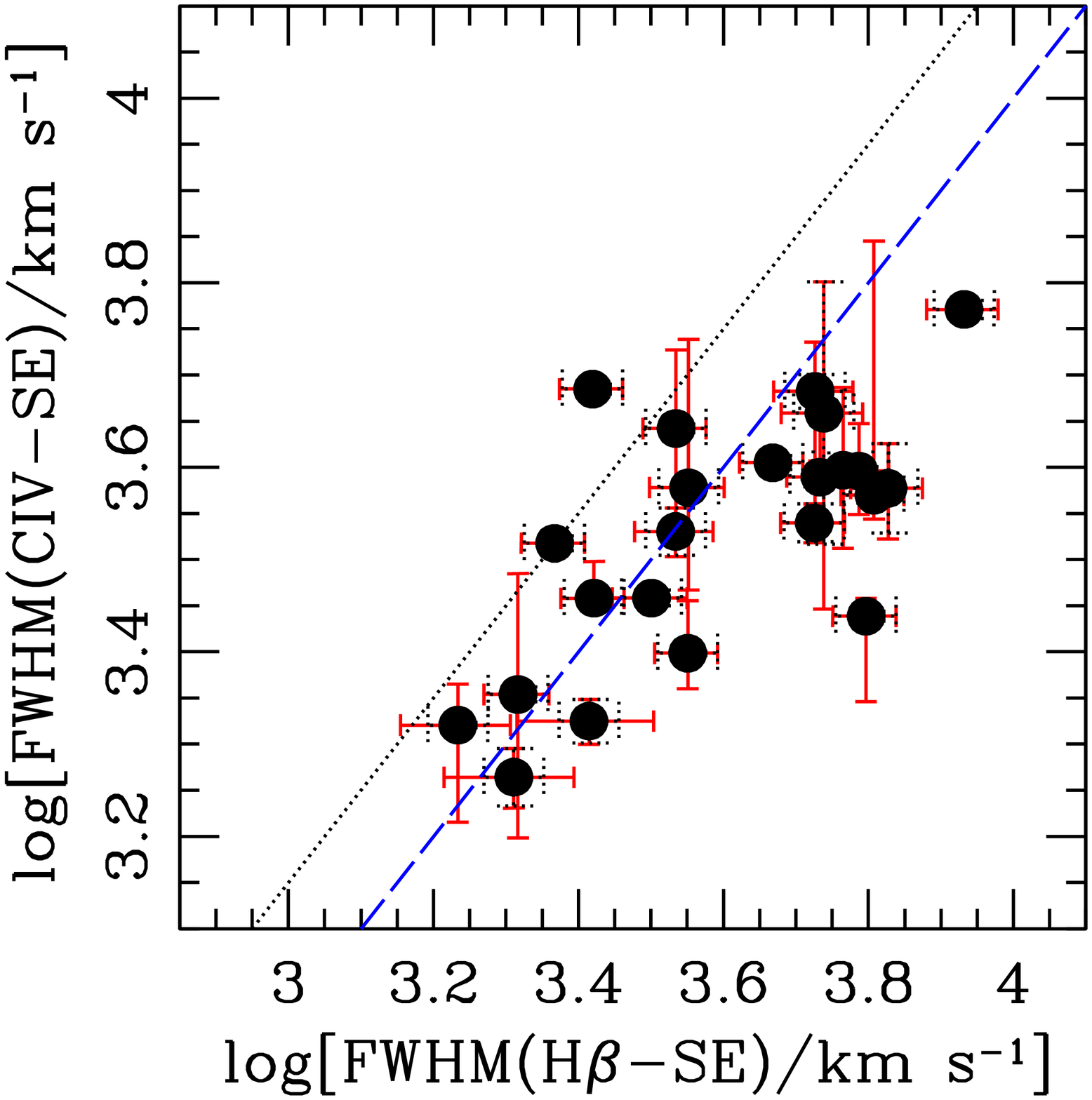}{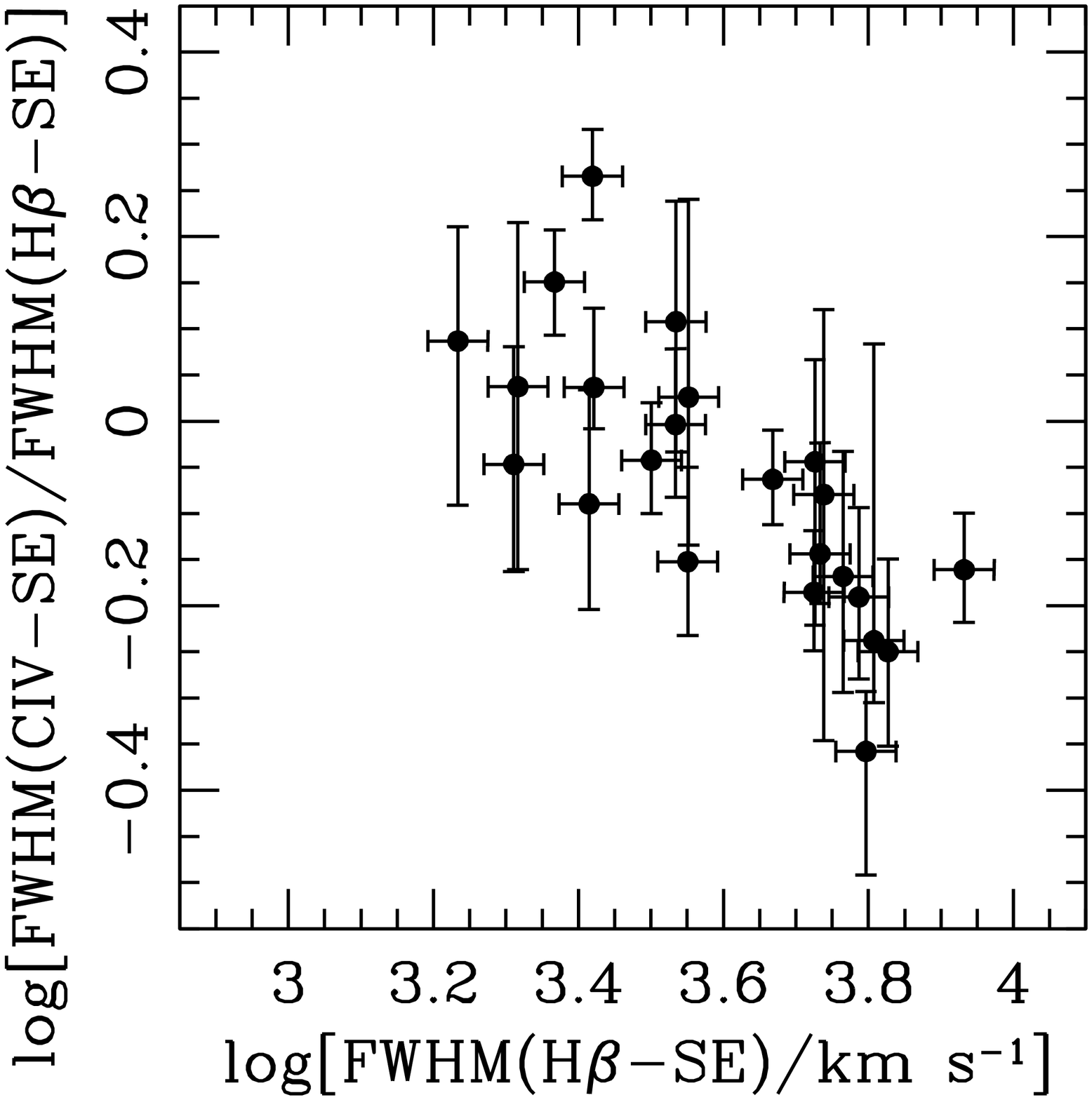}
\caption[]{
Relationship between the single-epoch \civ{} and single-epoch \hb{} 
line widths for the UV reverberation sample.
The data points are the weighted average of each FWHM entry in Tables
1 and 2.
({\it Left}) The FWHM(\civ) and FWHM(\hb) are compared directly.
The unity relationship (blue dashed line) is shown for comparison.
The (black) dotted line shows the expected relationship if FWHM(\civ)
is always $\sqrt2$ larger than FWHM(\hb). The black dotted error bars are
the errors on the weighted means, while the (red) solid error bars show the
range of line widths covered by the individual measurements and their
errors. The rms of the data points around the unity relation is
0.23\,dex and relative to the black dotted relationship is 0.33\,dex;
the average ``error bar'' is 0.06\,dex and 13\% of the errors exceed
0.1\,dex up to a maximum value of 0.41\,dex.
({\it Right}) The ratio of line widths are compared to FWHM(\hb).
The error bar for the ratio is based on the available ranges of linewidths
and measurement errors (red/solid error bars in the left diagram).
\label{fig11.fig}}
\end{figure}

\begin{figure}[ht]
\plottwo{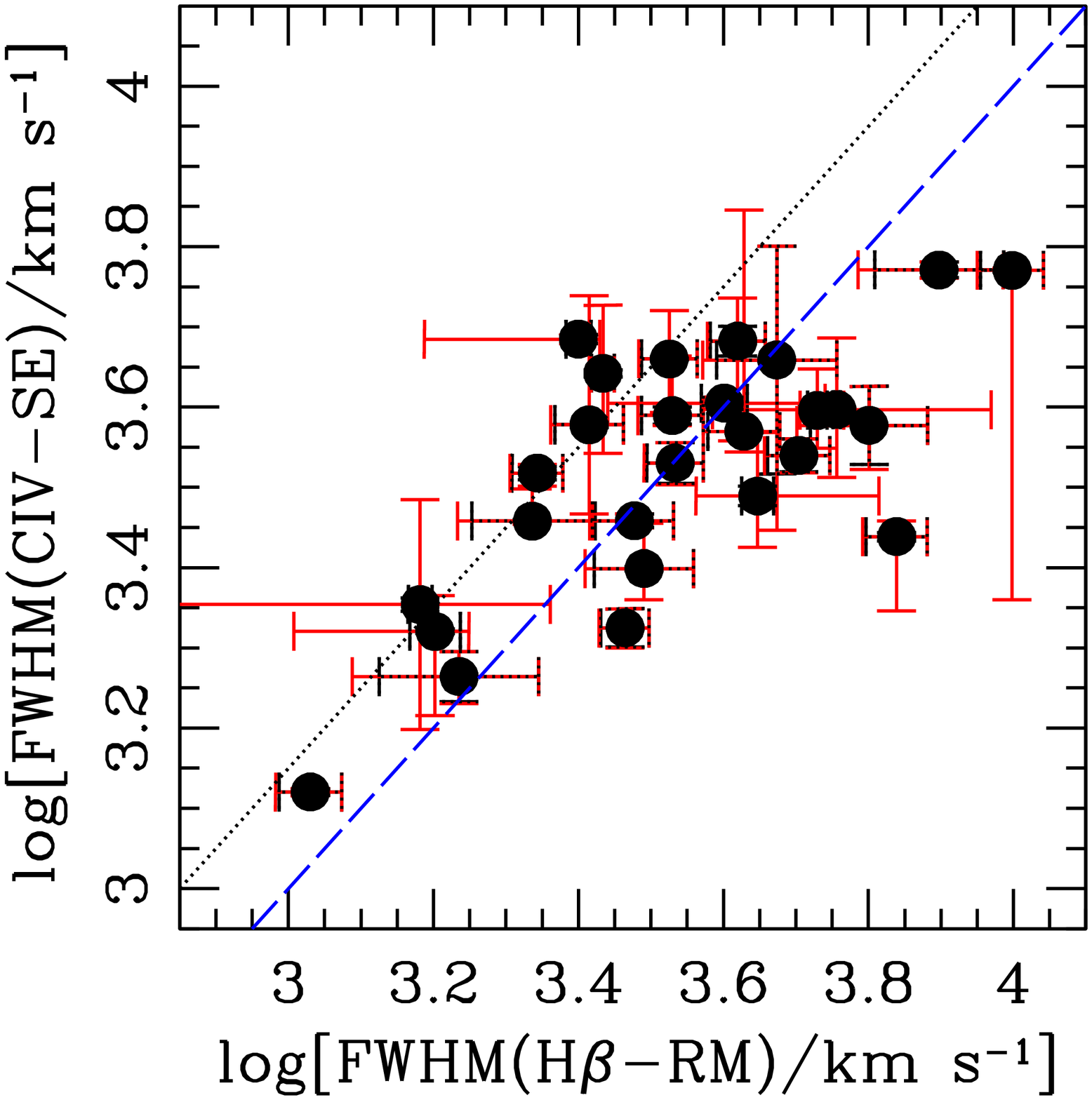}{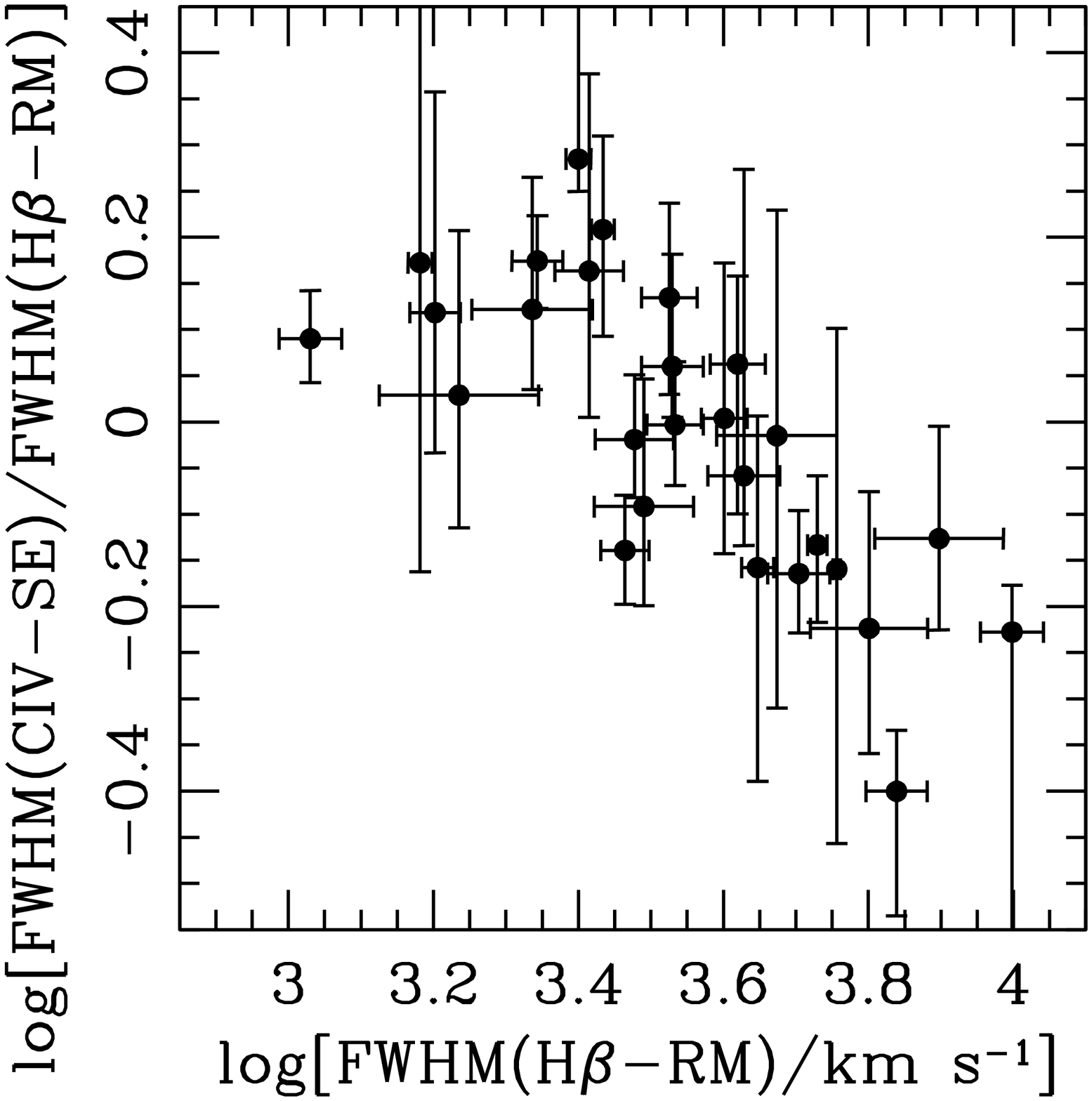}
\caption[]{
Relationship between the single-epoch \civ{} and \hb{} line widths for 
the current sample. The \hb{} data are the weighted average of the
FWHM of the individual epoch spectra listed by Peterson \et (2004).
The \civ\ data points are likewise the weighted average of each
FWHM entry in Table~2.
({\it Left}) The FWHM(\civ) and FWHM(\hb) are compared directly.
The rms of the data points around the unity relation is
0.16\,dex and relative to the black dotted relationship is 0.23\,dex;
the average ``error bar'' is 0.07\,dex and 33\% of the errors exceed
0.1\,dex up to a maximum value of 0.41\,dex.
({\it Right}) The ratio of line widths are compared to FWHM(\hb).
See Figure~\ref{fig11.fig} for symbols.
\label{fig12.fig}}
\end{figure}

For comparison, we examined the data in this work 
in a similar
manner.  Figure~\ref{fig11.fig} (left panel) compares the single-epoch line 
widths of \civ\ and \hb.  Given the multiple FWHM entries
for both \civ\ and \hb, we choose to plot only the weighted means for 
each object of the FWHM(\civ) and FWHM(\hb) values, respectively. 
The solid (red) error bars in the diagram represent the range in measured 
FWHM values based on the
individual epoch FWHM measurements and the measurement errors.
The best comparison is of line widths obtained at epochs in which a
continuum change has had time to propagate to both the \civ\ and 
\hb\ emitting regions. Since no statistically significant sample of
such data exists, we are limited to comparing the single-epoch measurements
regardless of observed epoch. This may introduce additional scatter.
In order to check whether the results based on 
single-epoch \hb\ line widths may be spurious, we also compare 
in Fig.~\ref{fig12.fig} (left panel) the single-epoch \civ\ line widths versus 
the \hb\ FWHM values obtained at each epoch of the monitoring data 
studied by Peterson \et (2004); the red error bars are larger in this
figure because the \hb\ error bars represent the actual measurement errors 
rather than a lower-limit estimate of 10\% as in the
case of the single-epoch \hb\ 
measurements in Fig.~\ref{fig11.fig}.  Figures~\ref{fig11.fig}
and~\ref{fig12.fig} show that, as in the case of the modified BQS sample,
the line widths of \civ\ and \hb\ are strongly correlated and that
some, but not all, of the scatter can
be attributed to measurement errors and the lack of simultaneity
of the \civ\ and \hb\ measurements (red error bars);
the rms values in these figures is similar to that of the
the modified BQS sample (see captions).

There remains, however, an inverse correlation between the
ratio of the \civ\ and \Hbeta\ line widths and FWHM(\Hbeta)
in both samples, as seen in the right-hand panels in 
Figs.\ \ref{Laorf2.fig}, \ref{fig11.fig}, and \ref{fig12.fig}.
This could be attributable to a number of effects, including,
for example, a small difference
in the slopes of the \Hbeta\ and \civ\ $R-L$ relationships.
We defer discussion of this issue to a forthcoming paper, since
for the present purposes, it is sufficient to note that
differences in \hb\ and \civ\ line widths
are now only of order $\pm$0.2\,dex, or a factor 1.6. 
This translates
to a difference in mass estimates of a factor 2.5, which is within the
uncertainties of the single-epoch mass estimates.

Our reanalysis suggests that the mass estimates based on \civ\
may not be as untrustworthy as 
Baskin \& Laor claim. Simply removing data points that are problematic
for a number of reasons greatly reduces the discrepancies between
\civ-based and \Hbeta-based mass estimates. This is, of course,
not to say that there are not systematic effects (which are
the subject of several current investigations), but only that these
effects are small relative to the level of accuracy 
to which we claim that we can measure the masses of AGN black holes at this time.

\section{Mass Estimates for the PG Quasars}

Table~\ref{BG92_ML.tab} lists the single-epoch mass estimates
of the PG quasars studied by Boroson \& Green (1992) without
mass measurements from reverberation mapping techniques. The mass
estimates are based on eq.\ (\ref{logMopt_L51.eq}) in the main text.


\begin{small}

\end{small}

\clearpage


\clearpage
\begin{deluxetable}{lllccllcllll} 
\vspace{-15cm}
\tablewidth{0pt}
\hspace{-14cm}
\rotate{}
\tablecaption{Optical Spectral Parameters and Masses \label{Opt_pars.tab}}
\tabletypesize{\tiny}
\tablehead{
\colhead{ Object} &
\colhead{ Alternative} &
\colhead{ } &
\colhead{ FWHM(H$\beta$)\tablenotemark{b}} &
\colhead{ } &
\colhead{ $\log\,[{}\lambda L_{\lambda}$(5100\AA){}} &
\colhead{ $\log\,[{}L(H\beta){}$} &
\colhead{ } &
\colhead{ log\,[ $M/M_{\odot}$ ]\tablenotemark{c}} &
\colhead{ log\,[ $M/M_{\odot}$ ]\tablenotemark{c}} &
\colhead{ log\,[ $M/M_{\odot}$ ]\tablenotemark{d}} &
\colhead{ } \\
\colhead{ } &
\colhead{ Name} &
\colhead{ $z$\tablenotemark{a}} &
\colhead{ (km s$^{-1}$)} &
\colhead{ Ref.} &
\colhead{ /ergs s$^{-1}${} ] } & 
\colhead{ /ergs s$^{-1}${} ] } & 
\colhead{ Ref.} &
\colhead{ (H$\beta$,$L_{\lambda}$,SE)} &
\colhead{ (H$\beta$,$L$(H$\beta$),SE)} &
\colhead{ (H$\beta$,rms)} \\ 
\colhead{(1)} &
\colhead{(2)} &
\colhead{(3)} &
\colhead{(4)} &
\colhead{(5)} &
\colhead{(6)} &
\colhead{(7)} &
\colhead{(8)} &
\colhead{(9)} &
\colhead{(10)} &
\colhead{(11)} \\ 
}
\tablecolumns{12}
\startdata
      Mrk335  & PG0003$+$199  &   0.02578 &  1585 & 1&  44.16$\pm$0.118&      \nodata   &4& 7.379$^{+0.099}_{-0.129}$&      \nodata   & 7.152$^{+0.101}_{-0.131}$&  \\
              &               &           &  1841 & 2&  43.71$\pm$0.011& 41.95$\pm$0.043&2& 7.192$^{+0.081}_{-0.099}$& 7.175$^{+0.083}_{-0.102}$&            &  \\
  PG0026$+$129&               &   0.14200 &  1821 & 1&  45.10$\pm$0.017&       \nodata  &3& 8.059$^{+0.098}_{-0.126}$&        \nodata &8.594$^{+0.095}_{-0.122}$&  \\
              &               &           &  2250 & 2&      \nodata    & 42.76$\pm$0.043&2&     \nodata       & 7.855$^{+0.091}_{-0.116}$&            &  \\
  PG0052$+$251&               &   0.15500&   5187 & 1&  45.03$\pm$0.017&       \nodata  &3& 8.926$^{+0.096}_{-0.123}$&         \nodata& 8.567$^{+0.081}_{-0.100}$&  \\
              &               &          &   5463 & 2&      \nodata    & 43.14$\pm$0.044&2&     \nodata       & 8.867$^{+0.101}_{-0.132}$&                 &  \\
  Fairall9    &               &   0.04702&   6261 & 2&      \nodata    & 42.23$\pm$0.041&2&     \nodata       & 8.413$^{+0.083}_{-0.103}$& 8.407$^{+0.086}_{-0.108}$&  \\
      Mrk590  &               &   0.02638&   2627 & 2&  44.01$\pm$0.009& 42.24$\pm$0.041&2& 7.690$^{+0.079}_{-0.097}$& 7.667$^{+0.083}_{-0.103}$& 7.677$^{+0.063}_{-0.074}$&  \\
       3C120  &               &   0.03301&   2328 & 2&  43.92$\pm$0.011& 42.26$\pm$0.039&2& 7.529$^{+0.080}_{-0.097}$& 7.572$^{+0.083}_{-0.103}$& 7.744$^{+0.195}_{-0.226}$&  \\
      Akn120  &               &   0.03230&   6120 & 2&  44.37$\pm$0.007& 42.81$\pm$0.042&2& 8.652$^{+0.082}_{-0.101}$& 8.758$^{+0.092}_{-0.118}$& 8.176$^{+0.052}_{-0.059}$&  \\
  PG0804$+$761&               &   0.10000&   3045 & 1&  45.06$\pm$0.014&      \nodata   &3& 8.479$^{+0.096}_{-0.124}$&      \nodata   & 8.841$^{+0.049}_{-0.055}$&  \\
              &               &          &   3276 & 2&      \nodata    & 42.94$\pm$0.037&2&      \nodata      & 8.299$^{+0.095}_{-0.120}$&                 &  \\
  PG0844$+$349&  TON951       &   0.06400&   2386 & 1&  44.49$\pm$0.012&       \nodata  &3& 7.909$^{+0.083}_{-0.103}$&         \nodata& 7.966$^{+0.150}_{-0.231}$&  \\
              &               &          &   2787 & 2&  44.38$\pm$0.010& 42.56$\pm$0.050&2& 7.975$^{+0.082}_{-0.101}$& 7.915$^{+0.089}_{-0.111}$&                 &  \\
      Mrk110  & PG0921$+$525  &   0.03529&   2079 & 1&  43.63$\pm$ 0.11&      \nodata   &4& 7.276$^{+0.100}_{-0.130}$&       \nodata  & 7.400$^{+0.094}_{-0.121}$&  \\
              &               &          &   2067 & 2&      \nodata    & 41.6$\pm$0.0465&2&      \nodata      & 7.050$^{+0.086}_{-0.107}$&                 &  \\
  PG0953$+$414&  K348$-$7     &   0.23410&   3111 & 1&  45.40$\pm$0.022&       \nodata  &3& 8.715$^{+0.107}_{-0.143}$&       \nodata  & 8.441$^{+0.084}_{-0.104}$&  \\
              &               &          &   3224 & 2&  45.07$\pm$0.011& 43.63$\pm$0.026&2& 8.536$^{+0.097}_{-0.125}$& 8.715$^{+0.115}_{-0.157}$&                 &  \\
     NGC3783  &               &   0.00973&   3555 & 2&  43.20$\pm$0.010& 41.52$\pm$0.041&2& 7.443$^{+0.090}_{-0.113}$& 7.474$^{+0.086}_{-0.107}$& 7.474$^{+0.072}_{-0.087}$&  \\
     NGC4151  &               &   0.00332&   6421 & 2&  42.58$\pm$0.016& 40.90$\pm$0.034&2& 7.566$^{+0.108}_{-0.143}$& 7.596$^{+0.099}_{-0.128}$& 7.124$^{+0.129}_{-0.184}$&  \\
  PG1226$+$023&      3C273    &   0.15830&   3500 & 1&  46.02$\pm$0.017&       \nodata  &3& 9.204$^{+0.128}_{-0.183}$&        \nodata & 8.947$^{+0.083}_{-0.103}$&  \\
              &               &          &   3627 & 2&  46.06$\pm$0.014& 44.27$\pm$0.043&2& 9.262$^{+0.130}_{-0.186}$& 9.222$^{+0.139}_{-0.206}$&                 &  \\
  PG1229$+$204&  TON1542      &   0.06301&   3335 & 1&  44.39$\pm$0.012&       \nodata  &3& 8.139$^{+0.082}_{-0.101}$&        \nodata & 7.865$^{+0.171}_{-0.285}$&  \\
              &               &          &   3504 & 2&  44.10$\pm$0.012& 42.34$\pm$0.029&2& 7.997$^{+0.080}_{-0.098}$& 7.978$^{+0.083}_{-0.102}$&                 &  \\
  PG1307$+$085&               &   0.15500&   5307 & 1&  45.01$\pm$0.028&       \nodata  &3& 8.930$^{+0.096}_{-0.123}$&        \nodata & 8.643$^{+0.107}_{-0.142}$&  \\
              &               &          &   5315 & 2&  44.73$\pm$0.024& 43.15$\pm$0.050&2& 8.756$^{+0.089}_{-0.111}$& 8.848$^{+0.102}_{-0.134}$&                 &  \\
      Mrk279  &               &   0.03045&   5411 & 2&  43.82$\pm$0.017& 42.07$\pm$0.045&2& 8.198$^{+0.080}_{-0.099}$& 8.183$^{+0.083}_{-0.103}$& 7.543$^{+0.102}_{-0.133}$&  \\
  PG1411$+$442&  PB1732       &   0.08960&   2640 & 1&  44.62$\pm$0.014&        \nodata &3& 8.080$^{+0.086}_{-0.107}$&        \nodata & 8.646$^{+0.124}_{-0.174}$&  \\
              &               &          &   2611 & 2&  44.39$\pm$0.010& 42.72$\pm$0.045&2& 7.924$^{+0.082}_{-0.101}$& 7.958$^{+0.091}_{-0.115}$&                 &  \\
     NGC5548  &               &   0.01717&   5822 & 2&      \nodata    & 41.53$\pm$0.038&2&       \nodata     & 7.907$^{+0.085}_{-0.107}$& 7.827$^{+0.017}_{-0.017}$&  \\
  PG1426$+$015&  Mrk1383      &   0.08647&   6808 & 1&  44.88$\pm$0.014&       \nodata  &3& 9.065$^{+0.092}_{-0.116}$&       \nodata  & 9.113$^{+0.113}_{-0.153}$&  \\
              &               &          &   6624 & 2&      \nodata    & 42.61$\pm$0.041&2&       \nodata     & 8.699$^{+0.088}_{-0.111}$&                 &  \\
      Mrk817  &               &   0.03145&   4657 & 2&  43.96$\pm$0.022& 42.15$\pm$0.048&2& 8.156$^{+0.080}_{-0.098}$& 8.106$^{+0.084}_{-0.104}$& 7.694$^{+0.063}_{-0.074}$&  \\
  PG1613$+$658&  Mrk876       &   0.12900&   8441 & 1&  44.84$\pm$0.018&        \nodata &3& 9.226$^{+0.091}_{-0.115}$&      \nodata   & 8.446$^{+0.165}_{-0.270}$&  \\
              &               &          &   8662 & 2&      \nodata    & 42.94$\pm$0.044&2&       \nodata     & 9.139$^{+0.096}_{-0.123}$&                 &  \\
  PG1617$+$175&  Mrk877       &   0.11240&   5316 & 1&  44.85$\pm$0.014&        \nodata &3& 8.830$^{+0.091}_{-0.115}$&       \nodata  & 8.774$^{+0.091}_{-0.115}$&  \\
              &               &          &   5636 & 2&      \nodata    & 42.47$\pm$0.027&2&       \nodata     & 8.471$^{+0.084}_{-0.105}$&                 &  \\
\tablebreak
  PG1700$+$518&               &   0.29200&   2185 & 1&  45.68$\pm$0.025&        \nodata &3& 8.585$^{+0.117}_{-0.160}$&       \nodata  & 8.893$^{+0.091}_{-0.103}$& \\
              &               &          &   2127 & 2&  45.47$\pm$0.010& 43.57$\pm$0.041&2& 8.427$^{+0.109}_{-0.146}$& 8.319$^{+0.114}_{-0.156}$&                            & \\
     3C390.3  &               &   0.05610&  40000 & 2&  43.82$\pm$0.017& 42.17$\pm$0.052&2& 8.893$^{+0.080}_{-0.099}$& 8.943$^{+0.085}_{-0.105}$& 8.458$^{+0.087}_{-0.110}$&  \\
      Mrk509  &               &   0.03440&   3424 & 2&      \nodata    & 42.72$\pm$0.042&2&     \nodata       & 8.194$^{+0.090}_{-0.114}$& 8.155$^{+0.035}_{-0.038}$
&  \\
  PG2130$+$099& II\,Zw\,136   &   0.06298&   2294 & 1&  44.54$\pm$0.012&        \nodata &3& 7.906$^{+0.084}_{-0.104}$&      \nodata   & 8.660$^{+0.049}_{-0.056}$&  \\
              &               &          &   2901 & 2&  44.28$\pm$0.011& 42.57$\pm$0.045&2& 7.947$^{+0.081}_{-0.099}$& 7.958$^{+0.088}_{-0.111}$&                 &  \\
     NGC7469  &               &   0.01632&   2639 & 2&  43.74$\pm$0.012& 42.01$\pm$0.045&2& 7.524$^{+0.081}_{-0.099}$& 7.521$^{+0.083}_{-0.103}$& 7.086$^{+0.047}_{-0.053}$&  \\
\enddata
\tablenotetext{a}{Redshifts are obtained from the NASA/IPAC Extragalactic Database.}
\tablenotetext{b}{FWHM(\hb) measured in the single-epoch spectrum in units of km s$^{-1}$.}
\tablenotetext{c}{The central mass (and uncertainties) estimated based on single-epoch optical
        spectroscopy.}
\tablenotetext{d}{The central mass (and uncertainties) determined from multi-epoch
        spectrophotometry and reverberation mapping techniques. All values are 
	adopted from  Peterson \et (2004).
}
\vspace{-0.2cm}
\tablerefs{
(1) Boroson \& Green 1992;
(2) Marziani \et 2003;
(3) Neugebauer \et 1987;
(4) Schmidt, \& Green 1983; Kellerman, \et 1989; Schmidt, Schneider, \& Gunn (1995)
}
\end{deluxetable}

\begin{flushleft}
\begin{deluxetable}{llclcllcccl} 
\tablewidth{630pt}
\hspace{-14cm}
\rotate
\tablecaption{Ultraviolet Spectral Parameters and Masses \label{UV_pars.tab}}
\tabletypesize{\tiny}
\tablecolumns{11}
\tablehead{
\colhead{ } &
\colhead{ Date} &
\colhead{ } &
\colhead{ Telescope/} &
\colhead{ Resolution} &
\colhead{ FWHM(C{\sc IV})\tablenotemark{a}} &
\colhead{ $\sigma$(C{\sc IV})\tablenotemark{b}} &
\colhead{ log\,[$\lambda$L$_{\lambda}$/erg s$^{-1}$]} &
\colhead{ log\,[M/M$_{\odot}$]\tablenotemark{c}} &
\colhead{ log\,[M/M$_{\odot}$]\tablenotemark{c}} &
\colhead{ } \\
\colhead{ Object\phantom{mmmmm}} &
\colhead{ Observed} &
\colhead{ $z$} &
\colhead{ Instrument} &
\colhead{ (\AA)} &
\colhead{ (km s$^{-1}$)} &
\colhead{ (km s$^{-1}$)} &
\colhead{ (1350\AA)} &
\colhead{ (FWHM(C{\sc IV}),SE)} &
\colhead{ ($\sigma$(C{\sc IV}),SE)} &
\colhead{ Note} \\
\colhead{(1)} &
\colhead{(2)} &
\colhead{(3)} &
\colhead{(4)} &
\colhead{(5)} &
\colhead{(6)} &
\colhead{(7)} &
\colhead{(8)} &
\colhead{(9)} &
\colhead{(10)} &
\colhead{(11)} \\
}
\startdata
%
Mrk\,335 &1989/10/29~-- \\
         & \phn --~1991/06/30 &0.02578&   \iue/SWP  &  6.0  &   2291$\pm$\phn\phn27 &   2116$\pm$\phn160 &  44.173$\pm$0.020  &  7.471$^{+0.018}_{-0.018}$ &  7.469$^{+0.062}_{-0.073}$ & \\
        &    1990/10/12  &           &    \hut      &  3.0  &   1741$\pm$\phn\phn99  &  1806$\pm$\phn360 &  44.291$\pm$0.078  &  7.295$^{+0.062}_{-0.073}$ &  7.394$^{+0.150}_{-0.231}$ & \\
         &   1994/12/16  &           &    \hst/FOS  &  1.4  &   2023$\pm$\phn\phn17  &  2140$\pm$\phn\phn93 &  44.262$\pm$0.013  &  7.410$^{+0.018}_{-0.019}$ &  7.526$^{+0.040}_{-0.044}$ & \\
PG\,0026$+$129 & 1994/11/27 & 0.14200 &    \hst/FOS &  1.4  &   1837$\pm$\phn 136  &  3364$\pm$\phn\phn70 &  45.165$\pm$0.025  &  7.805$^{+0.087}_{-0.108}$ &  8.397$^{+0.068}_{-0.080}$ & flg\\
PG\,0052$+$251 & 1992/06/29 & 0.15500 &   \iue/SWP  &  6.0  &   3983$\pm$\phn 370  &  5118$\pm$\phn486 &  45.265$\pm$0.037  &  8.530$^{+0.100}_{-0.130}$ &  8.815$^{+0.101}_{-0.132}$ & \\
               & 1993/07/22 &         &    \hst/FOS &  2.2  &   5192$\pm$\phn 251  &  5083$\pm$\phn437 &  45.176$\pm$0.041  &  8.713$^{+0.078}_{-0.095}$ &  8.761$^{+0.094}_{-0.120}$ & \\
Fairall\,9&  1993/01/22  & 0.04702   &    \hst/FOS  &  2.2  &   2593$\pm$\phn\phn65  &  2981$\pm$\phn197 &  44.470$\pm$0.028  &  7.736$^{+0.037}_{-0.040}$ &  7.924$^{+0.061}_{-0.071}$ & \\
         & 1994/04/28~--12/26 &         & \iue/SWP$^d$  &  6.0  &   2831$\pm$\phn\phn40  &  3532$\pm$\phn\phn92 &  44.582$\pm$0.011  &  7.871$^{+0.036}_{-0.039}$ &  8.131$^{+0.040}_{-0.044}$ & ref(1)\\
         &   1995/03/11  &           &    \hut      &  3.0  &   2370$\pm$\phn 151  &  2978$\pm$\phn508 &  44.759$\pm$0.126  &  7.811$^{+0.088}_{-0.111}$ &  8.076$^{+0.143}_{-0.214}$ & \\
Mrk\,590&    1991/01/14  & 0.02638   &    \iue/SWP  &  6.0  &   4839$\pm$\phn\phn59  &  3574$\pm$\phn141 &  44.119$\pm$0.029  &  8.091$^{+0.020}_{-0.020}$ &  7.895$^{+0.037}_{-0.040}$ & \\
3C\,120 &    1993/08/25  & 0.03301   &    \iue/SWP  &  6.0  &   3302$\pm$\phn\phn75  &  3199$\pm$\phn169 &  44.943$\pm$0.039  &  8.196$^{+0.059}_{-0.068}$ &  8.236$^{+0.070}_{-0.083}$ & \\
       & 1994/02/19,27;03/11 &       &    \iue/SWP  &  6.0  &   3278$\pm$\phn 105  &  3409$\pm$\phn286 &  44.617$\pm$0.056  &  8.017$^{+0.052}_{-0.059}$ &  8.119$^{+0.079}_{-0.097}$ & \\
Akn\,120& 1988/01/20;02/12; & 0.03230&    \iue/SWP  &  6.0  &   3989$\pm$\phn 451  &  3795$\pm$\phn165 &  44.634$\pm$0.021  &  8.197$^{+0.095}_{-0.121}$ &  8.221$^{+0.052}_{-0.059}$ & \\
        &    1991/01/13  &           &              &       &                  &                &                    &                   &                  & \\
        &    1995/07/29  &           &    \hst/FOS  &  1.4  &   3945$\pm$\phn\phn42  &  3240$\pm$\phn149 &  44.482$\pm$0.022  &  8.106$^{+0.031}_{-0.034}$ &  8.002$^{+0.048}_{-0.054}$ & \\
Mrk\,79 &    1978/04/15  & 0.02219   &    \iue/SWP  &  6.0  &   3182$\pm$\phn 521  &  3344$\pm$\phn222 &  43.879$\pm$0.039  &  7.600$^{+0.124}_{-0.175}$ &  7.710$^{+0.058}_{-0.067}$ & flg\\
        &    1979/11/14  &           &    \iue/SWP  &  6.0  &   3049$\pm$\phn 128  &  2971$\pm$\phn248 &  43.495$\pm$0.058  &  7.360$^{+0.053}_{-0.060}$ &  7.404$^{+0.077}_{-0.094}$ & flg\\
        &    1982/12/28  &           &    \iue/SWP  &  6.0  &   3113$\pm$\phn 122  &  3803$\pm$\phn388 &  43.726$\pm$0.065  &  7.500$^{+0.048}_{-0.054}$ &  7.741$^{+0.087}_{-0.109}$ & flg\\
Mrk\,110 &   1988/02/28  & 0.03529   &    \iue/SWP  &  6.0  &   2990$\pm$\phn\phn64  &  2601$\pm$\phn272 &  43.770$\pm$0.050  &  7.488$^{+0.034}_{-0.037}$ &  7.434$^{+0.086}_{-0.108}$ & \\
         &   1988/02/29  &           &    \iue/SWP  &  6.0  &   1638$\pm$\phn\phn59  &  2576$\pm$\phn231 &  43.876$\pm$0.081  &  7.022$^{+0.051}_{-0.057}$ &  7.482$^{+0.081}_{-0.100}$ & \\
PG\,0953$+$414 &1991/06/17; \\
        &   1992/11/04--05 & 0.23410 &    \hst/FOS  &  1.5  &   2873$\pm$\phn\phn57  &  3512$\pm$\phn361 &  45.588$\pm$0.031  &  8.418$^{+0.089}_{-0.111}$ &  8.659$^{+0.115}_{-0.157}$ & \\
NGC\,3516 &  1995/03/12 &  0.00884   &    \hut      &  2.0  &   4675$\pm$\phn 538  &  3311$\pm$\phn372 &  42.830$\pm$0.093  &  7.379$^{+0.115}_{-0.157}$ &  7.146$^{+0.114}_{-0.154}$ & abs\\
        &    1995/12/30 &            &    \hst/FOS  &  1.4  &   4875$\pm$\phn\phn17  &  3132$\pm$\phn\phn64 &  42.823$\pm$0.017  &  7.411$^{+0.066}_{-0.078}$ &  7.094$^{+0.068}_{-0.080}$ & abs\\
        &    1996/02/21 &            &    \hst/FOS  &  1.4  &   5147$\pm$\phn 103  &  3245$\pm$\phn\phn84 &  43.192$\pm$0.013  &  7.654$^{+0.049}_{-0.055}$ &  7.320$^{+0.051}_{-0.058}$ & abs\\
        &    1996/04/13 &            &    \hst/FOS  &  1.4  &   4729$\pm$\phn\phn28  &  3430$\pm$\phn\phn92 &  43.143$\pm$0.013  &  7.554$^{+0.049}_{-0.056}$ &  7.342$^{+0.053}_{-0.061}$ & abs\\
        &    1996/08/14 &            &    \hst/FOS  &  1.4  &   4525$\pm$\phn\phn97  &  3137$\pm$\phn\phn79 &  43.030$\pm$0.012  &  7.456$^{+0.057}_{-0.066}$ &  7.205$^{+0.058}_{-0.067}$ & abs\\
        &    1996/11/28 &            &    \hst/FOS  &  1.4  &   3940$\pm$\phn\phn18  &  2834$\pm$\phn\phn95 &  42.485$\pm$0.034  &  7.047$^{+0.084}_{-0.104}$ &  6.828$^{+0.088}_{-0.110}$ & abs,flg\\
        &    1998/04/13 &          & \hst/STIS/G140L& 0.88  &   4912$\pm$\phn\phn23  &  3973$\pm$\phn\phn36 &  42.793$\pm$0.012   &  7.402$^{+0.067}_{-0.080}$ &  7.284$^{+0.068}_{-0.080}$ & abs\\
NGC\,3783& 1991/12/21 -- \\
         & \phn -- 1992/07/29 &0.00973&  \iue/SWP$^d$& 6.0  &   2831$\pm$\phn\phn22  &  3273$\pm$\phn100  & 43.601$\pm$0.014   &  7.352$^{+0.025}_{-0.027}$ &  7.545$^{+0.035}_{-0.038}$ & ref(2) \\
         &   1992/07/27  & 0.00973   &   \hst/FOS   & 1.95  &   2308$\pm$\phn\phn17  &  3179$\pm$\phn185  & 43.744$\pm$0.022   &  7.250$^{+0.020}_{-0.021}$ &  7.595$^{+0.051}_{-0.058}$ & \\
NGC\,4051& 2000/03/25 & 0.00234    & \hst/STIS/E140M & 0.13 &   1319$\pm$\phn\phn13  &  1713$\pm$\phn227 &  41.373$\pm$0.058  &  5.507$^{+0.137}_{-0.201}$ &  5.801$^{+0.163}_{-0.263}$ & abs,ref(3) \\
NGC\,4151& 1993/11/27~--12/15&0.00332&  \iue/SWP$^d$& 6.0   &   6929$\pm$\phn\phn76   & 5220$\pm$\phn123  & 43.224$\pm$0.010   &  7.929$^{+0.045}_{-0.051}$ &  7.750$^{+0.048}_{-0.054}$ & abs,ref(5) \\
        & 1995/03/04--05 &           &    \hut      &  2.0  &   5418$\pm$\phn 150  &  4604$\pm$\phn249 &  43.340$\pm$0.019  &  7.777$^{+0.045}_{-0.050}$ &  7.703$^{+0.058}_{-0.067}$ & abs\\
        &     1995/03/07 &           &    \hut      &  2.0  &   5062$\pm$\phn\phn51  &  4651$\pm$\phn371 &  43.396$\pm$0.029  &  7.747$^{+0.039}_{-0.042}$ &  7.741$^{+0.073}_{-0.088}$ & abs\\
        &     1995/03/10 &           &    \hut      &  2.0  &   5246$\pm$\phn\phn44  &  4675$\pm$\phn397 &  43.396$\pm$0.031  &  7.778$^{+0.039}_{-0.043}$ &  7.745$^{+0.077}_{-0.093}$ & abs\\
        &     1995/03/13 &           &    \hut      &  2.0  &   5752$\pm$\phn 144  &  4585$\pm$\phn321 &  43.418$\pm$0.023  &  7.870$^{+0.041}_{-0.045}$ &  7.740$^{+0.066}_{-0.078}$ & abs\\
        &     1995/03/15 &           &    \hut      &  2.0  &   5173$\pm$\phn 593  &  4664$\pm$\phn475 &  43.354$\pm$0.044  &  7.744$^{+0.098}_{-0.126}$ &  7.721$^{+0.090}_{-0.113}$ & abs\\
        & 1998/02/10,06/01 &   & \hst/STIS/G140L & 0.88  &   3509$\pm$\phn\phn10  &  4384$\pm$\phn\phn66 &  43.038$\pm$0.006  &  7.239$^{+0.054}_{-0.062}$ &  7.500$^{+0.056}_{-0.064}$ & abs,ref(4) \\
3C\,273 & 1991/01/14,15,17& 0.15834  &     \hst/FOS &  1.5  &   3941$\pm$\phn 266  &  4027$\pm$\phn322 &  46.336$\pm$0.008  &  9.088$^{+0.130}_{-0.187}$ &  9.174$^{+0.134}_{-0.194}$ & \\
        & 1991/01/23;06/17&          &     \iue/SWP &  6.0  &   3673$\pm$\phn 420  &  3604$\pm$\phn954 &  46.467$\pm$0.026  &  9.097$^{+0.150}_{-0.230}$ &  9.147$^{+0.212}_{-0.432}$ & \\
        &     1991/07/09  &          &     \hst/FOS &  2.2  &   3693$\pm$\phn 843  &  3495$\pm$\phn501 &  46.089$\pm$0.026  &  8.901$^{+0.188}_{-0.338}$ &  8.920$^{+0.149}_{-0.228}$ & \\
        & 1991/12/07,12   &          &     \iue/SWP &  6.0  &   3834$\pm$\phn 155  &  3425$\pm$\phn556 &  46.309$\pm$0.018  &  9.050$^{+0.124}_{-0.174}$ &  9.019$^{+0.163}_{-0.264}$ & \\
        &     1992/01/05  &          &     \iue/SWP &  6.0  &   3645$\pm$\phn 447  &  3098$\pm$\phn811 &  46.323$\pm$0.026  &  9.014$^{+0.148}_{-0.226}$ &  8.940$^{+0.208}_{-0.414}$ & \\
\tablebreak
3C\,273&1992/01/18;02/01,15 & 0.15834&     \iue/SWP &  6.0  &   4671$\pm$\phn 127  &  3597$\pm$\phn560 &  46.377$\pm$0.015  &  9.258$^{+0.125}_{-0.176}$ &  9.098$^{+0.162}_{-0.262}$ & \\
       &1992/05/06;06/05  &          &     \iue/SWP &  6.0  &   4349$\pm$\phn 420  &  3643$\pm$\phn762 &  46.555$\pm$0.017  &  9.290$^{+0.147}_{-0.224}$ &  9.203$^{+0.190}_{-0.345}$ & \\
       &1992/06/05,21;07/05 &        &     \iue/SWP &  6.0  &   4167$\pm$\phn 178  &  3685$\pm$\phn434 &  46.575$\pm$0.011  &  9.263$^{+0.135}_{-0.198}$ &  9.224$^{+0.154}_{-0.242}$ & \\
       &1992/12/17,28,29  &          &     \iue/SWP &  6.0  &   3784$\pm$\phn 219  &  3445$\pm$\phn706 &  46.437$\pm$0.020  &  9.107$^{+0.133}_{-0.192}$ &  9.092$^{+0.185}_{-0.329}$ & \\
        &     1992/12/31  &          &     \iue/SWP &  6.0  &   4715$\pm$\phn 764  &  3669$\pm$\phn633 &  46.274$\pm$0.029  &  9.211$^{+0.162}_{-0.262}$ &  9.060$^{+0.167}_{-0.274}$ & \\
        &     1993/01/02  &          &     \iue/SWP &  6.0  &   4354$\pm$1952  &  3442$\pm$\phn725 &  46.012$\pm$0.052  &  9.003$^{+0.288}_{-1.228}$ &  8.866$^{+0.179}_{-0.309}$ & \\
        &     1993/01/03  &          &     \iue/SWP &  6.0  &   3434$\pm$\phn 238  &  1785$\pm$\phn574 &  46.213$\pm$0.045  &  8.903$^{+0.127}_{-0.180}$ &  8.402$^{+0.234}_{-0.543}$ & \\
        &1993/01/04--06,09 &         &     \iue/SWP &  6.0  &   4554$\pm$\phn 131  &  2953$\pm$\phn517 &  46.425$\pm$0.015  &  9.261$^{+0.127}_{-0.181}$ &  8.952$^{+0.172}_{-0.288}$ & \\
        &     1993/01/16  &          &     \iue/SWP &  6.0  &   3950$\pm$\phn 884  &  3714$\pm$\phn801 &  46.336$\pm$0.027  &  9.091$^{+0.191}_{-0.350}$ &  9.104$^{+0.187}_{-0.337}$ & \\
        &     1993/02/01a &          &     \iue/SWP &  6.0  &   4454$\pm$\phn 629  &  3740$\pm$\phn491 &  46.398$\pm$0.021  &  9.227$^{+0.157}_{-0.249}$ &  9.143$^{+0.153}_{-0.239}$ & \\
        &     1993/02/01b &          &     \iue/SWP &  6.0  &   3128$\pm$\phn 198  &  3398$\pm$\phn423 &  46.419$\pm$0.018  &  8.932$^{+0.133}_{-0.193}$ &  9.071$^{+0.151}_{-0.235}$ & \\
        & 1993/02/13;05/12,27; \\ 
        & 1993/12/14,27;       \\ 
        & 1994/01/08,14,24 &         &     \iue/SWP &  6.0  &   4599$\pm$\phn 163  &  3580$\pm$\phn394 &  46.395$\pm$0.015 &  9.253$^{+0.127}_{-0.180}$ &  9.103$^{+0.145}_{-0.220}$ & \\
       &1994/01/24;02/07,23 &        &     \iue/SWP &  6.0  &   4098$\pm$\phn 143  &  3793$\pm$\phn568 &  46.396$\pm$0.020  &  9.154$^{+0.127}_{-0.180}$ &  9.154$^{+0.160}_{-0.257}$ & \\
       &1994/02/07,23   &            &     \iue/SWP &  6.0  &   4617$\pm$\phn 145  &  3776$\pm$\phn943 &  46.292$\pm$0.033  &  9.203$^{+0.122}_{-0.171}$ &  9.095$^{+0.202}_{-0.390}$ & \\
       &1994/05/03,15     &          &     \iue/SWP &  6.0  &   4426$\pm$\phn 337  &  3685$\pm$\phn438 &  46.444$\pm$0.015  &  9.246$^{+0.137}_{-0.201}$ &  9.154$^{+0.150}_{-0.232}$ & \\
        &     1994/05/15  &          &     \iue/SWP &  6.0  &   3058$\pm$\phn\phn41  &  3015$\pm$\phn519 &  46.382$\pm$0.023  &  8.893$^{+0.124}_{-0.175}$ &  8.947$^{+0.169}_{-0.281}$ & \\
       &1994/05/30;06/15,20,27&         &  \iue/SWP &  6.0  &   4278$\pm$\phn\phn80  &  3697$\pm$\phn358 &  46.461$\pm$0.012  &  9.226$^{+0.128}_{-0.182}$ &  9.166$^{+0.143}_{-0.216}$ & \\
        &     1994/12/30  &           &    \iue/SWP &  6.0  &   4567$\pm$\phn 134  &  3795$\pm$\phn688 &  46.387$\pm$0.025  &  9.243$^{+0.126}_{-0.178}$ &  9.149$^{+0.174}_{-0.293}$ & \\
        & 1995/01/03,05-07,09; \\
        & 1995/01/12,29;02/14  &        &  \iue/SWP &  6.0 &    4344$\pm$\phn116  &  3730$\pm$\phn313 &  46.411$\pm$0.013  &  9.213$^{+0.126}_{-0.179}$ &  9.147$^{+0.138}_{-0.203}$ & \\
        & 1995/05/03,17,18,31; \\
        & 1995/06/14,27        &        &  \iue/SWP &  6.0  &   4387$\pm$\phn\phn95  &  3532$\pm$\phn350  & 46.337$\pm$0.014  &  9.182$^{+0.123}_{-0.171}$ &  9.061$^{+0.140}_{-0.207}$ & \\
       & 1995/12/21,26,31;  \\
        &1996/01/05,10,16,18,20; &      &           &       &         &          &               &              &                    & \\
        &1996/01/22,24,26,28,30 &       &  \iue/SWP &  6.0  &   4430$\pm$\phn237  &  3503$\pm$\phn614  & 46.380$\pm$0.018  &  9.213$^{+0.129}_{-0.185}$ &  9.076$^{+0.171}_{-0.285}$ & \\
        &    1996/01/20  &          &     \iue/SWP  & 6.0   &   4452$\pm$\phn641  &  3314$\pm$\phn998 &  46.222$\pm$0.040  &  9.134$^{+0.153}_{-0.239}$ &  8.944$^{+0.225}_{-0.492}$ & \\
PG\,1229$+$204 &1982/05,1983/06 & 0.06301& \iue/SWP & 6.0  &   3391$\pm$\phn205  &  3241$\pm$\phn457  & 44.654$\pm$0.028  &  8.066$^{+0.062}_{-0.073}$ &  8.094$^{+0.113}_{-0.154}$ & \\
PG\,1307$+$085 & 1993/07/21 & 0.15500 &  \hst/FOS   &  2.2  &   3465$\pm$\phn168  &  3687$\pm$\phn290  & 45.012$\pm$0.039  &  8.275$^{+0.071}_{-0.085}$ &  8.396$^{+0.085}_{-0.106}$ & \\
Mrk\,279&    1995/03/05  & 0.03045   &    \hut      &  2.0  &   4126$\pm$\phn487  &  3118$\pm$\phn414 &  43.795$\pm$0.118  &  7.781$^{+0.106}_{-0.141}$ &  7.605$^{+0.115}_{-0.157}$ & \\
        &    1995/03/11  &           &    \hut      &  2.0  &   3876$\pm$\phn\phn99  &  3286$\pm$\phn511 &  43.754$\pm$0.127  &  7.705$^{+0.067}_{-0.079}$ &  7.629$^{+0.130}_{-0.186}$ & \\
NGC\,5548&1988/12/14~-- \\
        & \phn --~1989/08/07&0.01717  & \iue/SWP$^d$&  6.0  &   4790$\pm$\phn\phn67  &  4815$\pm$\phn257 &  43.654$\pm$0.022  &  7.836$^{+0.026}_{-0.028}$ &  7.908$^{+0.049}_{-0.055}$ & ref(6)\\
        & 1993/04/19~--05/27 &        & \hst/FOS$^d$&  1.9  &   4096$\pm$\phn\phn14  &  3973$\pm$\phn\phn34 &  43.568$\pm$0.006  &  7.655$^{+0.026}_{-0.027}$ &  7.695$^{+0.026}_{-0.028}$ & ref(7)\\
         &     1995/03/14 &           &    \hut     &  2.0  &   3280$\pm$\phn\phn27  &  5050$\pm$\phn787 &  43.773$\pm$0.069  &  7.570$^{+0.038}_{-0.042}$ &  8.012$^{+0.122}_{-0.170}$ & \\
PG\,1426$+$015 & 1985/03/01--02 & 0.08647 & \iue/SWP&  6.0  &   3778$\pm$\phn448 &   4101$\pm$\phn391 &  45.295$\pm$0.023 &  8.500$^{+0.113}_{-0.154}$ &  8.638$^{+0.101}_{-0.133}$ & \\
Mrk\,817&   1981/11/06,07; \\
        &    1982/07/18    &0.03145   &    \iue/SWP &  6.0  &   4027$\pm$\phn\phn71  &  4062$\pm$\phn289 &  44.123$\pm$0.022  &  7.934$^{+0.020}_{-0.021}$ &  8.009$^{+0.059}_{-0.068}$ & \\
PG\,1613$+$658 & 1990/12/02,05,08,10; \\
         & 1991/02/25      & 0.12900   &   \iue/SWP &  6.0  &   5902$\pm$\phn136   & 3965$\pm$\phn215 &  45.221$\pm$0.023  &  8.848$^{+0.071}_{-0.085}$ &  8.570$^{+0.080}_{-0.098}$ & \\
PG\,1617$+$175 & 1993/05/13 & 0.11244  &   \iue/SWP  & 6.0  &   4558$\pm$1763  &  3383$\pm$1036  & 44.784$\pm$0.108  &  8.392$^{+0.253}_{-0.683}$ &  8.200$^{+0.214}_{-0.439}$ & flg\\
3C\,390 & 1994/12/31~-- \\
        & \phn --~1996/03/05 &0.05610 & \iue/SWP$^d$& 6.0  &   5895$\pm$\phn\phn32  &  4454$\pm$\phn\phn53 &  44.073$\pm$0.022  &  8.239$^{+0.013}_{-0.013}$ &  8.062$^{+0.016}_{-0.016}$ & ref(8)\\
        &    1996/03/31  &            &   \hst/FOS  & 1.4  &   4676$\pm$2386  &  4444$\pm$\phn263 &  43.909$\pm$0.035  &  7.951$^{+0.306}_{-7.951}$ &  7.973$^{+0.052}_{-0.059}$ & \\
\tablebreak
Mrk\,509&    1992/02/22  & 0.03440   &   \iue/SWP  &  6.0  &   5035$\pm$\phn298  &  3558$\pm$\phn205 &  44.641$\pm$0.029  &  8.403$^{+0.061}_{-0.072}$ &  8.168$^{+0.061}_{-0.070}$ & \\
        &    1992/06/21  &           &   \hst/FOS  &  2.0  &   4345$\pm$\phn\phn49  &  3426$\pm$\phn115 &  44.532$\pm$0.015  &  8.217$^{+0.033}_{-0.036}$ &  8.078$^{+0.042}_{-0.046}$ & \\
       & 1992/10/25,26,29 &          &   \iue/SWP  &  6.0  &   4973$\pm$\phn233  &  3647$\pm$\phn172 &  44.803$\pm$0.020  &  8.478$^{+0.060}_{-0.069}$ &  8.276$^{+0.060}_{-0.069}$ & \\
        &    1993/10/27;  \\
        &    1993/11/09   &          &   \iue/SWP  &  6.0  &   4961$\pm$\phn218  &  3127$\pm$\phn226 &  44.552$\pm$0.033  &  8.343$^{+0.050}_{-0.057}$ &  8.009$^{+0.068}_{-0.080}$ & \\
        &    1995/03/16  &           &   \hut      &  2.0  &   3716$\pm$\phn228  &  3174$\pm$\phn448 &  44.706$\pm$0.071  &  8.173$^{+0.071}_{-0.086}$ &  8.104$^{+0.118}_{-0.162}$ & \\
PG\,2130$+$099 & 1995/07/24 & 0.06298&   \hst/GHRS & 0.65  &   2113$\pm$\phn119  &  2390$\pm$\phn184  & 44.692$\pm$0.025  &  7.676$^{+0.061}_{-0.071}$ &  7.850$^{+0.073}_{-0.088}$ & \\
NGC\,7469 &  1996/06/10~-- \\
        & \phn --~07/29 & 0.01632    & \iue/SWP$^d$&  6.0  &   3094$\pm$\phn\phn53  &  3379$\pm$\phn182 &  43.774$\pm$0.016  &  7.520$^{+0.021}_{-0.022}$ &  7.664$^{+0.047}_{-0.052}$ & ref(9)\\
        &     1996/06/18  &          &   \hst/FOS  &  1.4  &   2860$\pm$\phn\phn12  &  3266$\pm$\phn110 &  43.679$\pm$0.015  &  7.402$^{+0.021}_{-0.022}$ &  7.584$^{+0.035}_{-0.038}$ & abs\\
\enddata
\tablenotetext{a}{Spectral resolution corrected line width FWHM(\civ) measured in the 
single-epoch spectrum in units of km s$^{-1}$.}
\tablenotetext{b}{Spectral resolution corrected line dispersion $\sigma$(\civ) measured 
in the single-epoch spectrum in units of km s$^{-1}$.}
\tablenotetext{c}{The central mass (and uncertainties; see text), listed in logarithmic 
units, estimated based on the single-epoch UV spectroscopic measurements and the calibrations
(equations~3 and~4) described in the text.}
\tablenotetext{d}{This spectrum is the average of the data obtained during an AGN Watch monitoring campaign. 
An average spectrum was used to avoid cluttering the data base.}
\tablecomments{abs --- Absorption in \civ{} profile is interpolated across; flg --- flagged objects. See main text (\S~\ref{omitteddata}) for details.}
\tablerefs{
(1) Rodriguez-Pascual et al.\ 1997;
(2) Reichert et al.\ 1994;
(3) Collinge et al.\ 2001; 
(4) Crenshaw et al.\ 2001 
(5) Crenshaw et al.\ 1996;
(6) Clavel, \et 1991; 
(7) Korista, \et 1995;
(8) O'Brien, \et 1998; 
(9) Wanders, \et 1997. 
}
\end{deluxetable}
\end{flushleft}

\begin{deluxetable}{lccccccc}
\tablecolumns{8}
\tabletypesize{\footnotesize}
\tablewidth{470pt}
\tablecaption{Regression Results
\label{regres.tab}}
\tablehead{
\colhead{} &
\colhead{} &
\multicolumn{2}{c}{\underline{\phantom{mmmm}BCES Fit\phantom{mmmm}}} &
\multicolumn{3}{c}{\underline{\phantom{mmmmmm}FITEXY Fit\phantom{mmmmmm}}} &
\colhead{} \\ 
\colhead{Unscaled Mass Estimate} &
\colhead{N} &
\colhead{Slope} &
\colhead{Intercept} &
\colhead{Slope} &
\colhead{Intercept} &
\colhead{Scatter\tablenotemark{a}} &
\colhead{Note} \\
\colhead{(1)} &
\colhead{(2)} &
\colhead{(3)} &
\colhead{(4)} &
\colhead{(5)} &
\colhead{(6)} &
\colhead{(7)} &
\colhead{(8)} 
}
\startdata
$\mu$[FWHM(\hb),L(5100\AA)]     & 34 & 1.14$\pm$0.11 & 6.69$\pm$0.15 & 1.19$\pm$0.19 & 6.64$\pm$0.24 & 51\% & \\ 
$\mu$[FWHM(\hb),L(\hb)]         & 28 & 1.04$\pm$0.13 & 6.60$\pm$0.18 & 1.11$\pm$0.21 & 6.52$\pm$0.28 & 50\% & \\
$\mu$[FWHM(CIV),L(1350\AA)]     & 85 & 0.94$\pm$0.05 & 6.71$\pm$0.10 & 1.03$\pm$0.07 & 6.55$\pm$0.10 & 42\% &\\
                             & 84 & 0.97$\pm$0.05 & 6.65$\pm$0.10 & 1.06$\pm$0.08 & 6.52$\pm$0.10 & 42\% & 1 \\ 
                             & 34 & 1.11$\pm$0.18 & 6.60$\pm$0.25 & 1.19$\pm$0.17 & 6.47$\pm$0.20 & 52\% & 2 \\ 
                             & 33 & 1.29$\pm$0.16 & 6.35$\pm$0.21 & 1.32$\pm$0.23 & 6.31$\pm$0.26 & 51\% & 1,2 \\
$\mu$[$\sigma$(CIV),L(1350\AA)] & 85 & 1.01$\pm$0.06 & 6.67$\pm$0.11 & 1.28$\pm$0.10 & 6.37$\pm$0.13 & 34\% & \\ 
                             & 84 & 1.04$\pm$0.06 & 6.62$\pm$0.11 & 1.33$\pm$0.11 & 6.30$\pm$0.12 & 33\% & 1 \\
                             & 34 & 1.21$\pm$0.19 & 6.47$\pm$0.26 & 1.37$\pm$0.16 & 6.27$\pm$0.20 & 42\% & 2 \\ 
                             & 33 & 1.41$\pm$0.17 & 6.19$\pm$0.22 & 1.52$\pm$0.20 & 6.08$\pm$0.25 & 40\% & 1,2 \\
\enddata
\tablenotetext{a}{Given in percent of the measurement value, $\mu{}$ and $M_{\rm BH}$ (Rev); symmetric in dependent and independent variable.}
\tablecomments{1 -- Object NGC\,4051 is excluded.  2 --  Weighted means. 
There are 34 entries because of the 27 
objects five have each an additional average entry and NGC 5548 has two additional average entries 
that are based on average monitoring data that are not included in 
the weighted means; see \S 3.3 for details.}
\tablecomments{
BCES results are those 
obtained from the bootstrapping pertaining to the bisector as they are 
considered less sensitive toward outliers (a few thousand realizations are made
in each fitting); the bootstrapping method tends to yield larger errors and 
sometimes steeper slopes than the formal BCES results, owing mostly to the
position of the NGC\,4051 data point. See text for details.
For each UV unscaled mass $\mu$ in column 1 four rows are given:
the first two rows are the results of fitting all the multiple data sets for
each object (including and excluding NGC 4051, see column 8), and the next 
two rows are the results where all data sets per object (excluding 
measurements based on mean monitoring spectra) were averaged, again excluding 
NGC 4051 in the second row as noted in column 8. 
}
\end{deluxetable}

\begin{deluxetable}{lcccc}
\tablecolumns{5}
\tablewidth{450pt}
\tablecaption{Mass Scaling Relationships --- Zeropoints
\label{zeropt.tab}}
\tablehead{
\colhead{} &
\colhead{} &
\colhead{Average Zeropoint} &
\colhead{Std. Dev.\ } &
\colhead{} \\
\colhead{Unscaled Mass Estimate} &
\colhead{N} &
\colhead{(dex)} &
\colhead{(dex)} &
\colhead{Note} \\
\colhead{(1)} &
\colhead{(2)} &
\colhead{(3)} &
\colhead{(4)} &
\colhead{(5)} 
}
\startdata
$\mu$[FWHM(\hb),L(5100\AA)]     & 34 & 6.907 $\pm$ 0.024 & 0.43 & \\
$\mu$[FWHM(\hb),L(\hb)]         & 28 & 6.674 $\pm$ 0.026 & 0.43 & \\
$\mu$[FWHM(CIV),L(1350\AA)]     & 85 & 6.659 $\pm$ 0.011 & 0.36 & \\
                             & 34 & 6.691 $\pm$ 0.012 & 0.43 & Weighted mean\\
$\mu$[$\sigma$(CIV),L(1350\AA)] & 85 & 6.726 $\pm$ 0.013 & 0.33 & \\
                             & 34 & 6.726 $\pm$ 0.015 & 0.37 & Weighted mean\\
\enddata
\tablecomments{For each UV unscaled mass $\mu$ in column 1 two rows are given:
first row is the fit results where multiple data sets were used for
each object, and the second row is the fit results where all data
sets per object (excluding measurements based on mean monitoring spectra)
were averaged.  See text for details.}
\end{deluxetable}

\begin{deluxetable}{lcccccc}
\tablecolumns{7}
\tablewidth{0pt}
\tablecaption{Probabilities of Mass Estimate Accuracies \label{probabilities.tab}}
\tabletypesize{\footnotesize}
\tablehead{
\colhead{} &
\multicolumn{5}{c}{\underline{\phantom{mmmmmmmmmmmmmm}Relative Accuracy\phantom{mmmmmmmmmmmmmm}}} &
\colhead{Absolute} \\
\colhead{Calibration } &
\colhead{Factor of 3} &
\colhead{Factor of 6} &
\colhead{Factor of 10} &
\colhead{1\,$\sigma$} &
\colhead{2\,$\sigma$} &
\colhead{Accuracy} \\
\colhead{} &
\colhead{(0.5 dex)} &
\colhead{(0.78 dex)} &
\colhead{(1.0 dex)} &
\colhead{(68\%)} &
\colhead{(95\%)} &
\colhead{Estimate\tablenotemark{a}} \\ 
\colhead{(1)} &
\colhead{(2)} &
\colhead{(3)} &
\colhead{(4)} &
\colhead{(5)} &
\colhead{(6)} &
\colhead{(7)} \\
}

\tablecolumns{7}
\startdata
FWHM(H$\beta$),L(5100\AA) & 22/34 $\approx$ 65\% & 34/34 = 100\% & 34/34 = 100\% & 0.52\,dex & 0.73\,dex & 0.70\,dex\\[5pt]
FWHM(H$\beta$),L(H$\beta$) & 20/28 $\approx$ 70\% & 28/28 = 100\% & 28/28 = 100\%  & 0.47\,dex & 0.70\,dex & 0.66\,dex \\[5pt]
FWHM(CIV), L(1350\AA) & 70/85 $\approx$ 82\% & 82/85 $\approx$ 96\% & 85/85 = 100\% & 0.32\,dex & 0.67\,dex & 0.56\,dex \\[5pt]
$\sigma$(CIV), L(1350\AA) & 73/85 $\approx$ 86\% & 83/85 $\approx$ 98\% & 85/85 = 100\% & 0.32\,dex & 0.62\,dex & 0.56\,dex \\[5pt]
\enddata
\tablenotetext{a}{The estimated upper limit in the absolute accuracy of the reverberation mapping masses
of a factor 2.9 (see text) is included here to provide an estimate of the absolute statistical uncertainty
of the single-epoch mass estimates.}
\end{deluxetable}

\begin{deluxetable}{lllll}
\tablewidth{300pt}
\tablecolumns{5}
\tabletypesize{\footnotesize}
\tablecaption{Modified Bright Quasar Survey Sample\tablenotemark{a}
\label{LaorModSample.tab}}
\tablehead{
\multicolumn{5}{c}{Objects} 
}
\startdata
0003$+$158 & 
0007$+$106 & 
0049$+$171 &
0052$+$251 & 
0804$+$761 \\
0838$+$770\tablenotemark{b} & 
0844$+$349 & 
0923$+$201\tablenotemark{b} & 
0947$+$396 & 
0953$+$414 \\
1049$-$006 & 
1100$+$772 & 
1103$-$006 & 
1116$+$215 & 
1121$+$422 \\
1149$-$110 & 
1151$+$117 & 
1216$+$069 & 
1226$+$023 & 
1229$+$204 \\
1259$+$593 & 
1302$-$102 & 
1307$+$085 & 
1309$+$355 & 
1310$-$108 \\
1322$+$659 & 
1352$+$183 & 
1415$+$451 & 
1416$-$129 & 
1425$+$267\tablenotemark{b} \\
1426$+$015 & 
1427$+$480 & 
1435$-$067 & 
1444$+$407 & 
1501$+$106 \\
1512$+$370 & 
1519$+$226 & 
1534$+$580 & 
1545$+$210 & 
1612$+$261 \\
1613$+$658 & 
1617$+$175\tablenotemark{b} & 
1626$+$554 & 
2112$+$059 & 
2130$+$099 \\
2308$+$098 & \\
\enddata
\tablenotetext{a}{This is the modified sample of PG
quasars, revisited in Appendix~\ref{Laor.app}, which
is considered to have more robust restframe 
UV spectral measurements.  }
\tablenotetext{b}{Borderline \iue{} quality data or
borderline absorption in \civ{} profile. These object
entries are flagged in Figure~\ref{Laorf2.fig}.
}
\end{deluxetable}

\begin{deluxetable}{llcccl} 
\vspace{-15cm}
\tablewidth{0pt}
\hspace{-14cm}
\tablecaption{Mass Estimates of Boroson \& Green PG Quasar Sample \label{BG92_ML.tab}}
\tabletypesize{\scriptsize}
\tablehead{
\colhead{ Object} &
\colhead{ } &
\colhead{ FWHM(H$\beta$)\tablenotemark{a}} &
\colhead{ log\,[ $\lambda L_{\lambda}$\tablenotemark{b}} &
\colhead{ log\,[ $M/M_{\odot}$ ]\tablenotemark{c}} &
\colhead{ } \\
\colhead{ } &
\colhead{ $z$} &
\colhead{ (km s$^{-1}$)} &
\colhead{ /ergs s$^{-1}${} ] } & 
\colhead{ (H$\beta$,$L_{\lambda}$,SE)} \\ 
\colhead{(1)} &
\colhead{(2)} &
\colhead{(3)} &
\colhead{(4)} &
\colhead{(5)} 
}
\tablecolumns{6}
\startdata
%
%
  PG\,0003$+$158  &  0.45000  &   4750.7  &   46.018$^{+0.033}_{-0.036}$  &   9.270$^{+0.088}_{-0.110}$ \\
  PG\,0007$+$106  &  0.08900  &   5084.6  &   44.816$^{+0.014}_{-0.015}$  &   8.728$^{+0.081}_{-0.099}$ \\
  PG\,0043$+$039  &  0.38400  &   5290.8  &   45.537$^{+0.030}_{-0.032}$  &   9.123$^{+0.085}_{-0.105}$ \\
  PG\,0049$+$171  &  0.06400  &   5234.3  &   44.004$^{+0.011}_{-0.011}$  &   8.347$^{+0.079}_{-0.097}$ \\
  PG\,0050$+$124  &  0.06100  &   1171.4  &   44.794$^{+0.097}_{-0.126}$  &   7.441$^{+0.093}_{-0.119}$ \\
  PG\,0157$+$001  &  0.16400  &   2431.9  &   44.975$^{+0.017}_{-0.018}$  &   8.166$^{+0.081}_{-0.100}$ \\
  PG\,0838$+$770  &  0.13100  &   2763.8  &   44.727$^{+0.015}_{-0.015}$  &   8.154$^{+0.080}_{-0.099}$ \\
  PG\,0923$+$129  &  0.02900  &   7598.4  &   43.860$^{+0.097}_{-0.125}$  &   8.598$^{+0.092}_{-0.117}$ \\
  PG\,0923$+$201  &  0.19000  &   1956.7  &   45.038$^{+0.018}_{-0.019}$  &   8.009$^{+0.082}_{-0.101}$ \\
  PG\,0934$+$013  &  0.05000  &   1254.3  &   43.875$^{+0.097}_{-0.126}$  &   7.041$^{+0.092}_{-0.117}$ \\
  PG\,0947$+$396  &  0.20600  &   4816.7  &   44.808$^{+0.020}_{-0.021}$  &   8.677$^{+0.081}_{-0.100}$ \\
  PG\,1001$+$054  &  0.16100  &   1699.8  &   44.741$^{+0.017}_{-0.017}$  &   7.738$^{+0.081}_{-0.099}$ \\
  PG\,1004$+$130  &  0.24000  &   6290.4  &   45.536$^{+0.022}_{-0.023}$  &   9.272$^{+0.084}_{-0.104}$ \\
  PG\,1011$-$040  &  0.05800  &   1381.0  &   44.259$^{+0.012}_{-0.012}$  &   7.317$^{+0.079}_{-0.097}$ \\
  PG\,1012$+$008  &  0.18500  &   2614.7  &   45.011$^{+0.021}_{-0.022}$  &   8.247$^{+0.082}_{-0.101}$ \\
  PG\,1022$+$519  &  0.04500  &   1566.4  &   43.696$^{+0.097}_{-0.126}$  &   7.145$^{+0.092}_{-0.117}$ \\
  PG\,1048$-$090  &  0.34400  &   5610.8  &   45.596$^{+0.027}_{-0.029}$  &   9.203$^{+0.085}_{-0.105}$ \\
  PG\,1048$+$342  &  0.16700  &   3580.9  &   44.708$^{+0.018}_{-0.019}$  &   8.369$^{+0.081}_{-0.099}$ \\
  PG\,1049$-$005  &  0.35700  &   5350.6  &   45.633$^{+0.028}_{-0.030}$  &   9.180$^{+0.085}_{-0.106}$ \\
  PG\,1100$+$772  &  0.31300  &   6151.2  &   45.575$^{+0.026}_{-0.027}$  &   9.272$^{+0.085}_{-0.105}$ \\
  PG\,1103$-$006  &  0.42500  &   6182.6  &   45.667$^{+0.033}_{-0.036}$  &   9.323$^{+0.086}_{-0.107}$ \\
  PG\,1114$+$445  &  0.14400  &   4554.4  &   44.734$^{+0.017}_{-0.017}$  &   8.591$^{+0.081}_{-0.099}$ \\
  PG\,1115$+$407  &  0.15400  &   1678.8  &   44.619$^{+0.017}_{-0.018}$  &   7.667$^{+0.080}_{-0.099}$ \\
  PG\,1116$+$215  &  0.17700  &   2896.9  &   45.397$^{+0.018}_{-0.019}$  &   8.529$^{+0.083}_{-0.103}$ \\
  PG\,1119$+$120  &  0.04900  &   1772.9  &   44.132$^{+0.012}_{-0.012}$  &   7.470$^{+0.079}_{-0.097}$ \\
  PG\,1121$+$422  &  0.23400  &   2192.3  &   44.883$^{+0.022}_{-0.023}$  &   8.030$^{+0.081}_{-0.100}$ \\
  PG\,1126$-$041  &  0.06000  &   2111.1  &   44.385$^{+0.012}_{-0.012}$  &   7.749$^{+0.080}_{-0.098}$ \\
  PG\,1149$-$110  &  0.04900  &   3032.2  &   44.107$^{+0.097}_{-0.126}$  &   7.924$^{+0.092}_{-0.117}$ \\
  PG\,1151$+$117  &  0.17600  &   4284.3  &   44.756$^{+0.020}_{-0.021}$  &   8.549$^{+0.081}_{-0.099}$ \\
  PG\,1202$+$281  &  0.16500  &   5036.4  &   44.601$^{+0.027}_{-0.029}$  &   8.612$^{+0.081}_{-0.099}$ \\
  PG\,1211$+$143  &  0.08500  &   1816.9  &   45.071$^{+0.014}_{-0.014}$  &   7.961$^{+0.082}_{-0.101}$ \\
  PG\,1216$+$069  &  0.33400  &   5179.9  &   45.721$^{+0.027}_{-0.028}$  &   9.196$^{+0.085}_{-0.106}$ \\
  PG\,1244$+$026  &  0.04800  &\phn720.6  &   43.801$^{+0.030}_{-0.032}$  &   6.523$^{+0.080}_{-0.099}$ \\
  PG\,1259$+$593  &  0.47200  &   3377.3  &   45.906$^{+0.034}_{-0.037}$  &   8.917$^{+0.087}_{-0.109}$ \\
  PG\,1302$-$102  &  0.28600  &   3383.4  &   45.827$^{+0.024}_{-0.026}$  &   8.879$^{+0.086}_{-0.107}$ \\
  PG\,1309$+$355  &  0.18400  &   2917.3  &   45.014$^{+0.019}_{-0.020}$  &   8.344$^{+0.082}_{-0.100}$ \\
  PG\,1310$-$108  &  0.03500  &   3606.0  &   43.725$^{+0.010}_{-0.011}$  &   7.884$^{+0.079}_{-0.097}$ \\
  PG\,1322$+$659  &  0.16800  &   2765.4  &   44.980$^{+0.098}_{-0.126}$  &   8.281$^{+0.094}_{-0.120}$ \\
  PG\,1341$+$258  &  0.08700  &   3013.9  &   44.344$^{+0.097}_{-0.126}$  &   8.037$^{+0.092}_{-0.117}$ \\
  PG\,1351$+$236  &  0.05500  &   6527.2  &   44.048$^{+0.011}_{-0.011}$  &   8.560$^{+0.079}_{-0.097}$ \\
  PG\,1351$+$640  &  0.08700  &   5646.1  &   44.835$^{+0.014}_{-0.015}$  &   8.828$^{+0.081}_{-0.099}$ \\
  PG\,1352$+$183  &  0.15800  &   3580.6  &   44.816$^{+0.017}_{-0.017}$  &   8.423$^{+0.081}_{-0.099}$ \\
  PG\,1354$+$213  &  0.30000  &   4126.7  &   44.977$^{+0.072}_{-0.086}$  &   8.627$^{+0.088}_{-0.110}$ \\
  PG\,1402$+$261  &  0.16400  &   1873.7  &   44.983$^{+0.017}_{-0.018}$  &   7.944$^{+0.081}_{-0.100}$ \\
  PG\,1404$+$226  &  0.09800  &\phn787.3  &   44.379$^{+0.017}_{-0.018}$  &   6.889$^{+0.080}_{-0.098}$ \\
  PG\,1415$+$451  &  0.11400  &   2591.2  &   44.561$^{+0.017}_{-0.018}$  &   8.014$^{+0.080}_{-0.098}$ \\
  PG\,1416$-$129  &  0.12900  &   6098.0  &   45.135$^{+0.037}_{-0.041}$  &   9.045$^{+0.083}_{-0.103}$ \\
  PG\,1425$+$267  &  0.36600  &   9404.7  &   45.761$^{+0.100}_{-0.130}$  &   9.734$^{+0.097}_{-0.126}$ \\
  PG\,1427$+$480  &  0.22100  &   2515.3  &   44.759$^{+0.021}_{-0.022}$  &   8.088$^{+0.081}_{-0.099}$ \\
  PG\,1435$-$067  &  0.12900  &   3156.9  &   44.918$^{+0.036}_{-0.040}$  &   8.365$^{+0.082}_{-0.102}$ \\
  PG\,1440$+$356  &  0.07700  &   1393.5  &   44.546$^{+0.014}_{-0.014}$  &   7.468$^{+0.080}_{-0.098}$ \\
  PG\,1444$+$407  &  0.26700  &   2456.5  &   45.203$^{+0.023}_{-0.024}$  &   8.289$^{+0.083}_{-0.102}$ \\
  PG\,1448$+$273  &  0.06500  &\phn814.7  &   44.482$^{+0.011}_{-0.011}$  &   6.970$^{+0.080}_{-0.098}$ \\
  PG\,1501$+$106  &  0.03600  &   5454.1  &   44.285$^{+0.010}_{-0.011}$  &   8.523$^{+0.079}_{-0.097}$ \\
  PG\,1512$+$370  &  0.37100  &   6802.7  &   45.602$^{+0.030}_{-0.032}$  &   9.373$^{+0.085}_{-0.106}$ \\
  PG\,1519$+$226  &  0.13700  &   2187.3  &   44.710$^{+0.019}_{-0.020}$  &   7.942$^{+0.081}_{-0.099}$ \\
  PG\,1534$+$580  &  0.03000  &   5323.5  &   43.687$^{+0.010}_{-0.011}$  &   8.203$^{+0.080}_{-0.097}$ \\
  PG\,1535$+$547  &  0.03800  &   1420.4  &   43.961$^{+0.010}_{-0.011}$  &   7.192$^{+0.079}_{-0.097}$ \\
  PG\,1543$+$489  &  0.40000  &   1529.2  &   45.445$^{+0.037}_{-0.041}$  &   7.998$^{+0.085}_{-0.105}$ \\
  PG\,1545$+$210  &  0.26600  &   7021.7  &   45.428$^{+0.023}_{-0.024}$  &   9.314$^{+0.084}_{-0.104}$ \\
  PG\,1552$+$085  &  0.11900  &   1377.0  &   44.704$^{+0.015}_{-0.015}$  &   7.537$^{+0.080}_{-0.099}$ \\
  PG\,1612$+$261  &  0.13100  &   2490.9  &   44.717$^{+0.026}_{-0.028}$  &   8.058$^{+0.081}_{-0.100}$ \\
  PG\,1626$+$554  &  0.13300  &   4473.8  &   44.580$^{+0.026}_{-0.028}$  &   8.498$^{+0.081}_{-0.099}$ \\
  PG\,1704$+$608  &  0.37100  &   6552.4  &   45.702$^{+0.030}_{-0.032}$  &   9.391$^{+0.086}_{-0.107}$ \\
  PG\,2112$+$059  &  0.46600  &   3176.4  &   46.181$^{+0.034}_{-0.037}$  &   9.001$^{+0.089}_{-0.112}$ \\
  PG\,2209$+$184  &  0.07000  &   6487.5  &   44.469$^{+0.012}_{-0.013}$  &   8.766$^{+0.080}_{-0.098}$ \\
  PG\,2214$+$139  &  0.06700  &   4532.0  &   44.662$^{+0.097}_{-0.126}$  &   8.551$^{+0.093}_{-0.118}$ \\
  PG\,2233$+$134  &  0.32500  &   1709.2  &   45.327$^{+0.027}_{-0.028}$  &   8.036$^{+0.083}_{-0.103}$ \\
  PG\,2251$+$113  &  0.32300  &   4147.2  &   45.692$^{+0.026}_{-0.028}$  &   8.989$^{+0.085}_{-0.106}$ \\
  PG\,2304$+$042  &  0.04200  &   6486.8  &   44.066$^{+0.097}_{-0.126}$  &   8.564$^{+0.092}_{-0.117}$ \\
  PG\,2308$+$098  &  0.43200  &   7914.3  &   45.777$^{+0.101}_{-0.131}$  &   9.592$^{+0.098}_{-0.126}$ \\
%
\enddata
\\[-8pt]
\tablenotetext{a}{FWHM(\hb) measured in the single-epoch spectrum in units 
of km s$^{-1}$.
Values are adopted from Boroson \& Green (1992) and are corrected
for spectral resolution as described in the text.}
\tablenotetext{b}{The luminosities at 5100\AA{} were computed based on the spectrophotometry
of Neugebauer \et (1987) and Schmidt \& Green (1983) as explained in Paper~I.} 
\tablenotetext{c}{The central mass (and uncertainties) 
estimated based on single-epoch optical
spectroscopy and the calibrated relationship of eq.\ (5).}
\tablecomments{Optical parameters and single-epoch estimates of the central black 
hole in the PG quasars studied by Boroson \& Green (1992) without robust mass measurements 
based on reverberation mapping techniques.} 
\vspace{-0.2cm}
\end{deluxetable}


\begin{thebibliography}{}{}
\bibitem[Akritas \& Bershady(1996)]{1996ApJ...470..706A} Akritas, M.\ G.\ 
	\& Bershady, M.\ A.\ 1996, \apj, 470, 706
\bibitem[Bachev et al.(2004)]{2004ApJ...617..171B} Bachev, R., Marziani, 
P., Sulentic, J.~W., Zamanov, R., Calvani, M., \& Dultzin-Hacyan, D.\ 2004, 
\apj, 617, 171

\bibitem[Baskin \& Laor(2005)]{2005MNRAS.356.1029B} Baskin, A., \& Laor, 
A.\ 2005, \mnras, 356, 1029
\bibitem []{} Bentz, M. C., Peterson, B. M., 
Pogge, R. W., Vestergaard, M., \& Onken, C. A. 2006, ApJ, submitted
\bibitem []{} Boroson, T. A., \& Green, R. F. 1992, \apjs, 80, 109 
\bibitem[Boroson(2003)]{2003ApJ...585..647B} Boroson, T.~A.\ 2003, \apj, 585, 647
\bibitem []{} Cardelli, J. A., Clayton, G. C., Mathis, J. S., 1989, \apj, 345, 245
\bibitem []{} Clavel, J., et al., 1991, ApJ, 366, 64
\bibitem[Crenshaw et al.(2001)]{2001ApJ...555..633C} Crenshaw, D.~M., 
Kraemer, S.~B., Bruhweiler, F.~C., \& Ruiz, J.~R.\ 2001, \apj, 555, 633 
\bibitem[Collinge et al.(2001)]{2001ApJ...557....2C} Collinge, M.~J., et 
al.\ 2001, \apj, 557, 2 
\bibitem[Dunlop(2004)]{2004cbhg.symp..342D} Dunlop, J.~S.\ 2004, 
Coevolution of Black Holes and Galaxies, ed. L. C. Ho, (Pasadena: Carnegie Observatories), 342
\bibitem[Ferrarese et al.(2001)]{2001ApJ...555L..79F} Ferrarese, L., Pogge, 
	R.~W., Peterson, B.~M., Merritt, D., Wandel, A., \& Joseph, C.~L.\ 2001, 
	\apjl, 555, L79
\bibitem[Gaskell(1982)]{1982ApJ...263...79G} Gaskell, C.~M.\ 1982, \apj, 263, 79
\bibitem []{} Gebhardt, K., \et 2000, \apj, 543, L5
\bibitem []{} Jester, S., \et 2005, AJ, 130, 873
\bibitem []{} Kaspi, S., Smith, P. S., Netzer, H., Maoz, D., Jannuzi, B. T., \&
	Giveon, U. 2000, \apj, 533, 631
\bibitem []{} Kaspi, S., Maoz, D., Netzer, H., Peterson, B. M., Vestergaard, M., \& Jannuzi, B. T.
	2005, \apj, 629, 61
\bibitem []{} Kellermann, K. I., Sramek, R., Schmidt, M., Shaffer, D. B., \& 
	Green, R. 1989, \aj, 98, 1195
\bibitem[]{1336} Kollatschny, W. 2003, A\&A, 407, 461

\bibitem[Korista et al.(1995)]{1995ApJS...97..285K} Korista, K.\ T., et
        al.\ 1995, \apjs, 97, 285
\bibitem[Laor(2000)]{2000ApJ...543L.111L} Laor, A.\ 2000, \apjl, 543, L111
\bibitem[Leighly(2000)]{2000NewAR..44..395L} Leighly, K.~M.\ 2000, New Astronomy Review,
        44, 395

\bibitem[]{1344}Maoz, D. 2002, astro-ph/0207295

\bibitem[Marziani et al.(1996)]{1996ApJS..104...37M} Marziani, P., 
Sulentic, J.~W., Dultzin-Hacyan, D., Calvani, M., \& Moles, M.\ 1996, 
\apjs, 104, 37
\bibitem[Marziani et al.(2003)]{2003ApJS..145..199M} Marziani, P., 
Sulentic, J.~W., Zamanov, R., Calvani, M., Dultzin-Hacyan, D., Bachev, R., 
\& Zwitter, T.\ 2003, \apjs, 145, 199

\bibitem[]{1354} McLure, R.J., \& Jarvis, M.J. 2002, MNRAS, 337, 109

\bibitem[Miyoshi et al.(1995)]{1356} Miyoshi, M., Moran, J., Herrnstein, J.,
Greenhill, L., Nakai, N., Diamond, P., Inoue, M. 1995,
Nature, 373, 127

\bibitem[]{1364} Nelson, C.H., Green, R.F., Bower, G., Gebhardt, K., \&
Weistrop, D. 2004, ApJ, 615, 652

\bibitem []{} Neugebauer, G., Green, R. F., Matthews, K., Schmidt, M., Soifer, 
	B. T., \& Bennett, J. 1987, \apjs, 63, 615
\bibitem[O'Brien et al.(1998)]{1998ApJ...509..163O} O'Brien, P.~T., et al.\ 
1998, \apj, 509, 163
\bibitem[Onken et al.(2004)]{2004ApJ...615..645O} Onken, C.~A., Ferrarese, 
L., Merritt, D., Peterson, B.~M., Pogge, R.~W., Vestergaard, M., \& Wandel, 
A.\ 2004, \apj, 615, 645

\bibitem[]{1375} Onken, C.A., \& Peterson, B.M. 2002, ApJ, 572, 746

\bibitem []{} Peterson, B. M., \& Wandel, A. 1999, \apj, 521, L95
\bibitem []{} Peterson, B. M., \& Wandel, A. 2000, \apj, 540, L13
\bibitem []{} Peterson, B. M. \et 2004, \apj, 613, 682
\bibitem []{} Peterson, B. M. \et 2005, \apj, 632, 799.
Erratum in press.
\bibitem[Press et al.(1992)]{1992nrfa.book.....P} Press, W.~H., Teukolsky, 
S.~A., Vetterling, W.~T., \& Flannery, B.~P.\ 1992, Numerical Recipes (2nd Ed.; Cambridge: University 
Press)
\bibitem[]{1390} Reichert, G.A., et al. (1994), ApJ, 425, 582
\bibitem[Richards et al.(2002)]{2002AJ....124....1R} Richards, G.~T., 
Vanden Berk, D.~E., Reichard, T.~A., Hall, P.~B., Schneider, D.~P., 
SubbaRao, M., Thakar, A.~R., \& York, D.~G.\ 2002, \aj, 124, 1
\bibitem []{} Rodriguez-Pascual, P.M., et al. 1997, ApJS, 110, 9
\bibitem []{} Schmidt, M., \& Green, R. F. 1983, \apj, 269, 352
\bibitem[Schmidt et al.(1995)]{1995AJ....110...68S} Schmidt, M., Schneider, 
   D.~P., \& Gunn, J.~E.\ 1995, \aj, 110, 68
\bibitem[Shemmer et al.(2004)]{2004ApJ...614..547S} Shemmer, O., Netzer, 
  H., Maiolino, R., Oliva, E., Croom, S., Corbett, E., \& di Fabrizio, L.\ 
  2004, \apj, 614, 547 
\bibitem[Tytler \& Fan(1992)]{1992ApJS...79....1T} Tytler, D., \& Fan, X.\ 
1992, \apjs, 79, 1
\bibitem[Vestergaard(2002)]{2002ApJ...571..733V} Vestergaard, M.\ 2002, \apj, 571, 733
(Paper I)
\bibitem[Vestergaard(2004)]{2004ApJ...601..676V} Vestergaard, M.\ 2004, \apj, 601, 676
\bibitem[Vestergaard \& Wilkes(2001)]{2001ApJS..134....1V} Vestergaard, M.,
        \& Wilkes, B.\ J.\ 2001, \apjs, 134, 1
\bibitem []{} Wandel, A., Peterson, B. M., \& Malkan, M. A. 1999, \apj, 526, 579
\bibitem []{} Wanders, I., et al. 1997, ApJS, 113, 69
\bibitem[Warner et al.(2003)]{2003ApJ...596...72W} Warner, C., Hamann, F., 
\& Dietrich, M.\ 2003, \apj, 596, 72
\bibitem[Wilkes(1984)]{1984MNRAS.207...73W} Wilkes, B.~J.\ 1984, \mnras, 
207, 73
\bibitem[Wills et al.(1993)]{1993ApJ...410..534W} Wills, B.~J., Netzer, H., 
Brotherton, M.~S., Han, M., Wills, D., Baldwin, J.~A., Ferland, G.~J., \& 
Browne, I.~W.~A.\ 1993, \apj, 410, 534
\bibitem[Wu et al.(2004)]{2004A&A...424..793W} Wu, X.-B., Wang, R., Kong, 
M.~Z., Liu, F.~K., \& Han, J.~L.\ 2004, \aap, 424, 793
\end{thebibliography}
\end{document}